# The Science Case for an Extended *Spitzer* Mission

Jennifer C. Yee[1], Giovanni G. Fazio[2], Robert Benjamin[3], J. Davy Kirkpatrick[4], Matt A. Malkan[5], David Trilling[6], Sean Carey[7], David R. Ciardi[8],
and
Dániel Apai[9], M. L. N. Ashby[2], Sarah Ballard[10], Jacob L. Bean[11], Thomas Beatty[12], Zach Berta-Thompson[13], P. Capak[14], David Charbonneau[2], Steven Chesley[15], Nicolas B. Cowan[16], Ian Crossfield[10], Michael C. Cushing[17], Julien de Wit[10], Drake Deming[18], M. Dickinson[19], Jason Dittmann[10], Diana Dragomir[10], Courtney Dressing[20], Joshua Emery[21], Jacqueline K. Faherty[22], Jonathan Gagné[23], B. Scott Gaudi[24], Michael Gillon[25], Carl J. Grillmair[7], Alan Harris[26], Joseph Hora[2], James G. Ingalls[7], Tiffany Kataria[15], Laura Kreidberg[2], Jessica E. Krick[7], Patrick J. Lowrance[7], William A. Mahoney[7], Stanimir A. Metchev[27], Michael Mommert[6], Michael "Migo" Mueller[28], Yossi Shvartzvald[15], Howard Smith[2], Kevin B. Stevenson[29], H. I. Teplitz[4], and S. P. Willner[2]

[1]Smithsonian Astrophysical Observatory
[2]Harvard-Smithsonian Center for Astrophysics
[3]University of Wisconsin-Whitewater
[4]Caltech/IPAC
[5]University of California Los Angeles
[6]Northern Arizona University
[7]Caltech/IPAC-Spitzer
[8]Caltech/IPAC-NExScI
[9]The University of Arizona
[10]Massachusetts Institute of Technology
[11]University of Chicago
[12]The Pennsylvania State University
[13]University of Colorado/Boulder
[14]Caltech/Spitzer Science Center
[15]Jet Propulsion Laboratory
[16]McGill University
[17]The University of Toledo
[18]University of Maryland at College Park
[19]National Optical Astronomy Observatory
[20]University of California Berkeley
[21]University of Tennessee
[22]American Museum of Natural History
[23]Carnegie Institution of Washington DTM
[24]The Ohio State University
[25]Université de Liège
[26]DLR
[27]The University of Western Ontario
[28]NOVA / Rijksuniversiteit Groningen / SRON

[29]Space Telescope Science Institute

October 11, 2017

# Contents

# Section 1 – Executive Summary

Robert Benjamin[1], David R. Ciardi[2], Giovanni G. Fazio[3], J. Davy Kirkpatrick[4], Matt A. Malkan[5], David Trilling[6], and Jennifer C. Yee[7]

[1]University of Wisconsin-Whitewater
[2]Caltech/IPAC-NExScI
[3]Harvard-Smithsonian Center for Astrophysics
[4]IPAC
[5]University of California Los Angeles
[6]Northern Arizona University
[7]Smithsonian Astrophysical Observatory

October 11, 2017

# 1 *Spitzer* Extended Mission

Although the final observations of the *Spitzer* Warm Mission are currently scheduled for March 2019, it can continue operations through the end of the decade with no loss of photometric precision. As we will show, there is a strong science case for extending the current Warm Mission to December 2020. *Spitzer* has already made major impacts in the fields of exoplanets (including microlensing events), characterizing near Earth objects, enhancing our knowledge of nearby stars and brown dwarfs, understanding the properties and structure of our Milky Way galaxy, and deep wide-field extragalactic surveys to study galaxy birth and evolution. By extending *Spitzer* through 2020, it can continue to make ground-breaking discoveries in those fields, and provide crucial support to the NASA flagship missions *JWST* and *WFIRST*, as well as the upcoming *TESS* mission, and it will complement ground-based observations by LSST and the new large telescopes of the next decade. This scientific program addresses NASA's Science Mission Directive's objectives in astrophysics, which include discovering how the universe works, exploring how it began and evolved, and searching for life on planets around other stars.

# 2 Characterizing and Discovering Exoplanets

Continued operations of *Spitzer* beyond the early-2019 shutdown will enable unique exoplanet science and support for NASA's ongoing and upcoming missions *K2*, *TESS*, *JWST*, and *WFIRST* that, simply put, may not be achievable without *Spitzer*. *Spitzer* enables characterization of planetary systems, especially those discovered by *K2* and *TESS*, beyond what is possible with current ground facilities and therefore also enables efficient and effective use of *JWST*. *Spitzer* is being used to refine the measured times of transits to high precision and at a time closer to observation with *JWST*. Without such timing measurements, many additional hours of *JWST* would be needed just to recover the transit, before the transits themselves could actually be observed. Further, *JWST* observations of the secondary eclipses of transiting planets will yield important information about the thermal structure of exoplanet atmospheres, but the times of secondary eclipses are unknown *a priori* because the orbital eccentricities are unknown. Spending a day of *Spitzer* to recover a planetary transit or constrain the time of the secondary eclipse could save many hours or even days of *JWST* time.

In addition to providing observational support for *JWST*, *Spitzer* is still performing critical science sooner and cheaper than is possible with *JWST*. *Spitzer* is enabling, and can continue to enable, studies of the exoplanet atmospheric structures sooner and at significantly less cost than with *JWST*. *Spitzer* provides a platform to observe for long periods of time, at great photometric precision, allowing us to search for and discover planets of special interest, such as those around TRAPPIST-1. Once discovered, *Spitzer* enables us to characterize the thermal properties of the planets in addition to determining accurate and precise orbital parameters, necessary for further detailed observations. Finally, the *Spitzer* microlensing parallax program is the most vibrant microlensing program in the US to study exoplanets. The mission and the community need to prepare for the microlensing experiment with *WFIRST*, and *Spitzer* uniquely provides the opportunity for such preparation so we can take full effective and efficient advantage of *WFIRST*.

# 3 Near Earth Objects

The *Spitzer* Space Telescope has been used to measure diameters and albedos (a proxy for composition) for nearly 2000 Near Earth Objects (NEOs) — about 15% of the known population. For most of these, partial lightcurves have also been measured, more than doubling the number of NEOs with good lightcurve constraints. A *Spitzer* Extended Mission could observe and categorize about $\sim$250 NEOs per year, and this number will increase by 20% per year or more with new NEO discoveries. At this rate, our catalog of characterized NEOs will continue to be 10% of the known population: physical characterization will keep up with discovery efforts, and the new observations would fill critical gaps in our understanding of small ($<$300 meters) NEOs. In addition, other newly discovered solar system targets of opportunity and interest (dead comets, etc.) can be readily incorporated into the baseline program. Our team of observing, modeling, and analysis experts has been working with *Spitzer* for its entire Warm Mission, and is fully prepared for a dedicated Extended Mission. *Spitzer* is the only existing facility that can make these measurements for a large number of small NEOs; the ability of *JWST* to observe NEOs will be limited by its maximum non-sidereal tracking rate of 30 milliarcsec per second, which NEOs will exceed when they are close to the Earth and *JWST*. The legacy value of these data is most important in the context of LSST, whose optical photometry and colors will be highly complementary to these *Spitzer* data and help provide a complete characterization of the observed NEOs.

# 4 Nearby Stars and Brown Dwarfs

The *Spitzer* mission has enabled huge advances in our knowledge of the solar neighborhood. Many of the most fundamental results have, in fact, taken place during the Warm Mission, once the extended photometric and astrometric capabilities of IRAC in its two short-wavelength channels were fully appreciated. These advances include measuring accurate distances to the nearest and coldest brown dwarfs, studying the atmospheric structure of gas giant exo-worlds, and identifying and characterizing an exo-solar system of seven earth-mass planets. Continuing the *Spitzer* warm mission through the end of the decade will allow other major advances in our knowledge of this sample: the discovery of additional nearby exo-Earths in the habitable zone, the continuous study of exoplanet-like cloud evolution and atmospheric dynamics, the establishment of the entire stellar-to-substellar mass function of the nearby census, a direct measurement of the low-mass cutoff of star formation using young moving groups, and a search for brown dwarfs even colder than the current $\sim$ 250K record holder. Each of these areas provides either new targets for critical *JWST* follow-up or enables prioritization of the current brown dwarf list so that *JWST* observations can be optimized.

# 5 Galactic Science with the *Spitzer* Extension

In the past decade, infrared all-sky and Galactic plane surveys, notably the *Spitzer* GLIMPSE and MIPSGAL programs, have been used to obtain a nearly complete census of over eight thousand Galactic HII regions and star formation complexes across the Milky Way Galaxy. Emission-line spectroscopy of associated molecular or ionized gas has been used to determine the distances and luminosity to a large and growing number of these objects; multiple groups find that a few dozen of these star formation complexes supply more than a third of the total ionizing luminosity of the Galaxy. Deep uniform follow-up observations of these particularly luminous complexes will provide detailed information on the star formation properties of objects that will be observed by *JWST* out to high redshift. In addition, the development of improved IRAC astrometry along with a decade-long baseline will allow for the measurement of proper motions of millions of Galactic plane sources that are not visible in optical wavelengths. This work will complement Gaia measurements of proper motions and parallax sources for foreground disk populations, constrain the mass density of the inner Milky Way, and prepare for astrometric measurements planned by *WFIRST*.

# 6 Birth and Evolution of Galaxies

One of IRACs unique capabilities, with no planned competition for the foreseeable future, is to carry out deep, wide extragalactic surveys at 3.6 and 4.5 $\mu$m. These bands have the unique power to measure the rest frame visible light of very distant galaxies and can thus yield redshift measurements via photometric techniques, and quantitatively determine stellar masses and possible AGN contributions. Extending *Spitzer*s lifetime would enable such surveys to go far beyond anything previously contemplated. The resulting data would be uniquely powerful for achieving several outstanding scientific goals, including: (1) understanding the formation and evolution of dusty star-forming galaxies from $1 < z < 4$ and massive galaxies throughout the entire epoch of cosmic reionization; (2) understanding the large-scale structure of galaxies, galaxy cluster formation at $z > 1.5$, and environmental effects on star formation; (3) understanding the origin of the cosmic infrared background radiation (CIBR) through observations of its spatial anisotropies and multi-wavelength correlations; and (4) identifying rare objects for *JWST* and ALMA (and eventually ELT) followups, including the coldest brown dwarfs and the first generation ($z > 7$) of luminous galaxies and quasars that re-ionized the Universe. For example, a baseline survey could encompass a hundred square degrees, with up to $\sim$1800 seconds per pixel to reach down to a sensitivity of $\sim$24 AB mag ($5\sigma$), which is several magnitudes deeper than *WISE*. This depth is an excellent complement to the deep new survey of near-IR imaging and spectroscopy to be made with *Euclid* and *WFIRST*. The legacy of these joint databases, would enable a tremendous range of frontier science.

# Overview

This document lays out the science cases for each topic in detail in Sections 2–6. A technical description of Spitzer's capabilities through 2020 is given in Section 7.

# Section 2 – *Spitzer* Exoplanet Science Beyond March 2019: Supporting NASA's Flagship Missions and Strategic Goals

David R. Ciardi[1], Jennifer C. Yee[2], Sarah Ballard[3], Jacob L. Bean[4], Thomas Beatty[5], Zach Berta-Thompson[6], David Charbonneau[7], Nicolas B. Cowan[8], Ian Crossfield[3], Drake Deming[9], Julien de Wit[10], Jason Dittmann[10], Diana Dragomir[3], Courtney Dressing[11], B. Scott Gaudi[12], Michael Gillon[13], Tiffany Kataria[14], Laura Kreidberg[7], Yossi Shvartzvald[14], and Kevin B. Stevenson[15]

[1]Caltech/IPAC-NExScI, Pasadena, CA USA
[2]Smithsonian Astrophysical Observatory, Cambridge, MA USA
[3]Dept. of Physics, Massachusetts Institute of Technology, Cambridge, MA USA
[4]Dept. of Astronomy & Astrophyiscs, University of Chicago, Chicago, IL USA
[5]Dept. of Astronomy & Astrophysics, The Pennsylvania State University, University Park, PA USA
[6]Dept. of Astrophysical and Planetary Sciences,University of Colorado, Boulder, Boulder, CO USA
[7]Harvard-Smithsonian Center for Astrophysics, Cambridge, MA, USA
[8]Dept. of Physics and Dept. of Earth & Planetary Sciences, McGill University, Montréal, QC Canada
[9]Dept. of Astronomy, University of Maryland at College Park, College Park, MD USA
[10]Dept. of Earth, Atmospheric, and Planetary Sciences, Massachusetts Institute of Technology, Cambridge, MA USA
[11]Dept. of Astronomy, University of California Berkeley, Berkeley, CA USA
[12]Dept. of Astronomy, The Ohio State University, Columbus, OH USA
[13]Space Sciences, Technologies and Astrophysics Research (STAR) Institute, Université de Liège, Liège, Belgium
[14]Jet Propulsion Laboratory, California Institute of Technology, Pasadena, CA USA
[15]Space Telescope Science Institute, Baltimore, MD USA

# 1 Introduction

The *Spitzer Space Telescope* has revolutionized our ability to characterize exoplanets through infrared observations during both primary transit and secondary eclipse events as well as through full phase curves and observations of microlensing parallax. These infrared observations have enabled us to begin the characterization of transiting super-Earth–sized and larger planets as part of our path to understanding the atmospheric composition of planets and the weather patterns associated with planets beyond our Solar System. In addition to the characterization planets, *Spitzer* has also been critical in the discovery of planets through long baseline observations enabling discovery of additional long period and very small transiting planets in systems where planets have been found from the ground or with *K2* through follow-up long-baseline observations. Additionally, *Spitzer* has been used to observe microlensing events identified from the ground to better characterize their planets. The unique combination of *Spitzer* and the ground observations will enable the first comparison of planets in the disk and the bulge in advance of *WFIRST*.

With the launch of the Transiting Exoplanet Survey Satellite (*TESS*) in 2018 and the impending discovery of thousands of transiting planets around bright nearby stars, NASA's next flagship mission the James Webb Space Telescope (*JWST*) is poised to be used to characterize extensively exoplanet atmospheres at levels that are currently unprecedented. Continued operation of *Spitzer* into the era of *TESS* and *JWST* will enable *Spitzer* to be used as a unique resource to characterize new systems around bright stars, discover additional planets beyond the sensitivity of *TESS*, and enable efficient and effective use of *JWST*.

Beyond *JWST*, NASA's next great exoplanet mission, the Wide-Field Infrared Survey Telescope (*WFIRST*), will utilize microlensing to explore exoplanetary systems beyond the snowline. *Spitzer* has been used to pioneer efforts to combine ground-based and space-based microlensing observations to characterize and discover new planets and measure their occurrence rate. *Spitzer* is playing a critical role in helping the community to prepare for the discoveries awaiting with *WFIRST*. *Spitzer* is uniquely able to determine the relative occurrence of planets in the disk and bulge environments of the galaxy and is setting the stage for *WFIRST*.

Finally, continued operation of *Spitzer* will enable exoplanet science not easily achievable with *JWST*, particularly when *Spitzer* data are combined with results from NASA's *TESS* and/or ESA's CHEOPS. Here we outline the exoplanet mission support and science that *Spitzer* could provide if extended beyond its scheduled early-2019 shutdown.

# 2 Preparing for and Supporting Exoplanet Characterization with *JWST* and *HST*

The James Webb Space Telescope (*JWST*), scheduled for launch in Spring 2019, offers dramatic new capabilities for characterizing exoplanets. With a suite of infrared spectrometers, *JWST* will enable highly detailed studies of exoplanet atmospheres. *JWST* will revolutionize our knowledge of the physical properties of dozens to possibly hundreds of exoplanets by making a variety of different types of observations – primarily of transiting exoplanets via transmission and emission spectroscopy.

However, *JWST* is a general purpose observatory with many equally compelling scientific endeavors competing for the same time as exoplanetary astrophysics. Thus, *JWST* must be used with great efficiency on only the exoplanet targets that the community believes have the best opportunity for returning significant scientific results. Preparatory observations will be crucial in making sure that the observations made with *JWST* are as effective and efficient as possible. *Spitzer* is currently playing a critical role in validating and characterizing targets discovered from the ground and with Kepler/K2 and can continue to play this unique and critical role long into the era of *JWST*. With the impending launch of *TESS*, a mission designed to find planets around bright stars suitable for observation with *JWST*, *Spitzer* is uniquely suited to make sure the targets for *JWST* are vetted and understood prior to observation with NASA's flagship mission.

*Spitzer*'s capability for precise, infrared time series observations that can span long time baselines makes it an ideal instrument for characterizing the atmospheres of exoplanets and discovering additional planets in known systems. In particular, *Spitzer* can provide strongly complementary observations to *JWST* in two key ways: vetting targets to guide the design of *JWST* observations and providing context for better interpreting *JWST* data.

Many of the exoplanets to be discovered by *K2* and *TESS* will be excellent candidates for *JWST* measurements, including atmospheric transmission spectra and dayside emission spectra. However, optimal planning of transmission and thermal emission measurements for newly discovered planets will be challenging because the planets' orbital eccentricities and global heat redistribution are not known. Long *Spitzer* time series observations will enable us to catch the secondary eclipse and pin down the orbital eccentricity, enabling precise timing for future observations. We will also obtain benchmark measurements of the planets' dayside temperatures, which are essential for predicting the signal-to-noise ratio for possible more detailed future observations with *JWST*. The dayside temperature is challenging to predict *a priori*, because it is dependent upon the radiative and advective processes in the atmosphere. *Spitzer* observations are needed to provide a first estimate of the expected thermal emission signal from the planet.

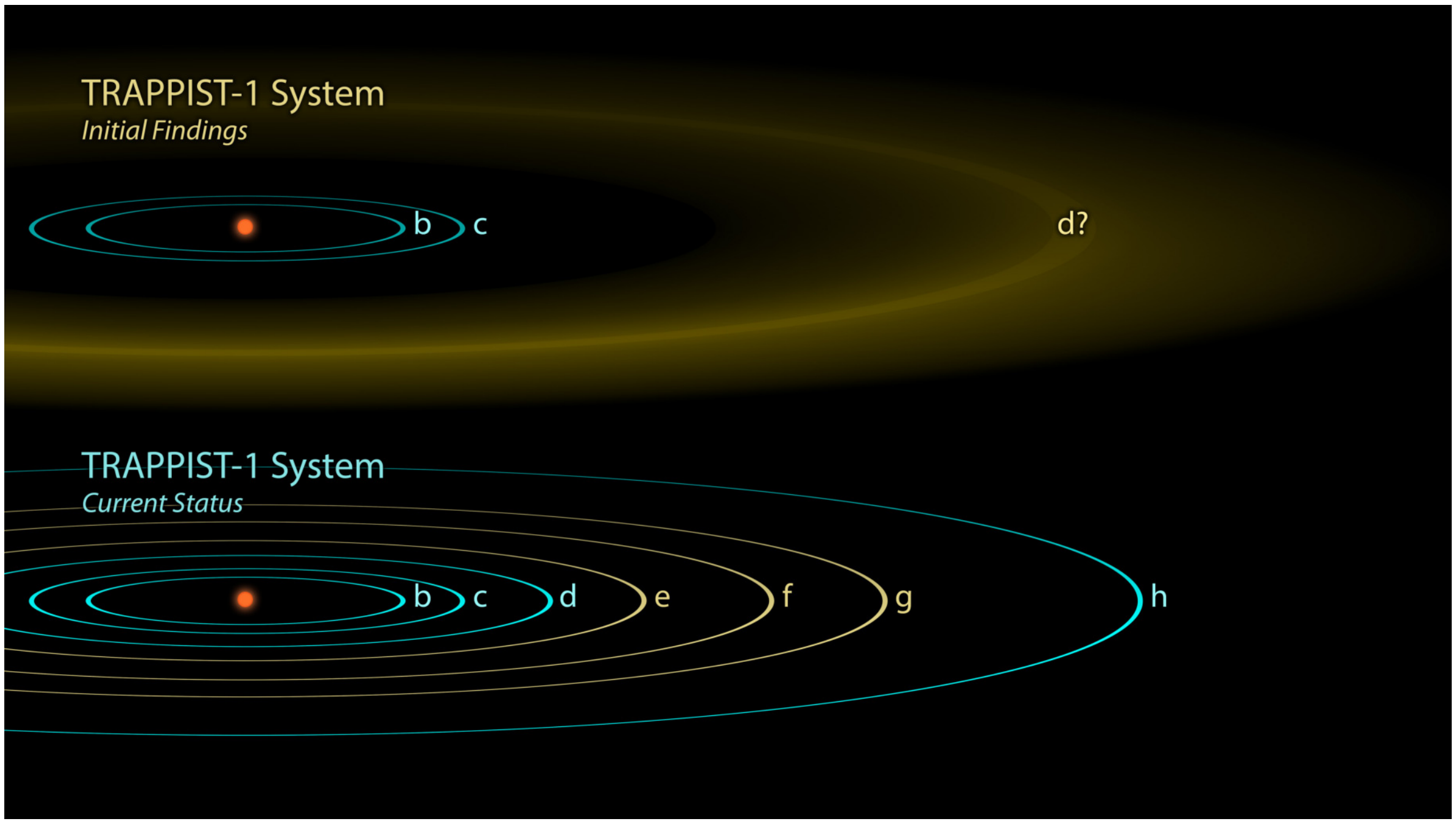

Figure 1: TRAPPIST-1 is an M-dwarf with 7 earth-sized planets. Ground-based observations discovered three planets (top) but it was with *Spitzer* that we realized that the richness of the system with 7 planets (bottom), 3 of which are orbiting within the habitable zone of the star. Credit: NASA/JPL/Caltech/R. Hurt (IPAC)

## 2.1 Discovering Targets for *JWST* and *HST*

### 2.1.1 *Spitzer*'s Discovery and Characterization of 7 Planets around TRAPPIST-1

The TRAPPIST-1 system was discovered from the ground and was initially thought to have 3 planets (Gillon et al. 2016), but follow-up observations of TRAPPIST-1 with *Spitzer* intended solely for the purpose of characterizing the third planet in the system discovered that there four other planets in the system for a total of 7 planets. That discovery, and the subsequent characterization of the 7 planets, was only possible because *Spitzer* was able to observe the TRAPPIST-1 system continuously for 21 days (500 hours).

The *Spitzer* observations showed that the TRAPPIST-1 system was significantly more interesting than anyone had realized. TRAPPIST-1d, the original planet discovered from the ground, would have been an interesting target for *JWST* because it is a small planet in the habitable zone of an M-star. But now we know, because of Spitzer, that there are 7 planets worthy of study – 4 of which are likely in the habitable zone (Gillon et al. 2017).

But more than that -– because *Spitzer* was able to study the system for such a long period of time, transit time variations were utilized to measure the masses of the planets, which are beyond the sensitivities of current radial-velocity instruments. The combination of the radii from the transit depths and the masses from the timing variations allow the determination of the bulk densities of the planets. As a result, we have a much more complete picture of the system and the individual planets within that system (Gillon et al. 2017; Luger et al. 2017, Grimm et al. submitted). In fact, the originally discovery had identified a "d" planet, but that discovery turned out to be glimpses of the "e", "f", and "g" planets and the actual "d" planet was significantly smaller and in a much shorter orbital period than previously thought. All of this was sorted out because the long time baseline observations afforded by *Spitzer*.

Since the initial discovery and monitoring by *Spitzer*, thousands of additional hours of *Spitzer* time are being used to monitor the TRAPPIST-1 system in even more detail. These observations will result in unprecedented constraints on its configuration and dynamic and on the planet masses. Masses known to better than 10% will inform us on their bulk compositions and volatile enrichment. In addition to finding these high-value *JWST* targets, their characterization with *Spitzer* contribute to maximizing the scientific return of *JWST* and its future atmospheric characterization programs (Morley et al., 2017).

Without *Spitzer*, we would never know just how fundamentally interesting and exciting TRAPPIST-1 is and *JWST* would just be observing just another M-star with a single planet of interest that we might be able to understand if we were to know the mass. Additionally, without *Spitzer*, we might have observed the planets discovered by the ground but have been confused by the other planets in the system. For example, TRAPPIST-1c is found to co-transit planets e and f which significantly alters the transit depth and structure (Figure 2). Without *Spitzer*, characterization of TRAPPIST-1c with *JWST* would have been severely compromised and biased.

Characterization of the planetary systems is critical before we utilize such a precious resource as *JWST*. Without such characterization, not only can we not fully interpret the *JWST* observations, but we may not even fully recognize what it is that we wish to observe with *JWST*. *Spitzer* provides a platform to do the background characterization necessary and that is simply not possible from the ground.

How many more TRAPPIST-1 systems are out there ready to be discovered by *Spitzer* and then characterized by *JWST*?

#### 2.1.2 More TRAPPIST-1 Systems

Over the past decade, the search for habitable zone earth-siszed planets has primarily centered around low-mass and ultra-cool dwarfs, because detection of these small planets is relatively easier when they orbit small stars. The discovery of TRAPPIST-1 opened the flood gates for the continued and earnest search for habitable zone planets around ultra-cool dwarfs discovered from the ground and with NASA's *TESS* mission. Two major lessons learned from the discovery and the existence of TRAPPIST-1 is that multi-planet earth-sized systems around low-mass stars may be common and that *Spitzer* is an essential and unique component in the discovery, characterization and understanding of these systems is necessary before detailed atmospheric studies can be performed.

*TESS* will uncover hundreds to thousands of planets transiting nearby stars. *TESS* will furnish approximately 2 dozen planets that (a) receive between 0.2 and 2 times the insolation of the Earth, (b) are likely rocky in composition ($< 2R_\oplus$, Rogers 2015), and (c) orbit stars brighter than $K \sim 9$ (Sullivan et al. 2015). A subset of these planets, all of which will orbit M dwarfs, will likely be the subject of extended study by *JWST*. The typical 27-day baseline for a *TESS* star translates to a very different completeness to transiting planets than NASA's *Kepler* mission. The known occurrence rates for planets derived from *Kepler* allows us to predict not only which planets *TESS* will detect, but also which ones it will miss, in systems with at least one other detected planet (Figure 3).

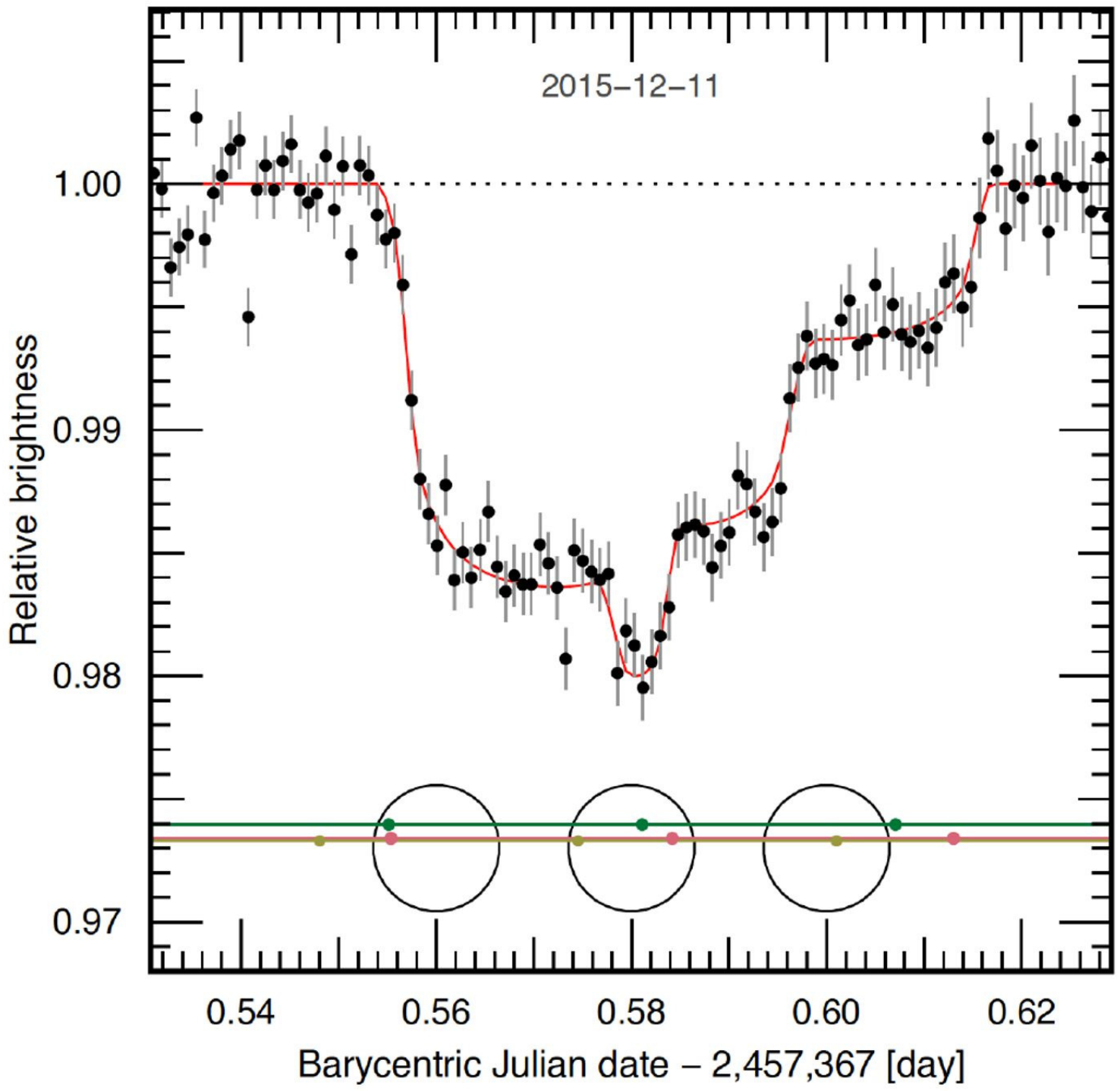

Figure 2: The triple transit of TRAPPIST-1c, 1e, and 1f. If *JWST* had observed TRAPPIST-1c without the knowledge of planets e and f provided by Spitzer, the interpretation of the planet c would have been significantly altered.

The sample simulated yield of planets orbiting M dwarfs from Sullivan et al. (2015) contains 11 planets in the approximate "habitable zone" (orbital periods within 30 and 60 days). With an estimation of *TESS*'s sensitivity in hand, and using the mixture model of Ballard & Johnson (2016) for M dwarf planets, we predict that there will exist an estimated 30 additional "habitable zone" planets orbiting known *TESS* host stars. These are stars with at least one, and even two, *TESS*-detected planets at shorter orbital periods. There is an approximately 1-in-10 chance that any *TESS* M dwarf exoplanet host, in which *TESS* identified a single transiting planet, hosts at least one other transiting planet. However, this likelihood jumps to *1-in-2* for host stars in which *TESS* identified two transiting planets.

A dedicated campaign on even a small handful of *TESS* stars, in which *TESS* detected at least two transiting planets, will furnish a sizeable contribution to the total pool of known habitable-zone transiting planets amenable to detailed follow-up. If *Spitzer* follows up 6 such systems, for example, the likely yield of 3 additional habitable-zone planets will be a significant fraction of the predicted 11 *TESS* detections.

In parallel to *TESS*, ground-based surveys for small planets around ultra-cool dwarfs are being conducted in earnest. Surveys like MEarth (Nutzman & Charbonneau, 2008) and SPECULOOS (Gillon et al., 2013) are surveying thousands of nearby cool and ultra-cool dwarfs for transiting planets. As with TRAPPIST-1, *Spitzer* will prove to be just as valuable, particularly given the number of systems that are expected to be found.

*Spitzer* is the only platform capable of characterizing these new systems in preparation for detailed observing with *JWST*: (1) it can stare for long, relatively uninterrupted periods at the same star (unlike *CHEOPS*, with a *HST*-type orbit in which Earth regularly occults the target star), and (2) it can point toward any part of the sky enabling it to follow-up any of the *TESS* host stars with multiple transiting planets.

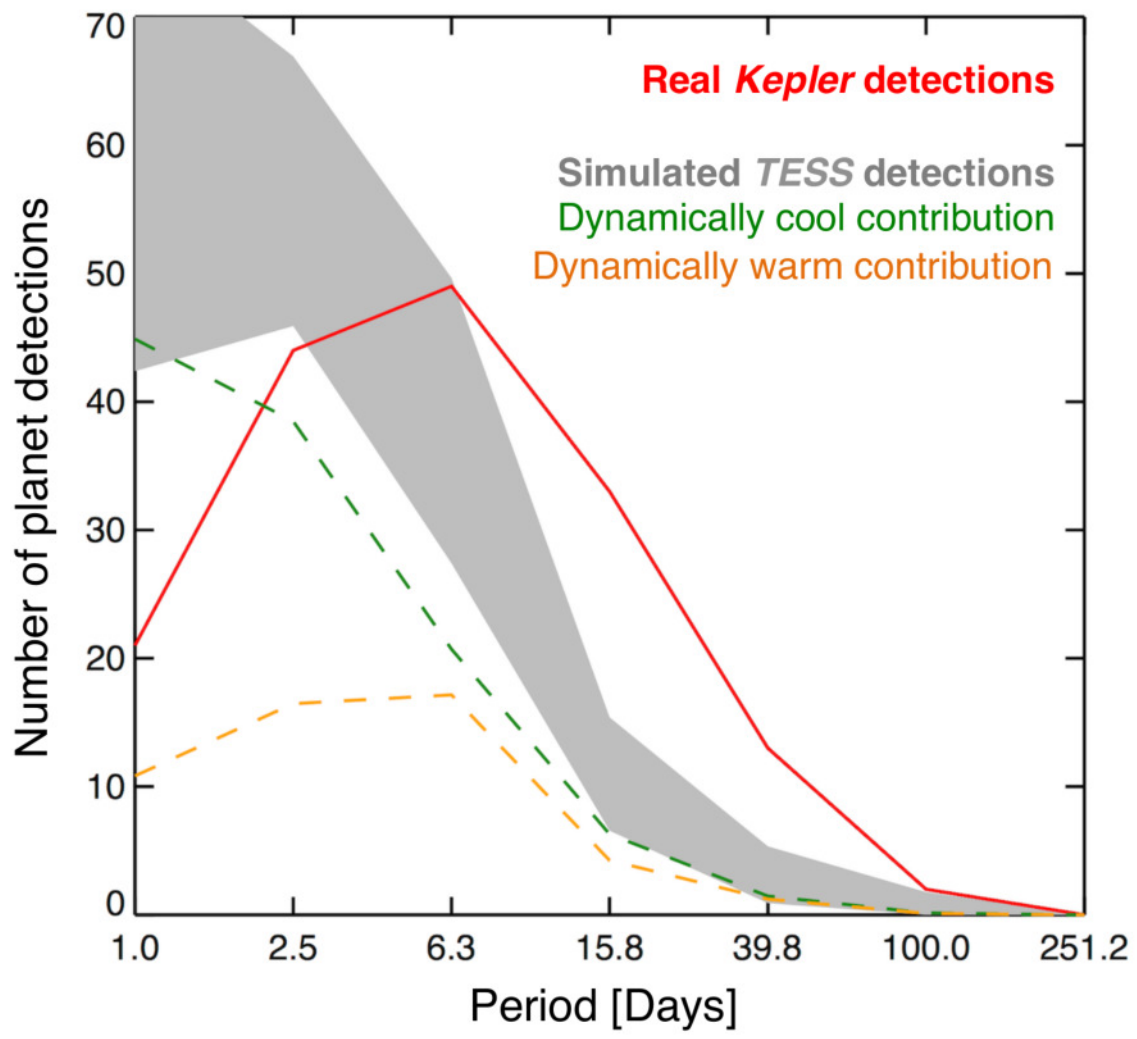

Figure 3: From *Kepler* we know that planets exist at a range of periods (red) with longer period planets often in the same systems as shorter period planets. *TESS* is biased toward detecting short period planets (gray) but may miss those on longer periods, including many HZ planets. By observing *TESS* systems with known short-period planets, *Spitzer* can discover these planets on longer orbits.

### 2.1.3 *Spitzer*'s Role: Preparing for *JWST* and *HST* Characterization

Without *Spitzer*, the TRAPPIST-1 system would still be half-explored and the planetary masses would not be accessible. The insights gained into systems - published in over 120 publications over the past year - ranging from the planetary bulk compositions to the complex resonance chains would not be within reach without *Spitzer*, yet this information is essential in order to prepare for detailed follow-up with *JWST*. Similarly, the atmospheric exploration of the planets with *HST* (de Wit et al. 2016, de Wit et al 2017 in review) would have been limited to the two inner planets.

Thanks to 1200 hours of *Spitzer* time, *JWST* will be ready to optimally explore the atmospheric properties of a whole system of 7 well-characterized temperate Earth-sized planets, 5 of which are already planned for observations via 4 *JWST* GTOs. *Spitzer* is uniquely tailored to enable the effective and efficient use of *JWST*. No other up coming facility (including *TESS* and *CHEOPS*) can take on this pivotal role, given *Spitzer*'s observing flexibility enabled by its heliocentric orbit and its high-sensitivity near-infrared instrumentation. Extending *Spitzer* into the era of *JWST* could, thus, ultimately make the difference between finding the first traces of an extrasolar biosphere within the next decade or not.

## 2.2 Orbital Determinations in Advance of *JWST* and *HST* Observations

Transit and secondary eclipse observations are extremely time consuming: the transit/eclipse events themselves are typically hours long and the required baseline before and after the events needs to be at least as long as the transits. Thus, accurate knowledge of the planetary orbit and the times of the transits and eclipses are absolutely critical for *JWST* used to observe a target efficiently.

Utilizing *Spitzer* to make the timing of a transit or secondary eclipse as precise as possible for *JWST* observations is extremely cost effective. As an example, the orbit of the planet within the K2-18 system was incorrectly determined from the *K2* data alone. K2-18b is a super-earth-sized planet orbiting within the habitable zone around a relatively nearby (and bright) cool M-dwarf - and as such, it is an excellent target for *JWST* to characterize the atmospheric properties of a small habitable zone planet. However, because of the relatively few number of transits observed, coupled with the low sampling afforded by *K2*, the transit mid-point was unknowingly measured incorrectly with the *K2* data. Consequently, when K2-18 was observed by *Spitzer* only a year after the *K2* discovery, the transit was 1.5 hours later than expected. Without *Spitzer*'s ability to do long time baseline observations which recovered the true transit time, the transit of K2-18b would have been 6-8 hours later than anticipated at the era of *JWST*. With the recovery observations with *Spitzer*, the transit time precision and accuracy were improved such that the uncertainty is less than 1 hour, projected to the time of potential *JWST* observations (see Figure 4).

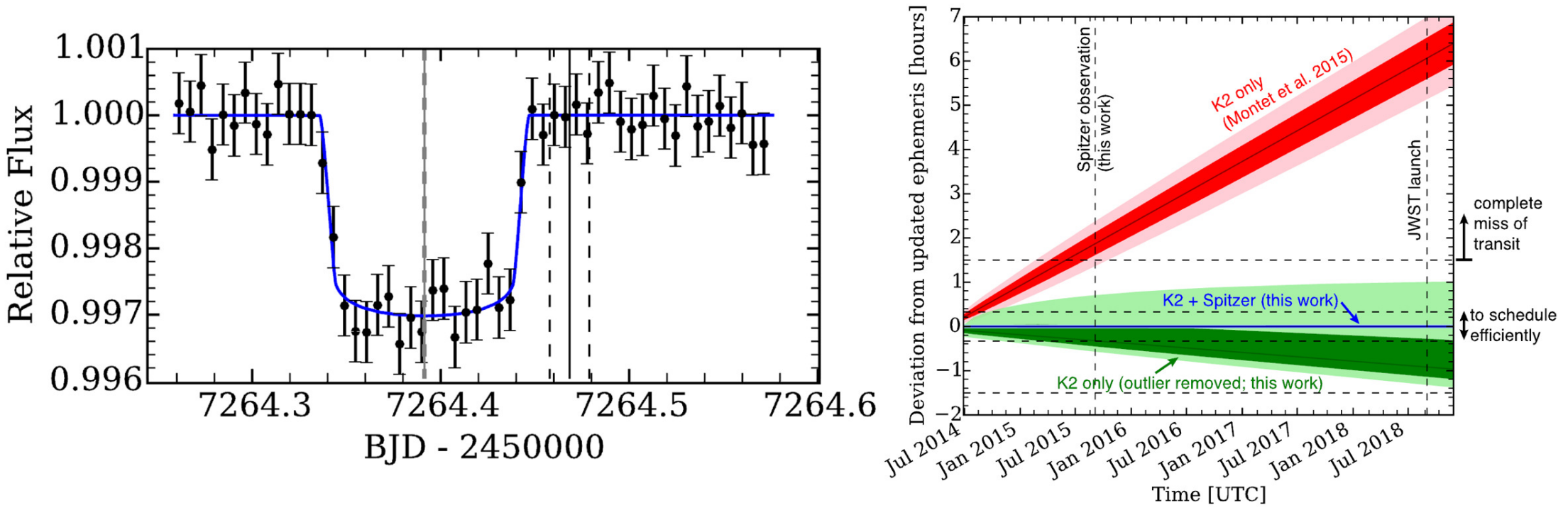

Figure 4: Left: The transit of K2-18b was observed by *Spitzer* approximately 1-year after the *K2*, but without the long time baseline afforded by *Spitzer*, the transit would have been missed by *JWST* because *K2* did not sample the light curve well enough to produce an accurate transit mid-point. Right: At the time of the *Spitzer* observations, the transit was already 1.5 hours later than anticipated and by the time of *JWST* observations, the transit would be 6-7 hours later than expected (Benneke et al. 2017).

Other systems would also have been lost if not for the recovery observations by *Spitzer*, including K2-3. Another habitable zone super-earth, K2-3d orbits a relatively bright cool dwarf with a period of only 10 days. *K2* observed more than 8 transits of K2-3d, and, yet, the derived ephemeris would have placed the transit out of a typical *JWST* observing window. But long time baseline observations with *Spitzer* recovered the orbit (Beichman et al. 2016).

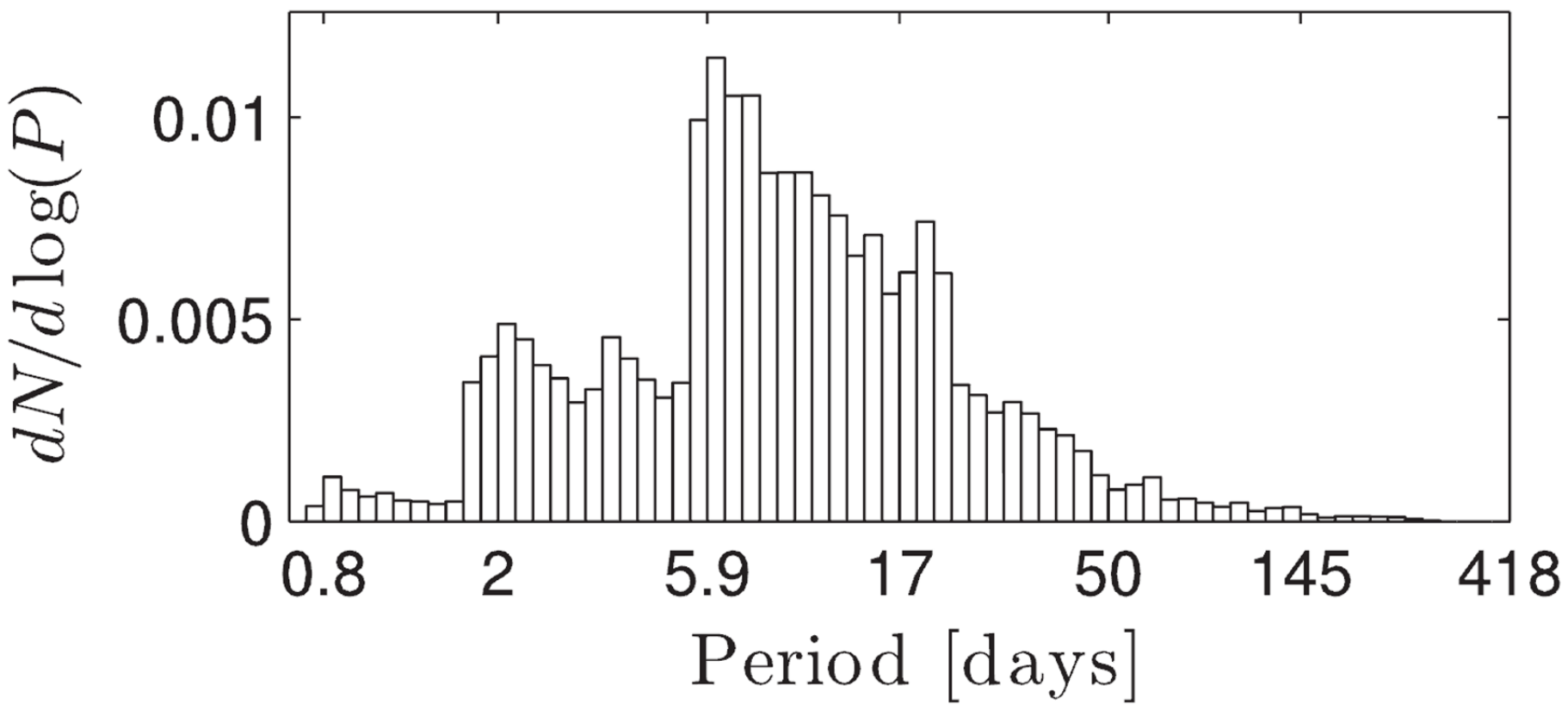

Figure 5: Predicted period distribution of planets detected by *TESS* (Sullivan et al. 2015)

When *TESS* launches in March 2018, it will find transiting planets around nearby bright stars suitable for more detailed studies by *JWST* – including, planets in the habitable zones of their stars, which will have periods of 20 days for M-star hosts and periods of hundreds of days for G-star hosts (Figure 5). With typical observing windows of only 30 – 60 days (except for the small continuous viewing zone at the ecliptic pole), *TESS* will mostly only observe 2 – 6 transits for planets and, thus, there is a high probability, just like for K2-18 and K2-3, that planning for observations of *TESS* planets with *JWST* will be uncertain by hours unless recovery observations are made.

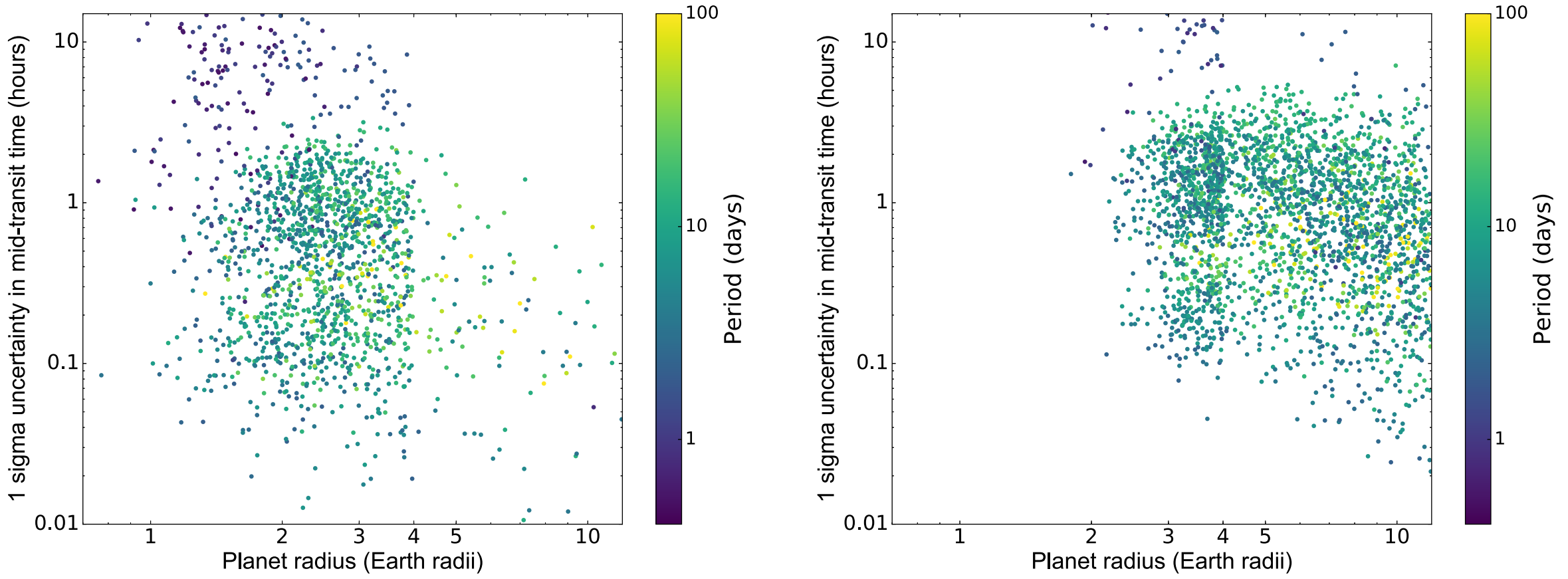

Figure 6: The uncertainty in mid-transit time for simulated planets one year after *TESS* observes them. *Left:* short cadence (2 minutes). *Right:* Long cadence (30 minutes).

Due to the photometric precision of *TESS*, and the relatively short observing baseline (compared to *K2* or *Kepler*), the ephemerides of most *TESS*-discovered planets will become "stale" very quickly. *K2* is producing hundreds of candidates, but *TESS* is expected to produce thousands of candidates (Sullivan et al. 2015); *Spitzer* is a vital characterization resource critical to choosing and understanding the systems to be observed with *JWST*. Figure 6 shows the uncertainty in the mid-transit time of all simulated *TESS*-planets with two or more transits, one year after the last transit is observed by *TESS* during the primary mission. Out of approximately 1640 2-minute cadence planets expected from *TESS*, 440 will have 1-$\sigma$ uncertainties on the mid-transit time greater than 1 hour, and 130 greater than 2 hours. For the 30-minute cadence, out of 3350 planets, 1360 will have uncertainties greater than 1 hour and 570 greater than two hours.

Ground-based resources will not be able to reliably recover transits shallower than ~1-3 mmag. Even for deeper transits, recovery of transits from the ground is very challenging if the mid-transit time uncertainties (1-sigma) are greater than ~4 hours, especially, when the Earth's diurnal schedule and weather patterns are coupled into the observability window functions.

*TESS* will discover over 500 planets with periods longer than 27 days that show at least two transits in the *TESS* light curves and have a total signal-to-noise ratio (SNR) greater than 7.3 (Sullivan et al. 2015). By also exploiting single-transit events, this yield can be more than doubled, with potential to discover up to 900 *additional* planets with periods > 27 days. In order to determine which of these single-transit events are true planets, the usual vetting process will need to be supplemented with extra steps. One of these steps is to capture a second transit. This will happen after an ephemeris has been obtained using RV monitoring of the system, and constraints from any additional, multi-transiting planets in the system. However, the uncertainty on the next mid-transit time is likely to be at least several hours, making it difficult to ensure a transit is captured from the ground. A space-based observatory such as *Spitzer* is critical to the confirmation of the vast majority of single-transiting *TESS* planets.

For secondary eclipse observations with *JWST*, the scenario is potentially even more dire. Without a proper understanding on the orbital parameters or a direct detection of the secondary eclipse, the eccentricity of the planetary orbits adds significant uncertainty to the time of the secondary eclipse (Figure 7). Direct observations of the secondary eclipse would eliminate this ambiguity – and as with transits, only space-based facilities like *Spitzer* are capable of making such detections. Without constraints on the orbit and/or the secondary eclipse location, the time of the secondary eclipse could differ from that expected of a circular orbit by many hours or even days. This is particularly critical for long period planets on eccentric orbits where the orbits are not understood at a precision well enough

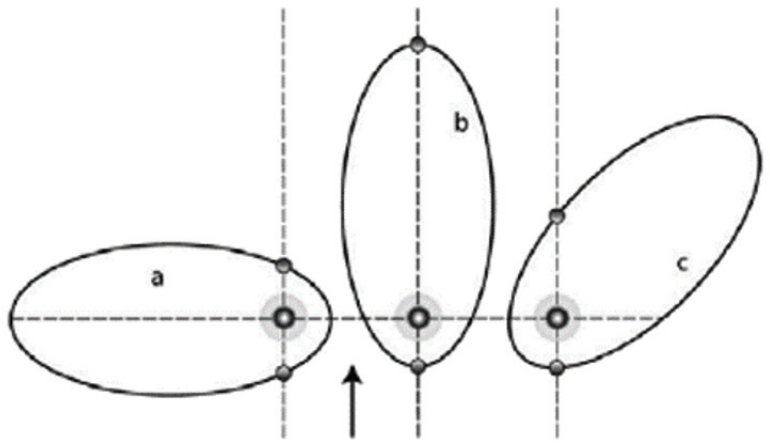

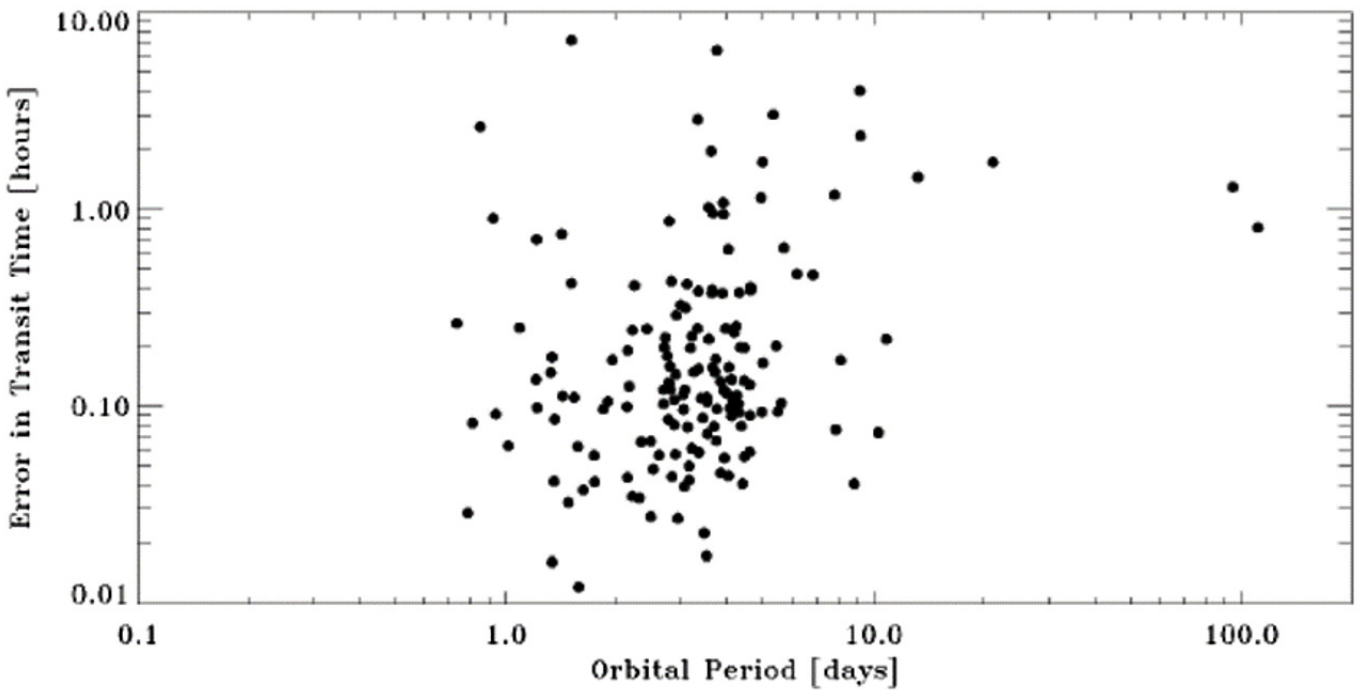

Figure 7: (Left) The geometry of primary and secondary transits is shown for different orbital configurations. Determining the exact timing for a secondary transit requires detailed radial velocity observations. When the planet's eccentricity and/or longitude of periastron are unknown or poorly known, the uncertainty in the secondary eclipse can be hours or even days. (Right) The uncertainty in future transit times in 2018 for a sample of transiting systems based on extrapolations from present day uncertainties in the parameters of simple Keplerian orbits. A few systems will have uncertainties as large as one hr or more by 2018, highlighting the need for continued monitoring to update orbital information through the lifetime of *JWST* (Kane & von Braun 2009; Beichman et al. 2014)

to accurately predict the transit and eclipses at the time of *JWST*.

For periods longer than about 5 days, exoplanet orbits become increasingly eccentric, with some reaching $e = 0.6$ at 20 days (Kipping, 2013). As a result, the simple expedient of assuming the secondary eclipses occur exactly 0.5 orbital periods after the observed transits no longer holds true. Indeed, at $e = 0.6$ it is possible for an eclipse to be 0.9 orbital periods after the transit. Other than extensive RV follow-up necessary to measure the eccentricities and arguments of periastron for these systems, the only way to feasibly determine the secondary eclipse times for these planets will be to catch them using long duration photometric monitoring with *Spitzer*.

There are currently three space-based observatories that can be used to recover transits and find secondary eclipses: *Spitzer*, *MOST* and *CHEOPS*. The *MOST* space telescope is not currently funded, and functions only if a user can purchase time. *MOST's* photometric precision is also lower than that of *TESS* (making it difficult to use for shallower single-transit events), and becomes equivalent to that of ground-based facilities for targets fainter than V mag of 11. The European Space Agency is launching CHaracterising ExOPlanets Satellite (*CHEOPS*) in late 2018 to obtain optical transits and phase curves of exoplanets, but only 20% of the time is available outside the guaranteed time and guest observations may not be accomplished if the observations conflict with the guaranteed time observations of the science team. Additionally, large portions of the *TESS* footprint is out of the field-of-view for *CHEOPS*, and *CHEOPS*'s orbit is similar to *HST*'s orbit, meaning that observations will be periodically interrupted by the Earth – making timing measurements more complicated and shorter transit events may be missed. Therefore, while *CHEOPS* may recover some of the transits, it cannot be relied upon entirely for transit recovery. Further, *CHEOPS* works in the optical and will have significant less sensitivity to late-type stars which have peak brightness in the infrared where both *Spitzer* and *JWST* operate. *Spitzer* is, by far, the most useful observatory for this science case both in terms of photometric precision and availability.

## 2.3 Complementing *JWST* and *HST* Observations

In addition to discovering planets and refining their ephemerides, *Spitzer* will provide essential planet *characterization* to support atmosphere studies with *HST* and *JWST*.

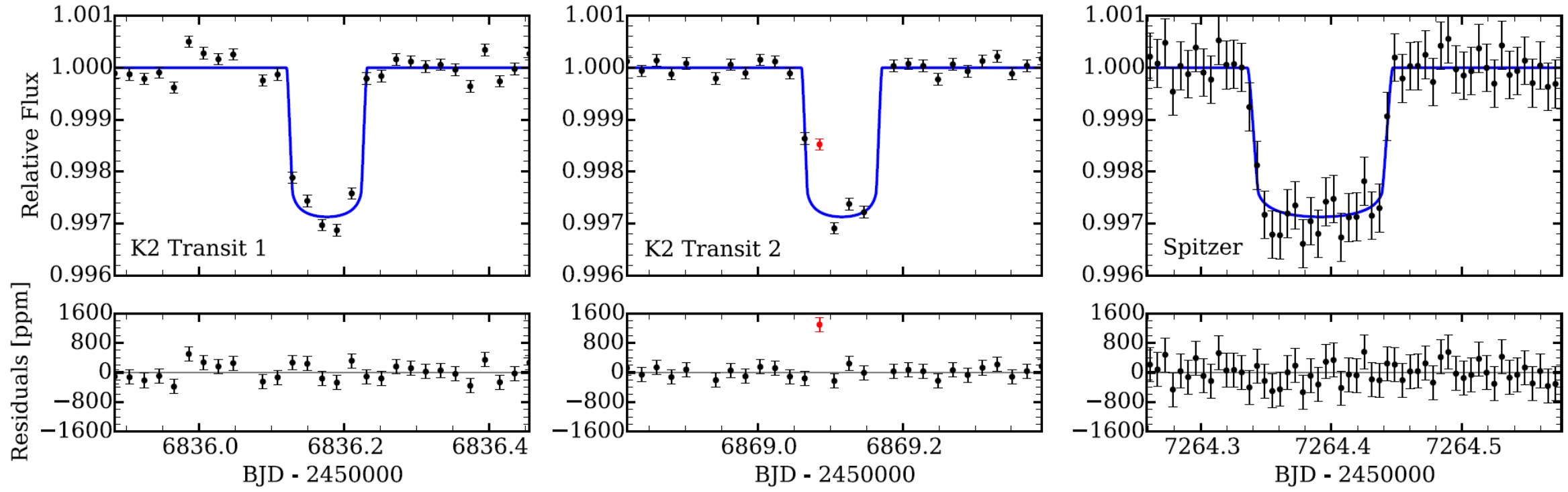

Figure 8: Transits of K2-18 as seen by *K2* and by *Spitzer*. The *Spitzer* transits are sampled significantly better than the *K2* transits, yielding a estimates of the planet radii that are a factor of 10 better than what was achieved with *K2* (Benneke et al. 2017).

#### 2.3.1 *Spitzer*'s Role in Characterizing Planets from *K2* and Eventually *TESS*

*Spitzer* has carried out an extensive program of follow up of exoplanet candidates from the *K2* mission. As discussed above, *Spitzer* has been utilized to ensure the accuracy of transit ephemerides of the transits are precise and accurate for later recovery by *JWST*. But *Spitzer* also has been vital in the physical characterization of *K2* planets, in addition to the orbital timing.

Because *K2* only observes in a 30 minute cadence, the transits, especially the ingress and egress, of moderately long period planets ($P \approx 30$ days or longer) are poorly sampled. As a result, the transit depths and planet radii are not always well constrained. The vast majority of planets discovered by *TESS* will have the same problem because they will be found in the 30 minute cadence full frame images (Sullivan et al. 2015). *Spitzer*, in contrast, can sample down to a few seconds or faster depending on the brightness of the star, since most stars are redder and brighter in the infrared. In these cases, *Spitzer* observations have been invaluable in improving the ephemeris and the quality of the planet radius measurements. As an example, the K2-18 planet radii derived from the *Spitzer* data were 10× more precise than the radii determinations from the *K2* data alone (see Figure 8).

#### 2.3.2 Masses for Planets from TTVs

One of the fundamental requirements for interpreting a planet's atmospheric composition is knowledge of the planet mass. The amplitude of features in the transmission spectrum scales linearly with the planet mass. As a result, the planetary mass is necessary to break the degeneracy between the compositional abundance and the presence of clouds.

Planets orbiting the coolest stars are generally the best targets for atmosphere characterization, thanks to their large planet-to-star radius ratios. However, mass measurements are challenging for these systems because the stars are faint in the optical, where radial velocity (RV) spectrographs operate. For example, the TRAPPIST-1 system has a V-band magnitude of 18, which is out of reach of current and planned RV instrumentation (Gillon et al., 2017). Fortunately, there is an alternate path to measuring the planet masses: transit timing variations (TTVs), which makes use of gravitational perturbations in planets' orbits. If one planet in the system is gravitationally tugged on by another, its transit time will deviate from expectations (e.g. Holman & Murray (2005)). These deviations are of order minutes, and so precise transit light curves are needed to detect and model TTVs.

*Spitzer*'s capability for precise, uninterrupted time-series observations makes it the ideal facility for TTV studies (Figure 9). It can provide sub-60 second timing precision on transit times for planets with sub-1% transit depths (Gillon et al., 2017). *JWST* is the only other facility capable of this (*CHEOPS* does not provide continuous time coverage), and *Spitzer* time is a bargain by contrast. Precise and robust mass determinations require the observation of dozens of transits (Gillon et al. 2017). *Spitzer* is often the only facility capable of measuring one of the most essential planet properties, mass, which is often not obtainable any other way for the most exciting, lowest-mass planets.

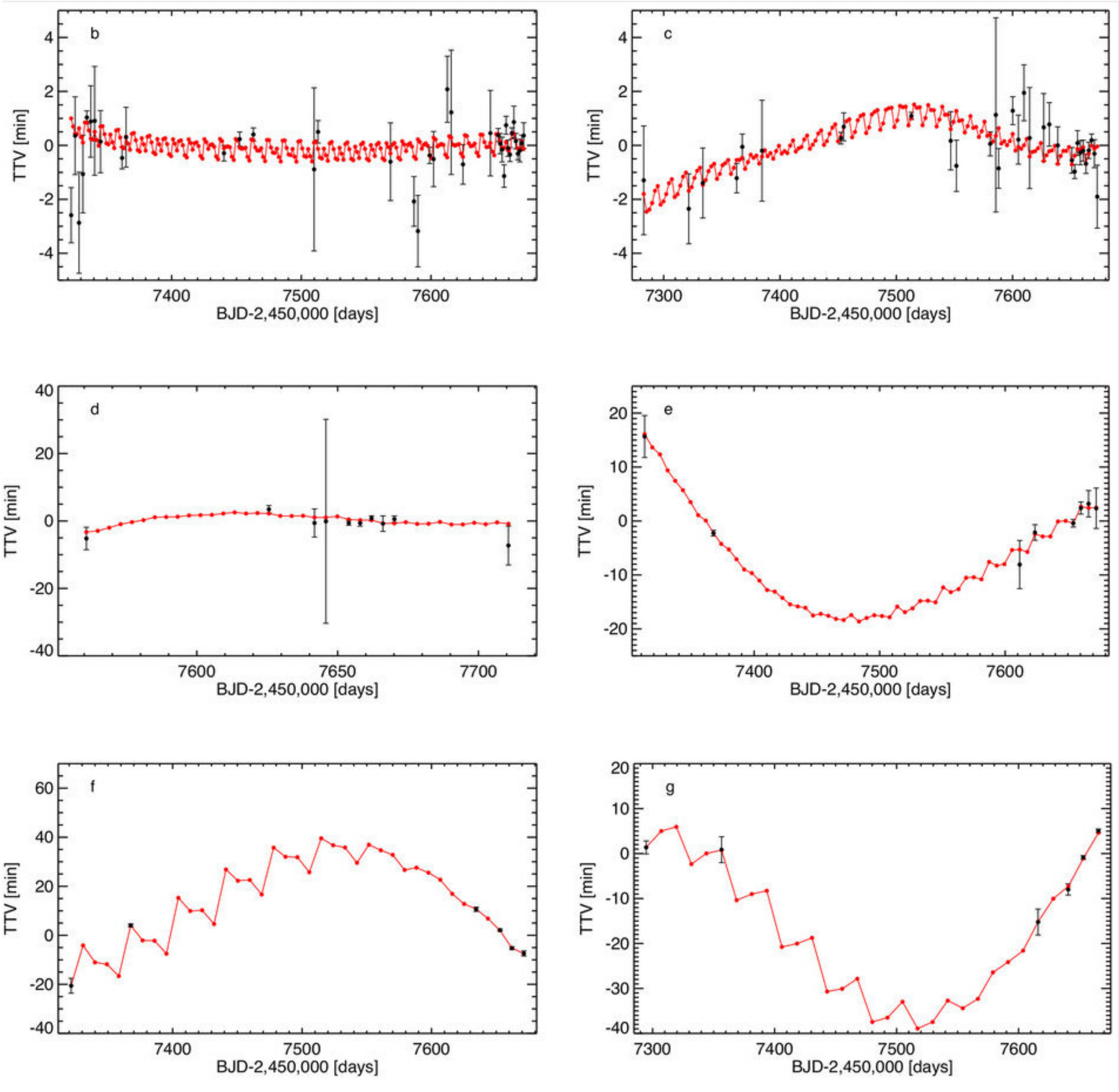

Figure 9: Transit timing variations of the planets in the TRAPPIST-1 system as determined from *Spitzer* observations of the transits. The TTVs yielded masses for the planets which are out of reach for today's radial velocity instruments (Gillon et al. 2017).

#### 2.3.3 Phase Curve Observations

Exoplanet atmospheres have complex physics and chemistry thanks to their asymmetric irradiation, rotation, and relatively low temperatures. Unlike the Solar System planets, for which rapid rotation and/or significant thermal inertia make 1D models a decent approximation, short-period planets are observed to have day-to-night temperature gradients of hundreds to thousands of degrees K (30–100%). In other words, understanding the dayside of the planet via emission or transmission spectroscopy is only half the story; the night side of the planet is the likely birthplace of advective clouds (e.g., Demory et al. (2013); Parmentier et al. (2016)) and is thought to be crucial for the cooling of planets, important for solving the inflated hot Jupiter problem (e.g., Thorngren et al. 2016), and for determining the width of the habitable zone (Yang et al. 2013, 2014).

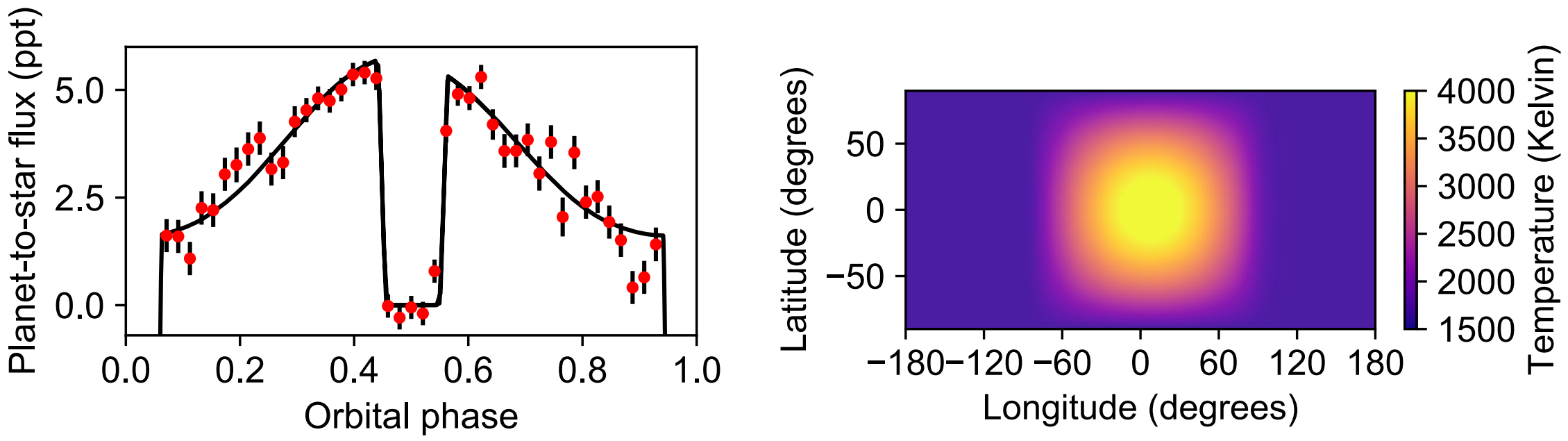

Figure 10: *Spitzer* 4.5 $\mu$m phase curve of the hot Jupiter WASP-103b, left, and retrieved temperature map (Kreidberg et al., in prep). Phase curve observations constrain the planet's longitudinal temperature structure, which is essential for estimating the temperature range probed by transmission and dayside emission spectroscopy with *HST* and *JWST*. Phase curves are also sensitive to possible cloud formation on a planet's nightside.

To fully understand a planet's three-dimensional structure, phase curve observations are needed. A full-orbit phase curve is a time-series observation of an exoplanet as it completes an entire orbit of its host star. Phase curves are sensitive to reflected light and thermal emission from the planet over its changing faces (assuming it is tidally locked). These measurements can put powerful constraints on the global thermal structure and cloud coverage of an exoplanet (as illustrated for the hot Jupiter WASP-103b in Figure 10). However, these measurements require long duration, continuous observations over an entire planetary orbit.

*JWST* is only expected to measure phase variations for a handful of planets, and likely, only those on the shortest periods. *Spitzer*, on the other hand, has been a workhorse for phase curve observations and has proven to be a stable platform for observing phase variations up to several days long (e.g. Lewis et al. 2013). There are currently approximately a dozen planets with *Spitzer* phase curves observations with an on-going *Spitzer* treasury program that will add ∼10 more systems (PI: Stevenson, Proposal ID 13038). We expect that number will quadrupled with the discoveries by *TESS*. It is expected that *TESS* find approximately 50 systems that are bright enough to obtain phase curves with *Spitzer* (Sullivan et al. 2015).

Obtaining phase curves of these additional systems will significantly expand the parameter space to a wider range of orbital properties, including planets with longer orbital periods and higher eccentricities. Eccentric planets have the potential to teach us a lot about short-period planets in general. For example, the rotation of the planet could bring the hotspot formed at pericenter in and out of view (so-called "ringing"), allowing us to infer the planet's rotational frequency (Cowan & Agol 2011 Kataria et al. 2013). More generally, the thermal emission of an eccentric planet as is passes through pericenter helps to illuminate the planet's radiative response time (Lewis et al. 2013; de Wit et al. 2016), a quantity otherwise degenerate with wind velocity. The seasons experienced by eccentric planets should produce changes in the planet's atmospheric temperature structure (e.g., transient temperature inversions; Machalek et al. 2010) and changes in cloudiness (Lewis 2017).

In addition to these synergies with *HST* and *JWST*, *Spitzer* phase curve observations will also be highly complementary to the *CHEOPS* observations, which will obtain optical phase curves of the similar quality to the infrared phase curves obtainable by *Spitzer*. Optical measurements are most sensitive to reflected light from clouds and hazes. By combining *Spitzer* and *CHEOPS* phase curves, we can disentangle the reflected versus emitted light components, to constrain the albedo and emissivity of the planetary atmospheres and provide information about the equatorial wind structure. For example, Demory et al. (2013) combined *Kepler* and *Spitzer* light curves to show that the lack of a secondary eclipse in the infrared and a strong phase-curve modulation in the optical must be the result of high-level clouds in the atmospheres, and the localized hot spot is shifted from the sub-solar point caused by strong equatorial winds.

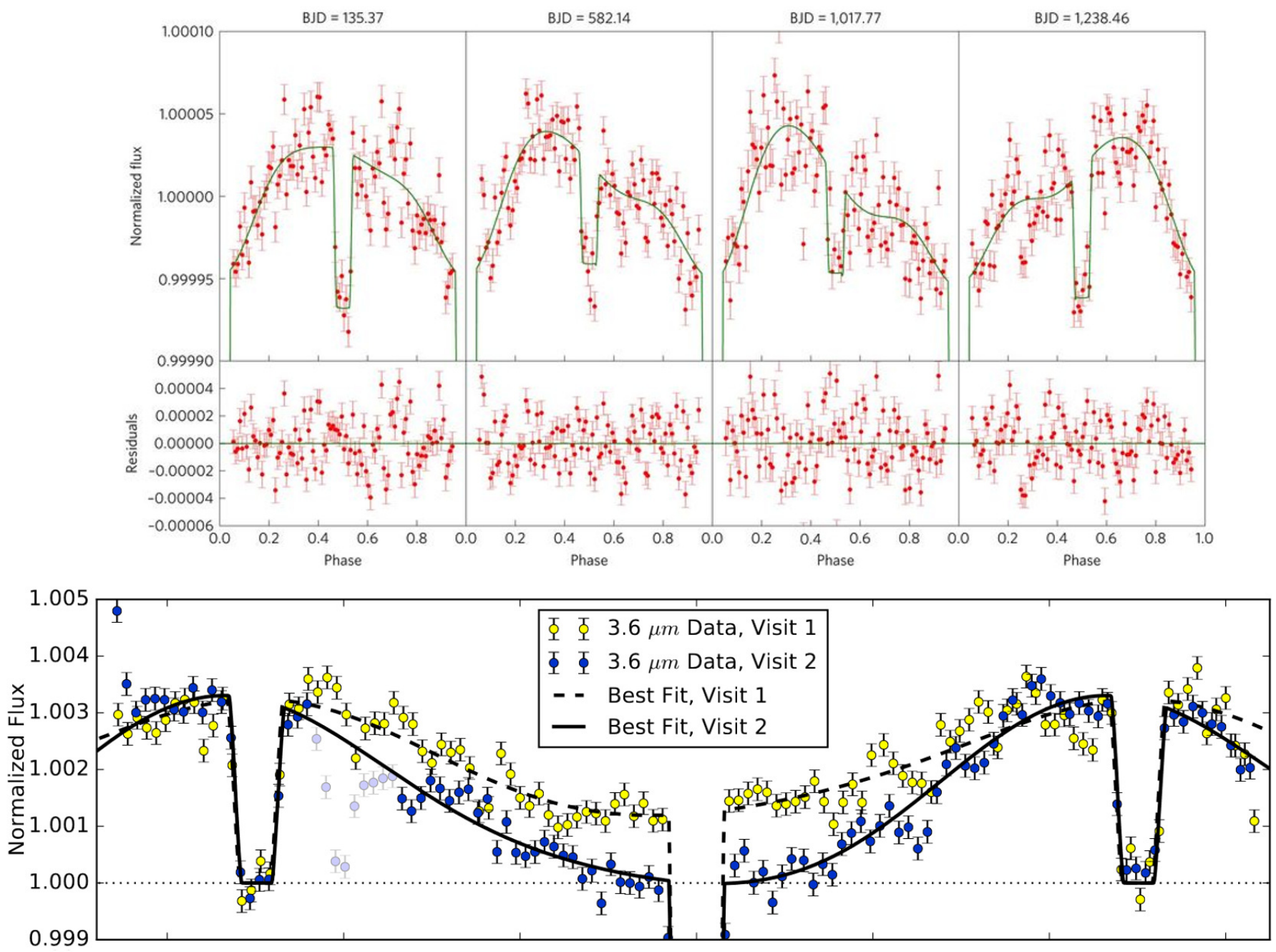

Figure 11: Optical phase curve variations of HAT-P-7b from *Kepler* (Armstrong et al. 2016) and in the infrared from *Spitzer* of WASP-43b Stevenson et al. 2017

### 2.3.4 Exoplanet Weather

Clouds and hazes (collectively known as condensates) are common in the Solar System and have been detected in many exoplanet atmospheres as well (e.g. Kreidberg et al. 2014, Sing et al. 2016, Knutson et al. 2014). Characterizing the prevalence and composition of condensates is an important component of exo-meteorology. In addition, condensates also pose a practical problem because they mask spectral features from the dominant molecules in the atmosphere, preventing observers from determining the atmospheric composition. Therefore, the best choice of targets and the optimal observing strategies for atmosphere characterization are contingent on the cloud and haze properties.

*Spitzer* provides a powerful diagnostic of whether condensates are present in a planet's atmosphere. By comparing *Spitzer* transit observations with *Kepler*/*TESS* optical transits, we can obtain a simple two-bandpass transmission "spectrum." The slope in the spectrum is sensitive to clouds and haze, because the optical properties of condensate particles vary as a function of wavelength, particularly between the optical and the infrared. This method has been used successfully to shed light on the condensate properties of hot Jupiters (e.g. Sing et al. 2016) and warm Neptunes (e.g. Morley et al. 2015). One result that has emerged from these studies is that planets have a diverse range of condensate properties that are not strongly correlated with other properties of the system (temperature, surface gravity). Because it is so challenging to predict whether condensates are present in an atmosphere, observational vetting is necessary. *Spitzer*'s coverage of long wavelengths therefore provides a unique capability to vet planets for condensates prior to observing them with *HST* or *JWST*.

In addition to screening for clouds, *Spitzer* observations will also be an important test of variable weather on exoplanets. In recent years, there has been increasing evidence that the structure of the exoplanet atmospheres may be variable and, thus, indicative of weather on the planets (e.g. Demory et al. 2016; Armstrong et al. 2016). Interpretations of single epoch observations of planetary transits, eclipses, or phase curves may not accurately represent the true atmospheric conditions, and multiple observations over many epochs may be necessary to accurately interpret transit and eclipse observations by *JWST* and to understand the weather phenomena occurring on exoplanet atmospheres. However, as with single phase curves, such a large time investment is unlikely to be available with *JWST* but is possible with a dedicated program on *Spitzer*.

Further, characterizing the variability over long time scales will enable more realistic estimates of the systematic instrumental induced and real variability of the compositional retrievals from *JWST* transit and eclipse observations.

Variability and weather phenomena can manifest itself into differing transit and eclipse depths and in changes in the phase curves of planets in both the optical and infrared (Figure 11). The variations (if real) affect our interpretations of the how exoplanet atmospheres are structured, how they behave, and the compositions we derive from the phase curves themselves and the transit and eclipse observations. Only with multi-epoch monitoring can these effects be resolved.

In order to understand the precision and accuracy of the atmospheric studies that will come out of the *JWST* observations, understanding the stability of the planetary atmospheres at the wavelengths of observations are critical. *Spitzer* is the only observatory that can provide both the wavelength coverage and the long time baseline observations to support the scientific interpretation of the *JWST* observations.

# 3 Preparing for *WFIRST* Microlensing

The *Spitzer*, microlens parallax program is the most vibrant, U.S. microlensing program to study exoplanets. As such, continued observations with *Spitzer* provides vital information and experience for the preparation of the community and operation of the *WFIRST* microlensing mission.

## 3.1 The Galactic Distribution of Planets

The primary goal of the *Spitzer*, microlensing program is to characterize microlensing exoplanets and make the first comparison of the exoplanet occurrence rates in the disk and the bulge of our galaxy. This measurement of the planet occurrence rates in these two very different environments will provide insights into planet formation processes and sets the stage for much more detailed *WFIRST* studies of the variations of planet occurrence rate with galactic environment. *Spitzer*, has revolutionized the field of microlensing by measuring parallaxes to over 500 microlensing events. As a result, this program has had a number of exciting and unanticipated results, including the characterization of an ultra-cool dwarf with an Earth-mass companion from the 2016 campaign (Shvartzvald et al., 2017), the lowest mass microlensing planet ever discovered.

The microlensing parallax effect is crucial for measuring the distances to the host stars, which is essential for determining which systems are in the disk as compared to the bulge. Observing this effect requires observing the same microlensing event from two different, widely separated locations. *Spitzer*, is uniquely suited to measuring this effect because its distance 1.5 AU from the Earth is well-matched to the fundamental microlensing scale of several AU and because observations from *Spitzer*, can be initiated within 3-10 days of identifying a transient, microlensing event from the ground. There is no other spacecraft with both of these capabilities. As such, no other systematic microlensing parallax survey will be possible until *WFIRST*. Furthermore, the *Spitzer*, microlensing program has become more powerful in recent years due to the start of the Korea Microlensing Telescope Network, a nearly-continuous, ground-based survey that allows good observational coverage of nearly all *Spitzer*, microlensing events, making it much easier to detect microlensing planets now than when the program started in 2014. In fact, KMTNet radically changed its observing strategy in 2016 to cover a wider area to support the *Spitzer*, microlensing campaigns and the scientific potential from combining data from the two experiments. Thus, continuing *Spitzer*, operations through September 2019 would permit an additional microlens parallax campaign for the microlens parallax program, which continues to make ground-breaking discoveries and has only become more valuable with time.

Distances to four planets have been published or submitted for publication based on the *Spitzer*, microlensing campaigns (Udalski et al., 2015; Street et al., 2016; Shvartzvald et al., 2017; Ryu, 2017). Three of the published planets are in the disk while one is in the bulge (Ryu, 2017). The distance distribution of all lenses shows that two thirds of the lens systems are in the disk with one third in the bulge (Zhu et al., 2017). The current number of planets is too small to draw statistical conclusions, but every Spitzer microlensing campaign contributes significantly to the total number of detected planets ($\sim 2$/year). Although the total number of planets will remain small ($\sim 12$ including all *Spitzer* campaigns through 2019 and *Kepler*, 2 Campaign 9), this is our only opportunity to compare the occurrence rate of planets in the bulge to planets in the disk until the launch of *WFIRST*. If no other bulge planets are detected, this will be enough to say that bulge planets are more rare than disk planets at 95% confidence. Even a preliminary measurement will improve our ability to frame the scientific questions that will be answered by the *WFIRST* microlensing survey.

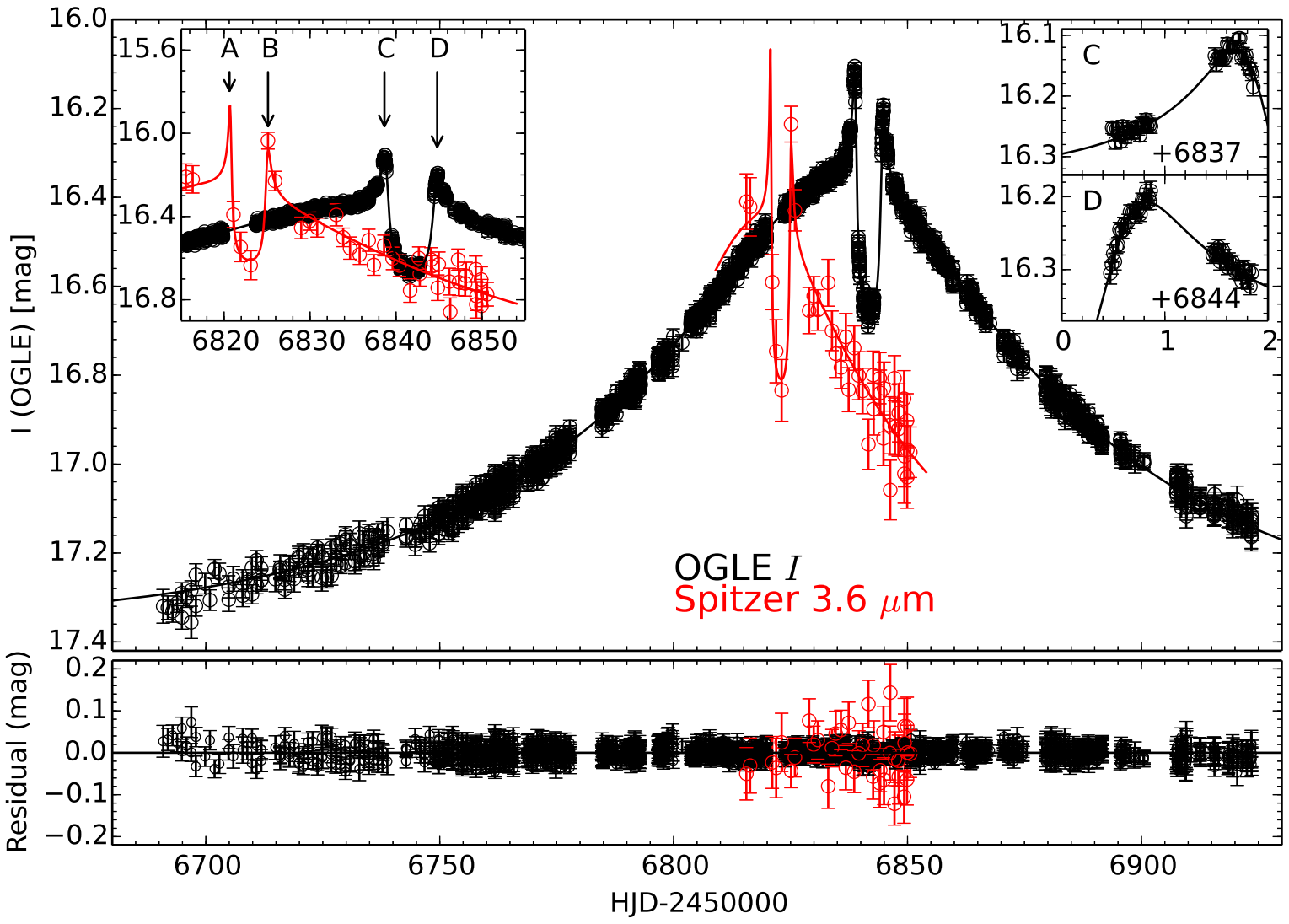

Figure 12: Light curve of OGLE-2014-BLG-0124, the first microlensing planet seen from both the ground and from space. *Spitzer*, (red) sees the planetary perturbation $\sim 20$ days earlier than it is seen from the ground (black). This difference is due to the parallax effect, which permitted a measurement of the planet's mass and distance from the Earth.

### 3.1.1 Measuring (Planet) Masses with *Spitzer*

*Spitzer* microlensing combined with other, higher-order effects, enables the complete characterization of a microlensing event. Figure 13 shows how, by definition in microlensing, if the distance to the lens system is measured, so too is the mass of the system (completely independent of the luminosity and distance to that system). Thus, *Spitzer* microlensing enables not only a measurement of the galactic distribution of planets, but also leads to better characterization of those planetary systems, including a direct mass measurement. Without this parallax measurement, we would only have been able to say that the planet characterized in Shvartzvald et al. (2017, see also Bond et al. 2017) was low mass. The *Spitzer* parallax was essential for identifying this as a $\sim 1$ Earth-mass planet orbiting a brown dwarf at $\sim 1$ AU.

Furthermore, *Spitzer* microlensing enables the discovery of dark objects such as brown dwarfs (Chung et al., 2017) and black holes and other stellar remnants, both isolated and in wide binaries (e.g. Shvartzvald et al., 2015). Thus, *Spitzer* microlensing also probes the intrinsic mass function of the galaxy, based on mass and independent of luminosity.

## 3.2 Development of *WFIRST* Microlensing

Microlensing is a major aspect of the WFIRST mission and to take full effective and efficient advantage of the WFIRST, the mission and the community need to prepare for *WFIRST*. In addition to influence the scientific questions to be answered by *WFIRST*, the *Spitzer* microlens parallax campaign is improving our understanding of the microlensing technique in a way that is vital to the success of the *WFIRST* microlensing mission. *WFIRST* is expected to detect many higher-order microlensing effects, such as parallax, that can be used to better characterize planetary systems. However, because microlensing events are so far away, it is generally difficult to impossible to independently confirm the properties of an individual system.

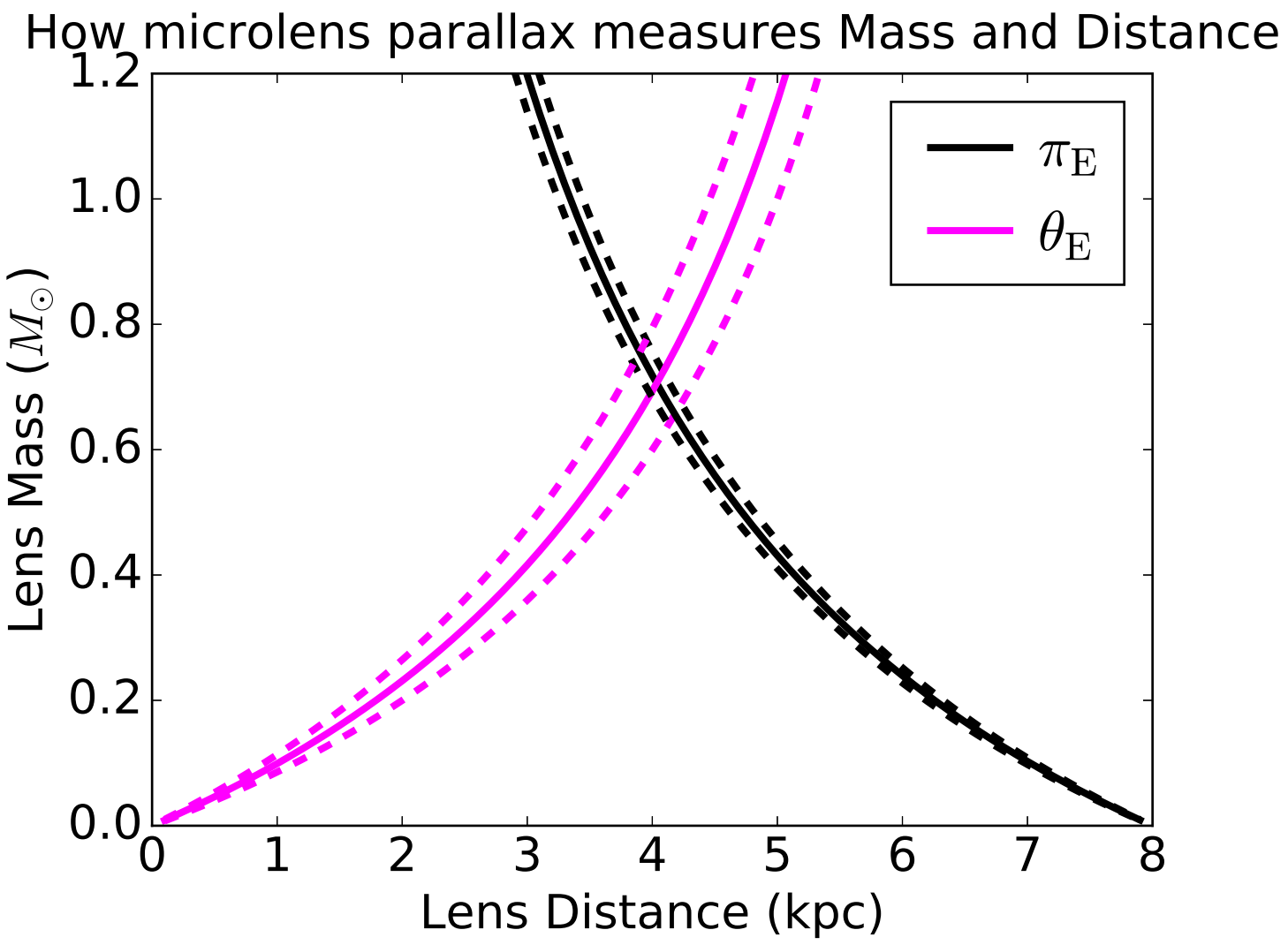

Figure 13: How the microlens parallax ($\pi_{\rm E}$) measured from *Spitzer* can combined with the measurement of the Einstein ring ($\theta_{\rm E}$) to yield the mass of and distance to the host star. The parallax constraint is the actual measurement for OGLE-2014-BLG-0124 and the magenta curves show the typical uncertainty on $\theta_{\rm E}$ (7%) for planetary events. When both quantities are measured, then both the mass of the lens system and its distance from the Earth are simultaneously measured.

The *Spitzer* microlensing campaign has enabled verification of these higher-order microlensing effects. With *Spitzer* microlensing, we have been able to measure the microlens parallax effect using multiple techniques, leading to a confirmation of the parallax measured from the orbital parallax effect, which is due to the motion of the observer about the Sun (Han et al., 2017; Shin et al., 2017). Such validation tests are crucial to reliably interpreting the *WFIRST* microlensing data.

The number of microlensing planets detected and characterized with *Spitzer* depends strongly on the amount of *Spitzer* time that is dedicated to this effort. Currently, the community is handling 100s of microlensing events per year typical yielding a handful of planets per year. A dedicated *Spitzer* effort to follow-up microlensing events could change the number of planets beyond the snowline with measured masses by an order of magnitude or more and help to definitively determine if planetary systems are more common in the Galactic Plane or the Galactic Bulge. Thus, *Spitzer* could be vital in helping us choose the fields in which to point *WFIRST* and interpret the *WFIRST* results within the context of previous Galactic distribution studies.

Additionally, *WFIRST* will yield thousands of microlensing events and being able to establish the processes and techniques for following up and interpreting the results is critical and the *Spitzer* microlensing experiments are laying the ground-work for the activities necessary to characterize these events at scale. The lessons learned and acquired by the *Spitzer* efforts are crucial to the *WFIRST* microlensing experiment succeeding, in an effective and efficient manner.

# Section 3 – Observations of Near Earth Objects in a *Spitzer* Extended Mission

David E. Trilling[1], Michael Mommert[1], Joseph Hora[2], Giovanni Fazio[2], Howard Smith[2], Steven Chesley[3], Joshua Emery[4], Alan Harris[5], and Michael "Migo" Mueller[6]

[1]Northern Arizona University
[2]Harvard-Smithsonian Center for Astrophysics
[3]Jet Propulsion Laboratory
[4]University of Tennessee, Knoxville
[5]DLR/Germany
[6]NOVA / Rijksuniversiteit Groningen / SRON, Netherlands

October 11, 2017

# 1 Introduction

Near Earth Objects (NEOs) are small Solar System bodies whose orbits bring them close to the Earth's orbit. NEOs lie at the intersection of Solar System evolution science, space exploration, and civil defense. They are compositional and dynamical tracers from elsewhere in the Solar System; the study of NEOs allows us to probe environmental conditions throughout the Solar System and the history of our planetary system, and provides a template for analyzing the evolution of planetary disks around other stars. NEOs are the parent bodies of meteorites, one of our key sources of detailed knowledge about the Solar System's development, and NEO studies are the essential context for this work. The space exploration of NEOs is primarily carried out through robotic spacecraft (NEAR, Hayabusa, Chang'e 2, Hayabusa-2, OSIRIS-REx). Energetically, some NEOs are easier to reach with spacecraft than the Earth's moon, and NEOs offer countless targets with a range of physical properties and histories. NEOs present advanced astronautical challenges for inevitable future robotic and manned missions as well as mining enterprises. Finally, NEOs are a civil defense matter: the impact threat from NEOs is real, as demonstrated in Chelyabinsk, Russia, in February, 2013. Understanding the number and properties of NEOs affects our planning strategies, international cooperation, and overall risk assessment.

The Spitzer Space Telescope is the most powerful NEO characterization telescope ever built. NEOs typically have daytime temperatures around 250 K. Hence, their thermal emission at 4.5 microns is almost always significantly larger than their reflected light (Figure 1). We can therefore employ a thermal model to derive NEO diameters and albedos; this process has been successfully tested, applied, and refined in our previous Spitzer work (Trilling et al., 2008, 2010, 2016; Mommert et al., 2014c,a,b, 2015; Harris et al., 2011; Mueller et al., 2011; Thomas et al., 2011). Measuring the size distribution and albedos and compositions of a large fraction of all known NEOs allows us to understand the scientific, exploration, and civil defense-related properties of the NEO population. Spitzer's sensitivity at 4.5 microns is unparalleled by any other current facility. JWST, through its larger aperture, will be more sensitive, when it eventually is operational, but can only observe NEOs when they are far away due to tracking rate and saturation limits. Even the slowest moving NEOs will not be good JWST targets because the slew overhead is so large compared to the necessary integration time for these targets. A large-scale survey of NEOs can *only* be carried out with Spitzer.

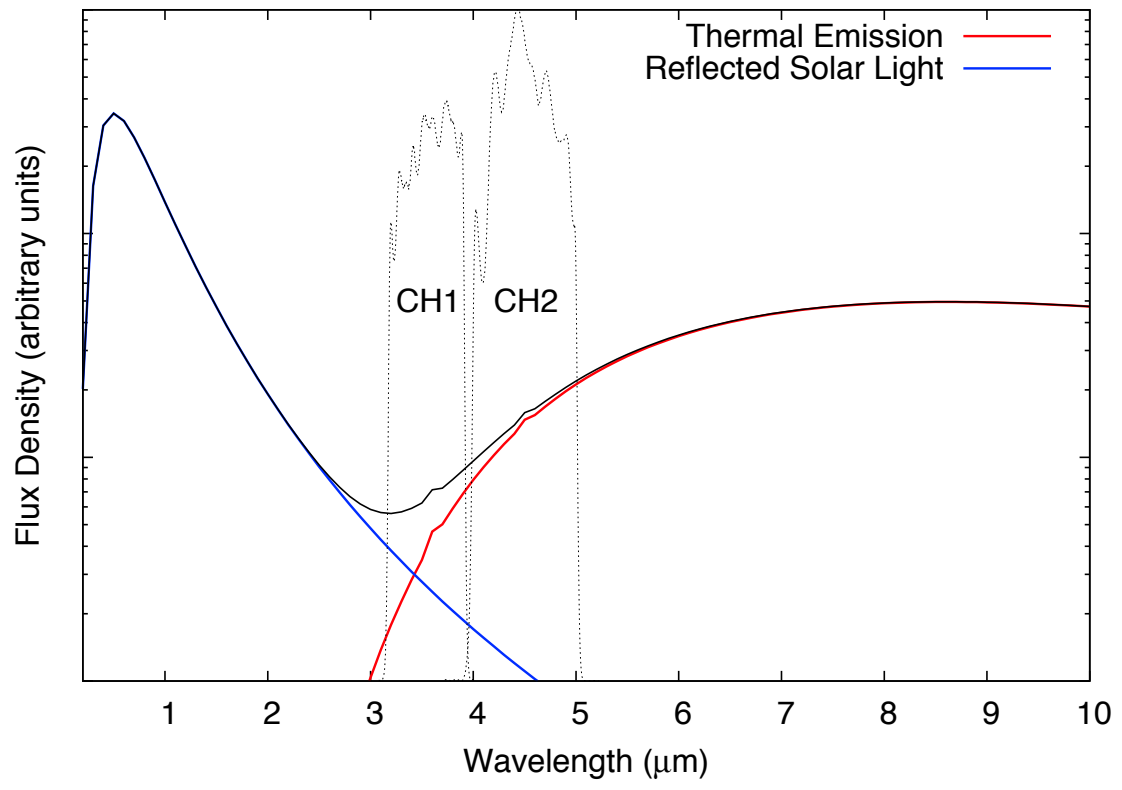

Figure 1: Spectral energy distribution of an arbitrary NEO. The black line (total observed radiation) is the sum of reflected light (blue line) and thermal emission (red line). In IRAC CH1, reflected light and thermal emission are comparable, making interpretation difficult. In IRAC CH2, thermal emission dominates, and that measured flux can readily be used for thermal modeling. The IRAC bandpasses are also shown for reference.

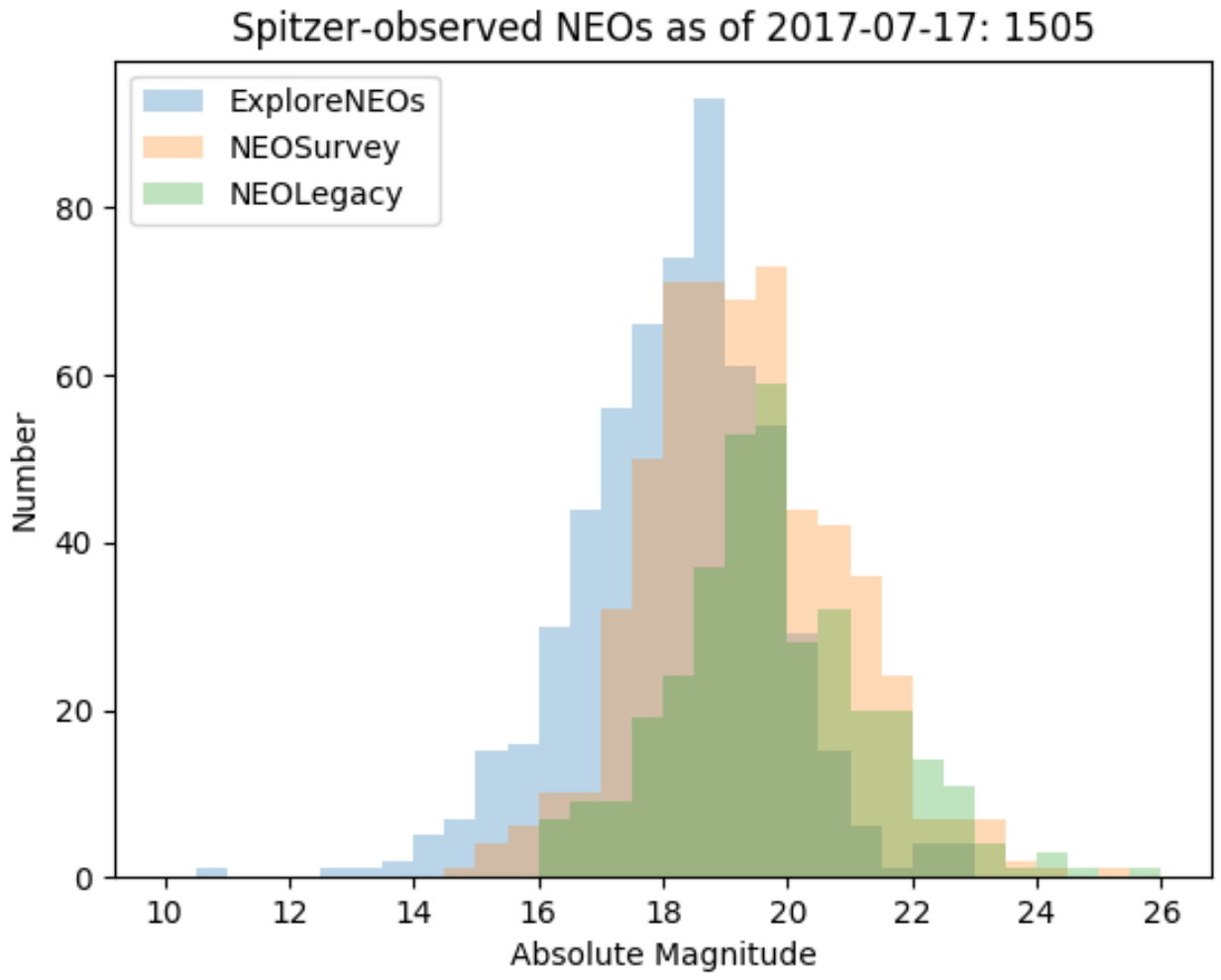

Figure 2: Histogram of Solar System absolute magnitudes (H magnitudes) for the three largest Spitzer NEO programs. H magnitudes of [17.5,20,22.5] correspond roughly to diameters of [1000,300,100] meters.

## 2 Estimate of yield

To estimate the number of NEOs available to Spitzer we carried out the following steps. It is important to recognize that NEOs have unusual visibilities as seen from Spitzer, so detailed calculations are necessary for each known object. The Spitzer Spice kernel (that is, the orbit of the Spitzer Space Telescope) is not calculated for dates after the current end-of-mission in March, 2019, so we use as our fiducial 12 month period January – December, 2018. The results should be very similar for any other 12 month period.

For every known NEO we calculate the expected flux in both CH1 (3.6 microns) and CH2 (4.5 microns) during its visibility windows; the details of these flux estimates are given in Trilling et al. (2016). We remove all objects that have already been observed by Spitzer, WISE or NEOWISE, and Akari. We next calculate the number of NEOs discovered in 2016 that are available in 2017 (one year later), and the number of NEOs discovered in 2017 that are available in 2017 (same year). The sum of these three populations allows us to estimate, given the current NEO catalog and discovery rate, the number of available targets in 2018 (our fiducial case). Our detection limit is 2.4 $\mu$Jy at 4.5 $\mu$m; all objects brighter than this can be detected at SNR$\geq$5 in 10,000 seconds integration time (around 3 hours clock time).

At the time of writing (September, 2017), we find that there are $\sim$250 NEOs that can be observed during that 12 month period. Many of these are small NEOs (<300 m), which are the most numerous and the least well studied members of the NEO population.

Because the number of known NEOs is increasing at some 20% per year, the actual number of available targets is likely to be 20% greater in 2019, and greater again in 2020. As additional NEO discovery assets come online (ATLAS this year, and by 2020 LSST first light data may also be contributing), that number may again increase.

Thus, observing *all* NEOs available to Spitzer in any given year is possible. The total time to carry out that work is likely to be on the order of 1,000 hours/year.

## 3 Science yield

For each observed NEO, we derive diameter and albedo. (Albedo is correlated with taxonomy and hence composition.) Figures 2 and 3 show the population of NEOs observed to date with Spitzer. The diameter and albedo results allow a high fidelity measurement of the NEO size distribution as well as the compositional diversity of this population. Furthermore, our observing approach allows us to derive (partial) lightcurves over around 3–4 hours, based on our observing strategy; one example is shown in Figures 4 and 5. Individual lightcurves reveal the shape and, in some cases, internal properties of those NEOs. The ensemble collection of lightcurves reveals the overall strength of the NEO population, as well as information about the collisional history of these bodies. At present there are around 1,000 NEOs with good lightcurve determinations, so hundreds/year is a significant contribution. Furthermore, most of the objects that are available in an extended Spitzer mission are smaller than around 300 meters, a population that is very undersampled.

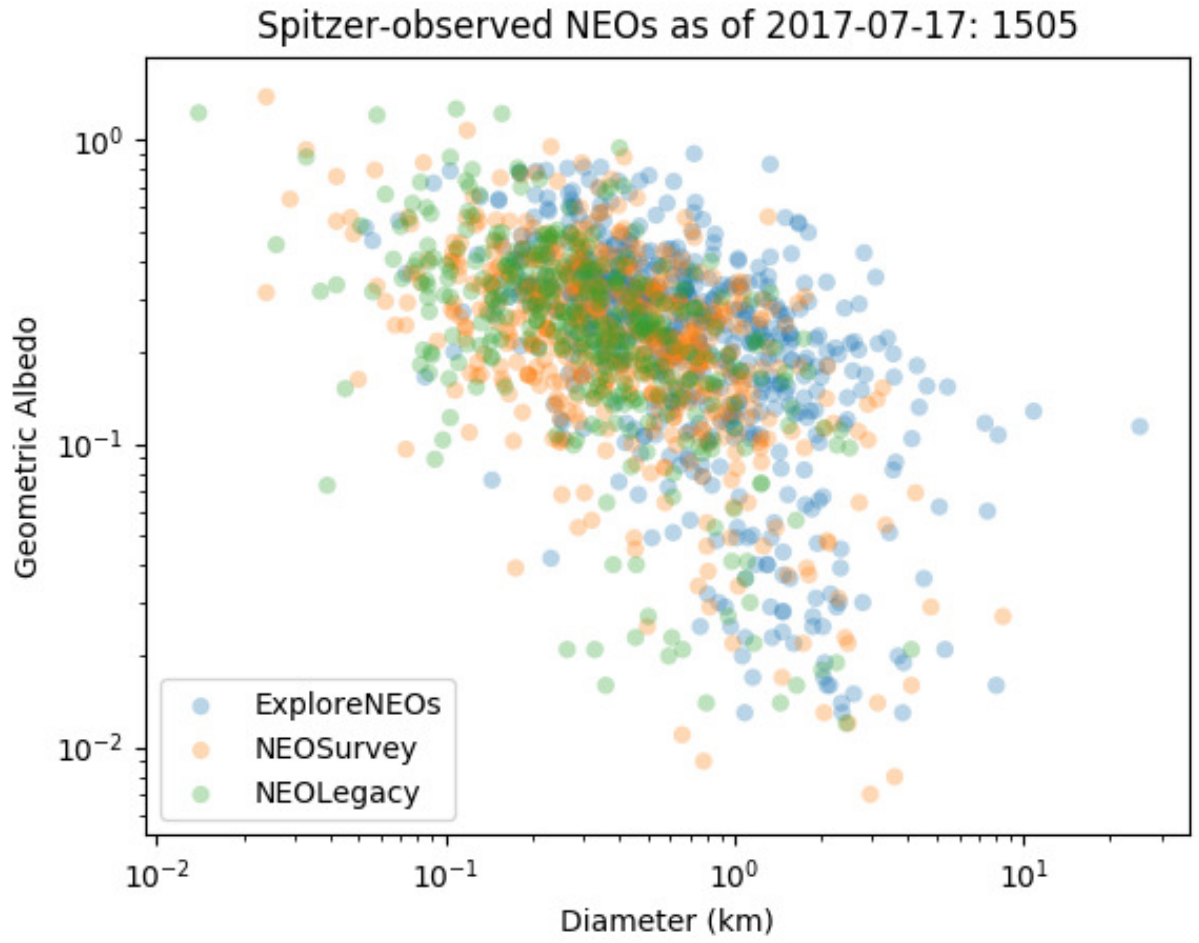

Figure 3: Diameter and albedo for the ∼2,000 NEOs that have been observed by Spitzer to date.

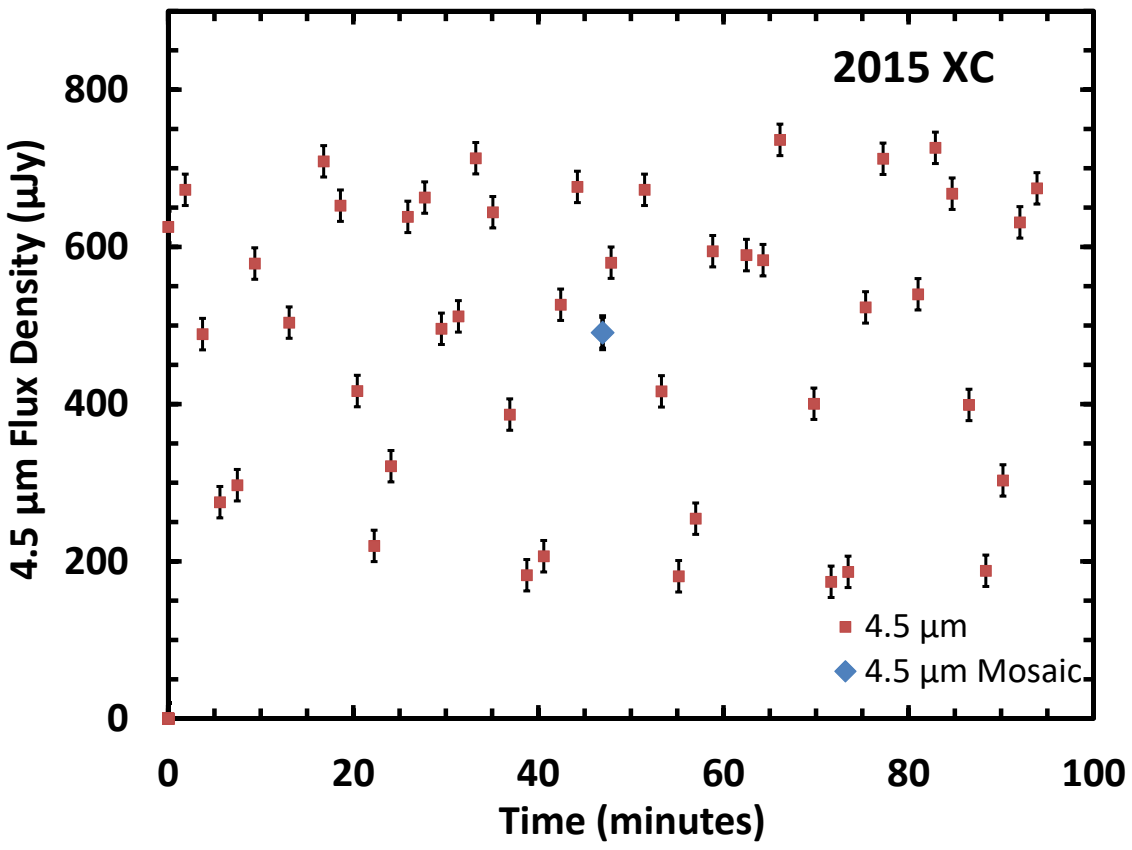

Figure 4: Spitzer lightcurve of NEO 2015XC, which was observed using 100 second frames over a period of ∼90 minutes.

At present, around 2,000 NEOs are discovered each year. Thus, by observing ∼200 NEOs each year with Spitzer we maintain the fraction of objections with known diameter and albedo at around 10% of the total population. Furthermore, continuing to observe NEOs with Spitzer would extend the total number of NEOs that have measured diameters and albedos, growing our overall catalog. Finally, as with any large catalog, rare objects only are discovered when the underlying catalog is large.

# 4 Feasibility and heritage

Our team has unsurpassed experience in planning and analyzing Spitzer NEO observations. This group includes the PI for IRAC (which is the usable Spitzer instrument) as well as the IRAC Program

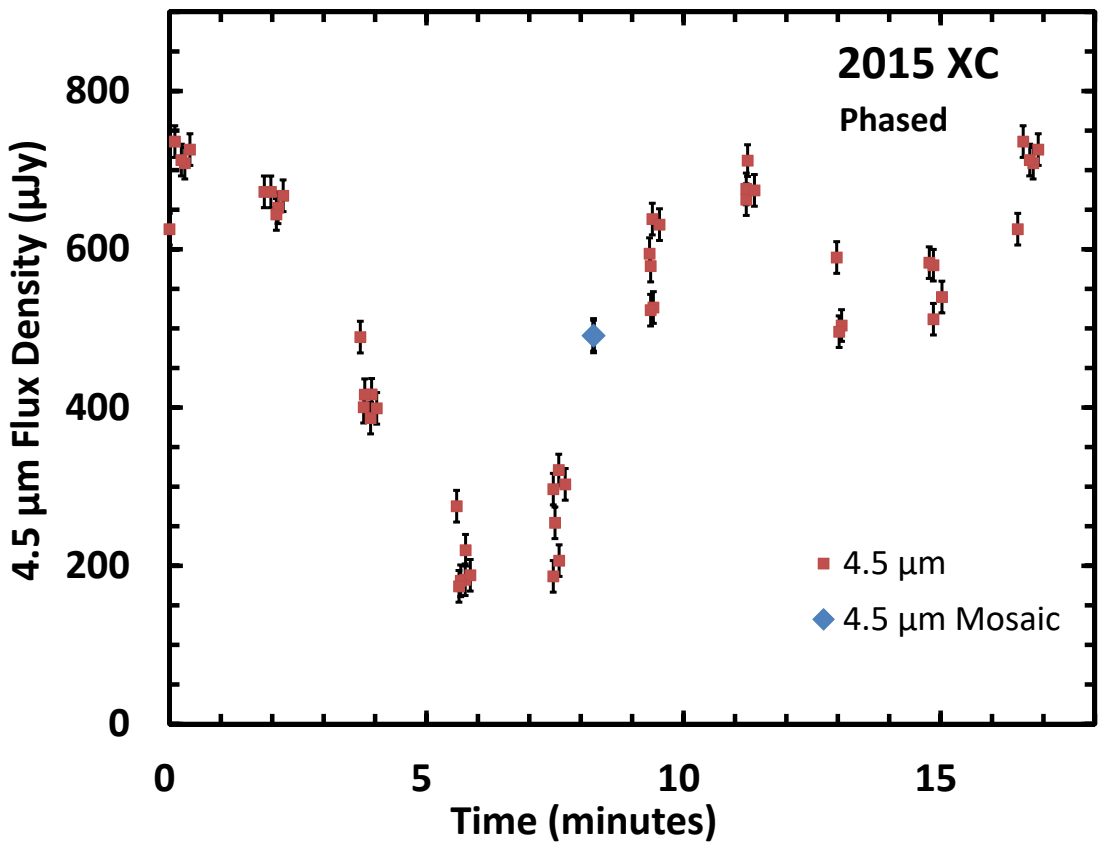

Figure 5: The same data as in Figure 4 except phased at a period of 16.5 minutes. More than 5 complete rotational periods were observed and the lightcurve shows a large amplitude, implying a lower limit on the ratio of major to minor axis of ∼4.

Scientist, who is also the chief data analyst for our team (and, not incidentally, the most experienced person in the world at processing Spitzer observations of NEOs).

Spitzer is drifting away from the Earth, and there are therefore data rate constraints for the observatory. However, because our program uses Spitzer's longest frame time, there are no data rate issues for our project.

Historically, the Spitzer NEO duty cycle has been roughly one target per day — around three hours per day. However, each target has a visibility window of a few days, and it has rarely happened in ten years that an NEO was not schedulable. Since, in general, there is nothing particularly special about any individual NEO, even in the case that occasional NEOs cannot be observed, that does not make a significant impact on the science yield from this program.

# 5 Legacy of these data

JWST will not, in general, be able to observe NEOs, as it cannot slew fast enough to observe any objects closer to the sun than 1.4 au. Furthermore, because of the large overheads associated with each individual JWST measurement, that facility is better suited for long, detailed characterizations of faint astrophysical objects than it is for a broad, shallow survey of hundreds of thousands of NEOs. Therefore, it is likely impractical — and in many cases impossible — for JWST to carry out a program that characterizes hundreds of NEOs each year. Spitzer, by comparison, is ideally suited for these observations, having both a simple imaging camera and efficient slewing capability. Finally, the amount of time that would likely be available in an extended Spitzer mission is commensurate with the time needed to carry out this project; the same is not true for JWST. Thus, even in the era of JWST, there is still a need and a role for Spitzer to carry out a survey of NEO characterization. Finally, we point out that the NEOWISE mission will no longer be functional after 2017/18 (Masiero et al., 2017) so Spitzer is indeed the only facility that is capable of making these measurements.

The single greatest legacy value of the diameters, albedos, and shapes that would be derived from Spitzer NEO observations is related to LSST. LSST will observe some 100,000 NEOs (LSST Science Collaboration et al., 2009), including almost all NEOs that have been observed by Spitzer. Combining Spitzer data (diameter, albedo, partial lightcurve) with LSST data (optical colors, partial lightcurves, phase curves) would allow for a very complete understanding of the physical properties of a large number of asteroids. This in turn will allow us to understand the evolution of NEOs but in their formation locations (generally the main belt) as well as their subsequent evolution in near-Earth space.

## 6 Spitzer's uniqueness

Spitzer is the only existing facility that can measure the diameters and albedos of the vast majority of these NEOs.

Furthermore, because of Spitzer's unique location in the Solar System, some NEOs are accessible to Spitzer that would not be accessible to any other observatory. This geometry was used to great strategic success in our observations of 2009 BD and 2011 MD, where our Spitzer characterizations helped inform a planned NASA mission to these asteroids (now canceled) long after the asteroids were no longer accessible to ground-based telescopes.

It is also important to note that no other observatory is being planned that could duplicate these unique capabilities. NEO studies by their nature require large object sets with multiple pointings, and even possible general future space observatories will have competitive programs and not easily be able to dedicate time to such large sample NEO studies.

## 7 Conclusion

Spitzer is an incredibly powerful NEO characterization tool. The observatory is functioning and producing unparalleled measurements of NEO diameters, albedos, and lightcurves. There are no technical obstacles to carrying out this work, as has been demonstrated by our ten-year ongoing survey. No other existing or planned facility can carry out this work.

# Section 4 – The Case for an Extension of the *Spitzer* Mission: Nearby Stars and Brown Dwarfs

J. Davy Kirkpatrick[1], Dániel Apai[2,3], Michael C. Cushing[4], Jacqueline K. Faherty[5], Jonathan Gagné[6], and Stanimir A. Metchev[7]

[1]IPAC, Mail Code 100-22, Caltech, 1200 E. California Blvd., Pasadena, CA 91125, USA; davy@ipac.caltech.edu
[2]Steward Observatory, The University of Arizona, Tucson, AZ 85721, USA; apai@arizona.edu
[3]Lunar and Planetary Laboratory, The University of Arizona, Tucson, AZ 85721, USA
[4]The University of Toledo, 2801 West Bancroft Street, Mailstop 111, Toledo, OH 43606, USA; michael.cushing@utoledo.edu
[5]Department of Astrophysics, American Museum of Natural History, Central Park West at 79th Street, NY 10024, USA; jfaherty@amnh.org
[6]Carnegie Institution of Washington DTM, 5241 Broad Branch Road NW, Washington, DC 20015, USA; jgagne@carnegiescience.edu
[7]The University of Western Ontario, Department of Physics and Astronomy, 1151 Richmond Avenue, London, ON N6A 3K7, Canada; smetchev@uwo.ca

# 1 Introduction

The characterization of the Solar Neighborhood, defined here to be objects within a few tens of parsecs of the Sun, has greatly benefited from the unique combination of instrumentation available onboard the *Spitzer Space Telescope*. Observations during *Spitzer's* cryogenic phase included the photometric characterization of M through T dwarfs (Patten et al. 2006) with the Infrared Array Camera (IRAC), establishment of the 5.5-38 $\mu$m spectral sequence of M through T dwarfs (Cushing et al. 2006) with the Infrared Spectrograph (IRS), and the identification of debris disks around young stars in nearby moving groups (Rebull et al. 2008; Chen et al. 2005) using the the Multiband Imaging Photometer for *Spitzer* (MIPS).

After cryogenic observations ended in early 2009, IRAC was still capable of doing high-quality science in it two short-wavelength channels (3.6 and 4.5 $\mu$m), and as such was well positioned to perform follow-up on discoveries coming from NASA's *Wide-field Infrared Survey Explorer* (*WISE*), which launched later that year. As the characterization of IRAC matured, its full astrometry- and photometry-measuring potential became clearer, enabling research in areas not anticipated before. Examples of warm mission observations include the discovery of the second coldest brown dwarf (Luhman et al. 2011) along with confirmation of the coldest example (now recognized as the fourth closest stellar system to the Sun; Luhman 2014), distance measurements to many of the coldest brown dwarfs (Dupuy & Kraus 2013; Luhman & Esplin 2014; Martin et al., in prep.), time-series analysis of weather-related phenomena in cold atmospheres (Heinze et al. 2013; Metchev et al. 2015; Cushing et al. 2016; Apai et al. 2017), and the characterization of a solar system with seven Earth-mass planets around a red dwarf just 12 pc away (Gillon et al. 2017).

Currently, the last *Spitzer* call for General Observer programs was Cycle 13, which includes observations through 2018 Sep 30. However, this end date is set by a lack of continued funding, rather than a technological or age-related issue preventing operations into future years. It has been shown that *Spitzer* can continue to operate with most of its present capability until 2020 Sep. In this white paper, we therefore highlight some of the areas in nearby star and brown dwarf research that would be enabled by a two-year *Spitzer* extension. Some of these programs have direct impacts on uncovering objects ripe for follow-up by the *James Webb Space Telescope* (*JWST*) while others exploit features of *Spitzer* that cannot be duplicated by observations with *JWST*.

# 2 Science Cases

## 2.1 Transiting Planets around Brown Dwarfs

The discovery and characterization of habitable exo-Earths is a top priority in astrophysics. *Kepler* photometry has shown that low-mass M-type stars tend to harbor multiple rocky planets (Dressing & Charbonneau, 2015). Confirmations of this trend include the recent discoveries of Earth-sized planets around Proxima Centauri (Anglada-Escudé et al., 2016) and in the Trappist-1 planetary system (Gillon et al., 2016, 2017). It remains to be seen whether smaller stars having smaller and more numerous planets is a tendency that continues below the hydrogen-burning limit: among brown dwarfs.

There is however every indication that this should be the case. First, empirical measurements of the stellar and substellar mass function show it to be continuous across the hydrogen burning limit (Chabrier, 2003, and references therein). Second, circumstellar disks are present at least as commonly around substellar objects as around stars (e.g., Testi et al., 2016, and references therein). And third, given that the $0.080 \pm 0.007 M_{\odot}$ host star of the seven-planet Trappist-1 system is very close to the hydrogen burning limit, there are reasons to expect that rocky planets form and exist in abundance around brown dwarfs, too.

Radial velocity and transit monitoring have been the most successful exoplanet detection methods. The effects that they measure are greater for low-mass stars than for Sun-like stars. For planetary transits in particular, the change of the star's brightness depends on the relative sizes of the planet and the star, and on having a nearly edge-on view of the planetary orbit. As in our solar system, the orbits of most transiting exoplanets are aligned with the spin of their host stars (Winn & Fabrycky, 2015). Because brown dwarfs settle to an approximately constant degeneracy pressure-supported radius of 0.85–1.0 Jupiter's ($R_{\rm Jup}$) beyond ages of 500 Myr (e.g., Chabrier et al., 2000), a transiting Earth-sized planet will attenuate the flux of the brown dwarf by an evolution-independent ≈1%: easily detectable by current ground- and space-based telescopes. Thus, brown dwarfs with spin axis inclinations of ≈90° (equator-on view) are among the best candidates to search for Earth-sized transiting planets.

The orbital periods of planets within the "habitable zone" around brown dwarfs are shorter than for stars, and so require less time to observe multiple transits or Doppler oscillations. For mature 500–2300 K dwarfs the habitable zones: (i) range from 0.001–0.025 AU (Bolmont et al., 2011; Barnes & Heller , 2013), (ii) offer higher probability of transits compared to low-mass stars, and (iii) correspond to orbital periods of ∼4 days or less (Fig. 1). This gives multiple (40+) opportunities per year to observe transits and secondary eclipses, and to build signal during transit or eclipse spectroscopy. Crucially, planets in edge-on orbits around brown dwarfs offer the best prospects for the detection of atmospheric biomarker gas (methane, oxygen, ozone) with *JWST* because of the larger feature contrast during transit and eclipse spectroscopy (Kaltenegger, 2017).

Undoubtedly, the details and likelihood of the origin and evolution of life around a continuously dimming brown dwarf are distinct from those around main sequence stars. The first detections of rocky planets around brown dwarfs will help address this question in earnest.

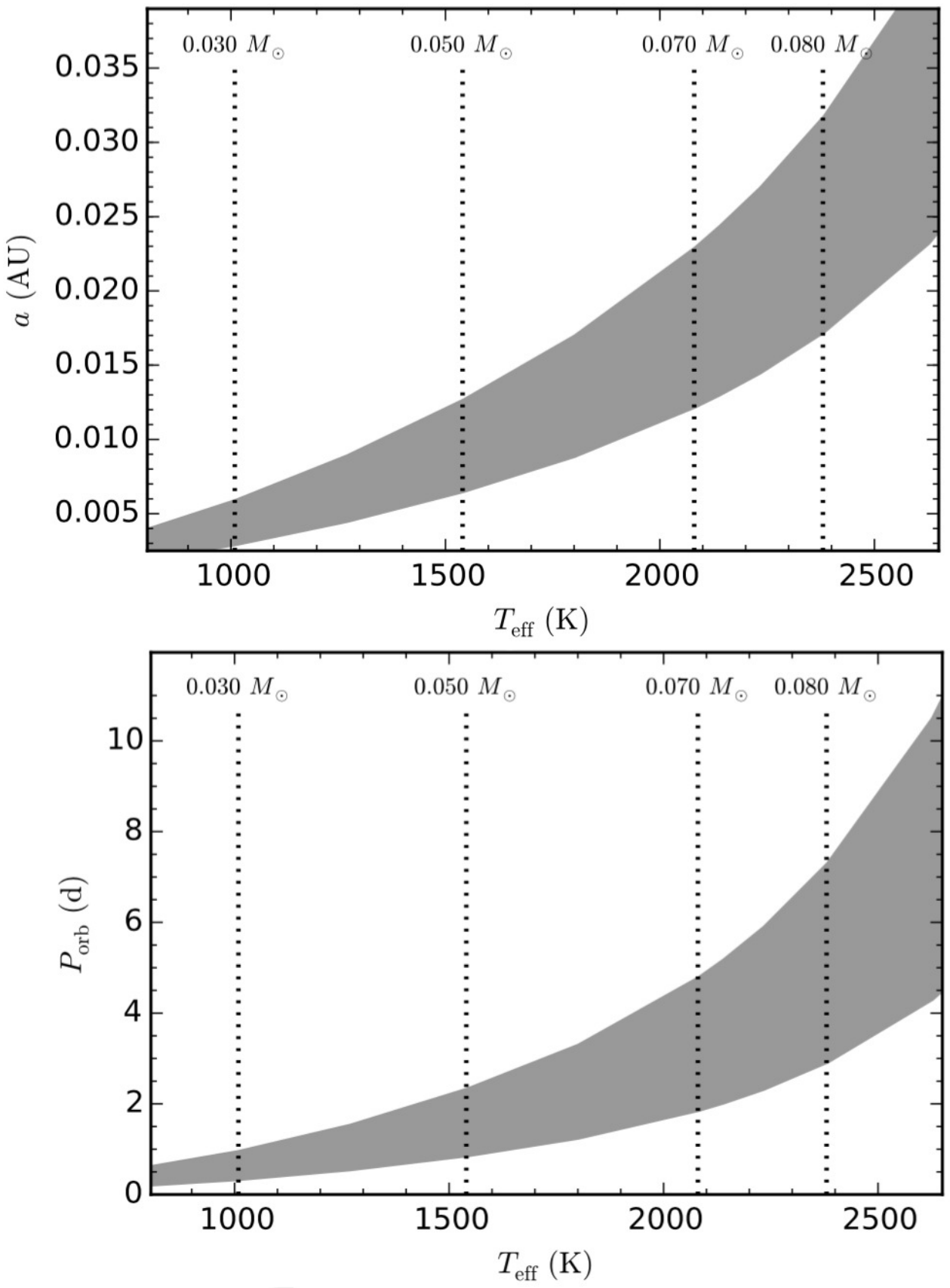

Figure 1: **Top:** Semi-major axis ($a$) for the habitable zone of 1 Gyr-old ultra-cool dwarfs as a function of $T_{\rm eff}$, using the dependency of the habitable zone with $T_{\rm eff}$ and luminosity from Barnes & Heller (2013) and evolutionary models from Baraffe et al. (2003). **Bottom:** Orbital period for habitable-zone planets with $\leq 15 M_\oplus$ as a function of $T_{\rm eff}$. Typical host-star masses at 1 Gyr are indicated in each panel.

The notion to target low-mass stars and brown dwarfs as promising hosts for the detection and characterization of habitable exo-Earths with *Spitzer* is not new. Triaud et al. (2013) developed this idea and estimated that 5,400 *Spitzer* hours would be needed to yield a 90% probability of finding one or more transiting systems in a sample of 120 ≥M3 dwarfs, assuming random planetary orbital orientations. Such a sizable request for *Spitzer* time is perhaps too risky for the anticipated yield. However, by refining the parent sample to focus solely on L and T dwarfs—which have tighter habitable zones than M dwarfs—and then only on the subset seen nearly equator-on, the detection efficiency can be increased by at least an order of magnitude.

The probability of a transit can be approximated as the ratio of the host star's radius to the orbital semi-major axis, and is about 5% for a planet in the habitable zone around a 1500 K brown dwarf: a factor of 3 greater than for the Trappist-1 host star (M8, $T_{\rm eff} = 2550$ K; Gillon et al., 2016). Further restricting the parent sample to high-inclination ($i > 80°$) rotators can drive the probability of transit close to 100%. While the inclinations of the spin axes of brown dwarfs are not a priori known, ongoing photometric monitoring and high-dispersion spectroscopic campaigns to measure $v \sin i$ are now furnishing these for an increasingly large sample of L and T dwarfs. An indirect measure of the viewing geometry of L and T dwarfs may also be contained in their intrinsic near-infrared colors, with redder objects appearing closer to equator-on than bluer objects (Metchev et al., 2015; Vos et al., 2017).

A photometric monitoring campaign with *Spitzer* focused on L and T dwarfs viewed equator-on can thus have an outsized yield of rocky habitable-zone exo-Earths. Dressing & Charbonneau (2015) estimate $0.16^{+0.17}_{-0.07}$ Earth-sized (1–1.5 $R_\oplus$) planets per M dwarf habitable zone. If the same statistics hold for brown dwarfs, a sample of 20 L and T dwarfs will reveal at least one transiting habitable-zone exo-Earth in a 100 hr per-target observation—in 2000 *Spitzer* hours—at the ~95% confidence level. Ground-based and *JWST* follow-up will then refine the orbital periods and spectroscopically characterize the transit planets' atmospheres.

Smaller planets will also be detectable, and may well comprise the bulk of the transit planet yield around brown dwarfs. Triaud et al. (2013) find that individual transits of even Mars-sized (0.5 $R_\oplus$) planets would be detectable with *Spitzer* (Figure 2). With the exoplanet size distribution showing at least equal numbers of Mars-sized as Earth-sized planets around M dwarfs (Dressing & Charbonneau, 2015), the projected yield of any kind of rocky planets around brown dwarfs could well be at least double the above estimates.

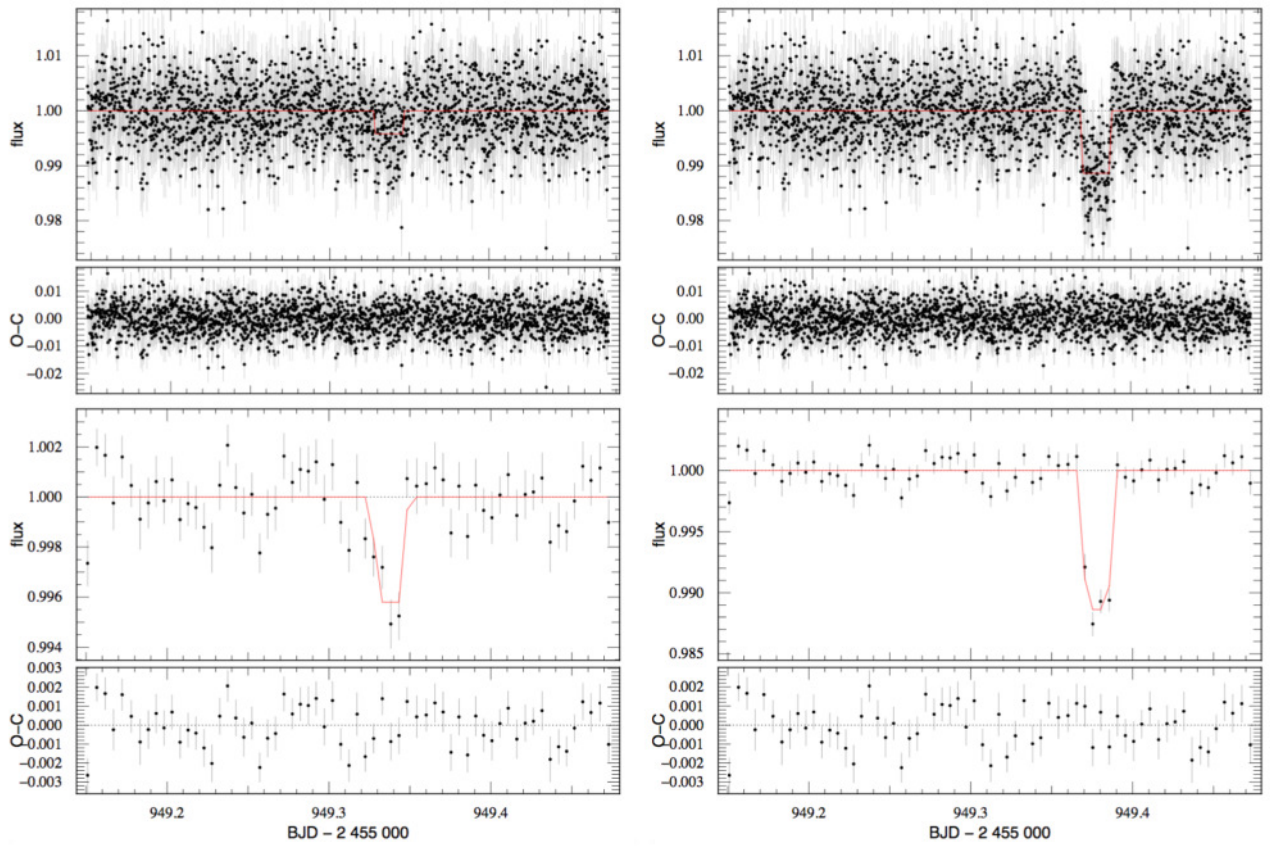

Figure 2: **Top 4 panels:** Archival *Spitzer* IRAC [4.5] data injected with simulated transits, and O-C residuals. **Bottom 4 panels:** Same as above, with the data and model binned for visual clarity. *Left panels:* A $1R_{\rm Mars} = 0.5R_\oplus$ planet transiting a $0.9R_{\rm Jup}$ brown dwarf. *Right panels:* A $1R_\oplus$ transit around a $0.9R_{\rm Jup}$ brown dwarf. Figure from Triaud et al. (2013).

*Spitzer*, with its long continuous staring capability and unparalleled sensitivity over the 3–5 $\mu$m spectral peak of L and T dwarfs, is uniquely capable of detecting habitable exo-Earths around brown dwarfs. The continuous staring is essential to detect the transits, as they will last only tens of minutes. Even if the expected ≈1% dimming is easily detectable on bright stars with ground-based telescopes, the vast majority of brown dwarfs are too faint for such precise photometry from the ground. That is, *Spitzer* truly opens the monitoring of brown dwarfs as planet hosts. It is unlikely that the requisite thousands of observing hours will be dedicated to exoplanet transit searches, or to any single science question with *JWST*.

## 2.2 Cloud Physics and Atmospheric Dynamics

At atmospheric temperatures below roughly 2,000 K, atoms and molecules form condensates which gravitationally settle to form clouds. With decreasing effective temperature, more and more species condense out, forming multi-layered cloud decks. Such clouds are not only typical of brown dwarfs, but also fundamentally impact the atmospheres of hot jupiters and (directly imaged) giant exoplanets. Condensate clouds affect ultracool atmospheres in three ways: (1) They regulate the emergent spectrum – even controlling the transition between the M and L dwarfs ($T_{\rm eff} \approx 2000$ K) and L and T dwarfs ($T_{\rm eff} \approx 1400$ K). (2) Clouds also impact the efficiency of radiation transport through the atmospheres and, therefore, the cooling and luminosity evolution of brown dwarfs and directly imaged exoplanets. (3) The presence of clouds impacts the local pressure-temperature profiles of the atmospheres, thus likely impacting atmospheric circulation.

Therefore, the ability to model condensate clouds is critical to the understanding of the properties and evolution of brown dwarfs and gas giant planets. Because of the superior data quality obtainable for brown dwarfs, it is expected that brown dwarf cloud models will also provide a key pathway to modeling cloud decks in transiting exoplanets and in directly imaged exoplanets.

Although ultracool model atmospheres have proven adept at reproducing the gross spectral changes of the MLTY spectral sequence (e.g., Cushing et al., 2008; Leggett et al., 2009) and the spectra of gas giant exoplanets (e.g., Barman, 2008; Konopacky et al., 2013; Macintosh et al., 2015), these models are static and one-dimensional. Real atmospheres are dynamic and the formation of clouds is inherently a three-dimensional process; parcels of gas are lofted high in the atmosphere, at which point cloud particles condense and gravitationally settle into clouds. If the clouds are vertically or horizontally heterogeneous due to atmospheric circulation, then the integrated light of a brown dwarf or planet may be modulated by changes in the level of heterogeneity and/or the rotation period of the object.

Indeed observations of L, T, and Y dwarfs have shown that they are variable from red optical to mid-infrared wavelengths. These observations provide strong evidence that some kind of breakup of the iron and silicate cloud decks is responsible for the transition between the L and T dwarfs (Radigan et al., 2014; Apai et al., 2013) and that the clouds are both horizontally and vertically heterogeneous (Buenzli et al., 2012; Apai et al., 2013; Metchev et al., 2015).

In the following we will summarize the most exciting prospects for a *Spitzer* extension for two sets of science questions focusing on cloud physics and on atmospheric circulation.

## 2.3 Rotational Mapping: The Physics of Cloud Decks

Comparative studies within the brown dwarf population demonstrated that ultracool atmospheres, unsurprisingly, are more complex than hotter stellar atmospheres: no single parameter captures their spectral and color diversity. Although effective temperature is the dominant parameter, the diversity seen among brown dwarfs with nearly identical temperatures is likely attributable to the combined effects of multiple secondary parameters (surface gravity, cloud structure, bulk composition, vertical mixing, inclination, atmospheric circulation, etc.), some of which are not readily observable. However, pinning down the effects of these secondary parameters through the traditional approach of identifying trends between the spectra of the brown dwarfs and a given potential secondary parameter would require studies in six- or seven-dimensional parameter space, impossible with the current size of the brown dwarf samples. Therefore, isolating the effects of cloud cover from other secondary parameters – essential for testing cloud models – remains very challenging.

An alternative approach has showed considerable success – in large part due to *Spitzer* – since 2012. Time-resolved high-precision observations of rotational modulations in brown dwarfs has provided *spatial information* from temporal modulations. The spatial information then can be used to study different cloud structures within the same brown dwarf, greatly reducing or eliminating the complicating effects of the other (global) secondary parameters (surface gravity, bulk composition, vertical mixing). Rotational modulation studies has allowed the *Spitzer* community to explore many aspects of condensate clouds.

Rotational modulations have been used to assess the occurrence rate of heterogeneous clouds in brown dwarfs in ground-based studies (e.g., Radigan et al. 2014), with *Hubble Space Telescope* spectroscopy (Buenzli et al., 2014) and, on the yet largest sample, with *Spitzer* (Metchev et al., 2015).

These studies have all provided consistent results and several exciting statistical insights into cloud properties across the L/T spectral types. *Spitzer* has remained uniquely capable of extending rotational modulation studies to the even cooler Y-dwarfs (Cushing et al., 2016; Leggett et al., 2016).

Time-resolved, multi-wavelength observations (*Spitzer*/IRAC Ch1 and Ch2; and simultaneous *Spitzer*/IRAC and *HST*/WFC3 spectroscopy) have enhanced the diagnostic power of the rotational modulations even further: Because different wavelengths penetrate to different depths in ultracool atmospheres, the longitudinal 'scans' can be applied to multiple layers in a single rotation. For example, combined *Spitzer+HST* observations have provided six-layer scans of L, L/T, and T-type brown dwarfs (Buenzli et al., 2012; Yang et al., 2016). Detailed modeling of the emission source function in the atmospheres and the color and spectral signatures of the rotational modulations have confirmed that they originate from cloud layers (relatively gray absorbers) located between 0.3 and 10 bars, with evidence for different cloud structures for objects with different spectral types.

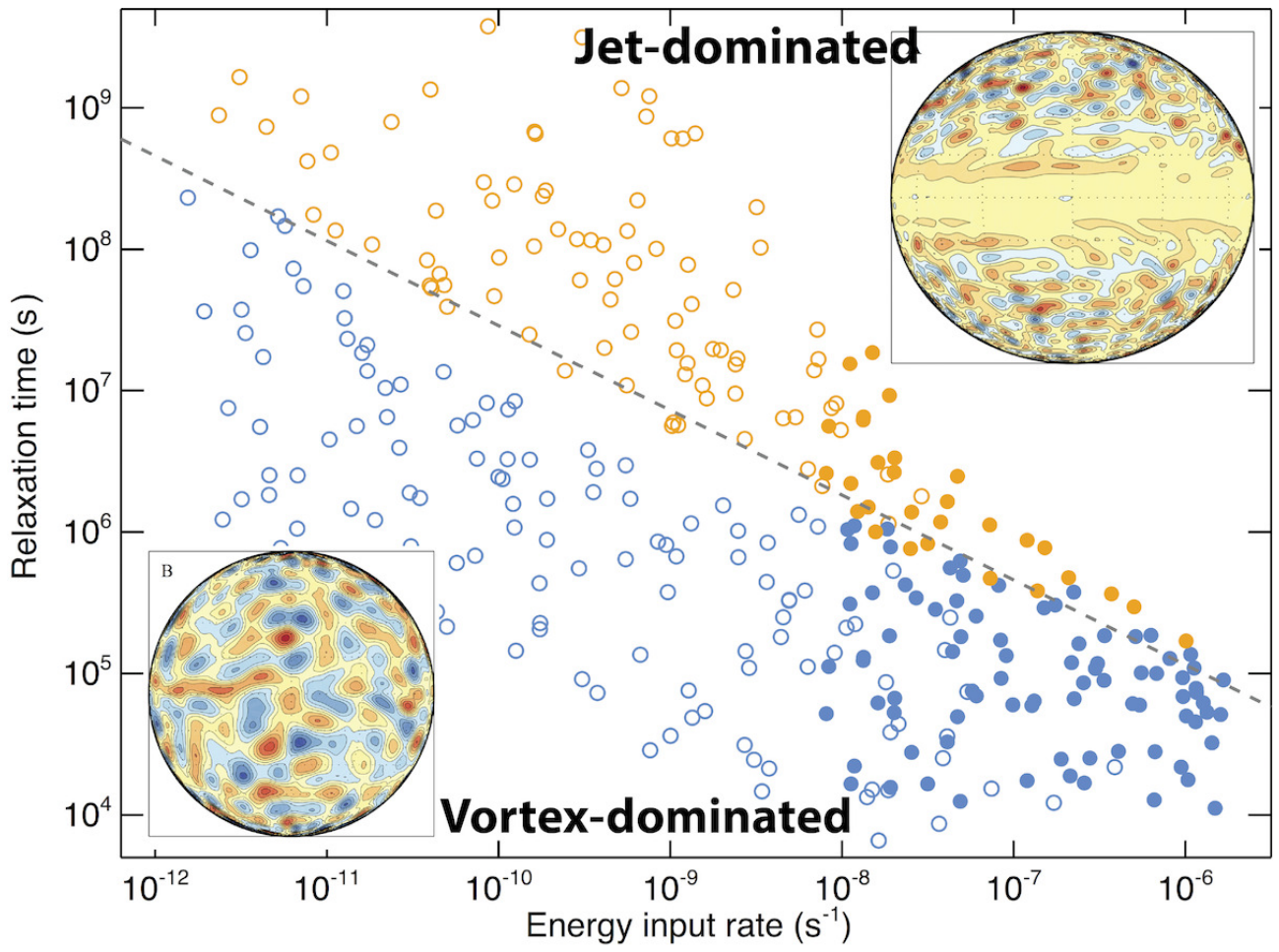

Figure 3: Simple shallow water simulation predict two fundamentally different atmospheric circulation regime: jet-dominated and vortex dominated. Jets appear where the rotation period is short compared to the typical heat transfer timescale; all Solar System planets have jet-dominated (zonal) circulation. Vortex-dominated circulation is predicted for slowly rotating objects with rapid heat transfer rates. Modified from Zhang & Showman (2014).

Simultaneous *HST* and *Spitzer* observations have demonstrated the power of multi-wavelength rotational modulation measurements. In the era of *JWST*, powerful multi-wavelength, high-precision, time-resolved spectroscopy will be available, but for only a small fraction of the several thousand brown dwarfs that could be studied. Identifying the ideal sample for *JWST* spectroscopy is an important, but very challenging task: ground-based observations provide samples that are strongly biased toward high-amplitude, short-period, bright variables. Ground-based observations, for example, remain blind to almost all rotational modulations in mid-to late-T-type brown dwarfs. An extension of the Spitzer mission would enable a large pre-*JWST* survey to identify the ideal set of variable brown dwarfs to provide detailed, quantitative characterization of cloud cover as a function of fundamental atmospheric properties through a *JWST* time-resolved spectroscopic survey.

### 2.3.1 Spitzer Monitoring: Atmospheric Circulation in Ultracool Atmospheres

*Spitzer*, and then *HST* have revealed that a fundamental property of hot jupiters is a powerful super-rotating equatorial jet (e.g. Knutson et al. , 2007). This pattern was predicted prior to the observations as an outcome of the extreme forcing due to the dayside injection of the stellar flux. However, the vast majority of exoplanets in the universe will be unlike hot jupiters: most planets receive far lower levels of irradiation. In contrast to hot jupiters, for example, directly imaged exoplanets are essentially unaffected by their host stars. The spectra of such directly imaged exoplanets remains poorly understood and often difficult to fit with one-dimensional atmospheres (e.g., Marois et al. 2008; Marley et al. 2010; Skemer et al. 2012).

It is expected that most exoplanet atmospheres will be heterogeneous and their spectra and atmospheric structures will be fundamentally impacted by atmospheric circulation (Figure 3). Atmospheric circulation will redistribute heat, define regions of updraft and downdraft, influencing the patterns of condensate clouds within the atmospheres. Understanding the cloud distribution will require understanding atmospheric circulation – a quest that is likely to be at the forefront of exoplanet science over the next decade or two.

Just as they do for cloud physics, brown dwarfs also provide a shortcut for understanding atmospheric circulation: data with superb quality can be collected for atmospheres with low temperatures that would be inaccessible in exoplanet systems. In the recently completed *Extrasolar Storms* Exploration Science program 1,144 hours of *Spitzer* Space Telescope time was used for the first comprehensive photometric monitoring of brown dwarfs.

In the *Extrasolar Storms* program *Spitzer* Ch1 and Ch2 observations have probed brown dwarf lightcurve evolution over 21 different temporal baselines – from a single rotation to up to a thousand rotations (covering more than one year; Figure 4). When the program began, the timescales over which light curve evolution would occur, and whether it would commonly occur, was unknown. *Extrasolar Storms* revealed continuous lightcurve evolution in all six of its targets and showed that lightcurves evolve quickly, over timescales of just a few rotational periods (Yang et al. 2016; see Figure 5). Apai et al. (2017) showed that the light curve evolution in the high-amplitude targets is consistent with the effects of planetary-scale waves (k=1 and k=2 wavenumbers) propagating through the atmospheres with slightly different apparent periods, likely due to differential rotation in a zonal circulation regime (Figures 6 and 7). These observations provide the first comprehensive view of atmospheric circulation in non-irradiated ultracool atmospheres and exemplify the power of the quasi-continuous high-precision photometric monitoring that *Spitzer* can provide.

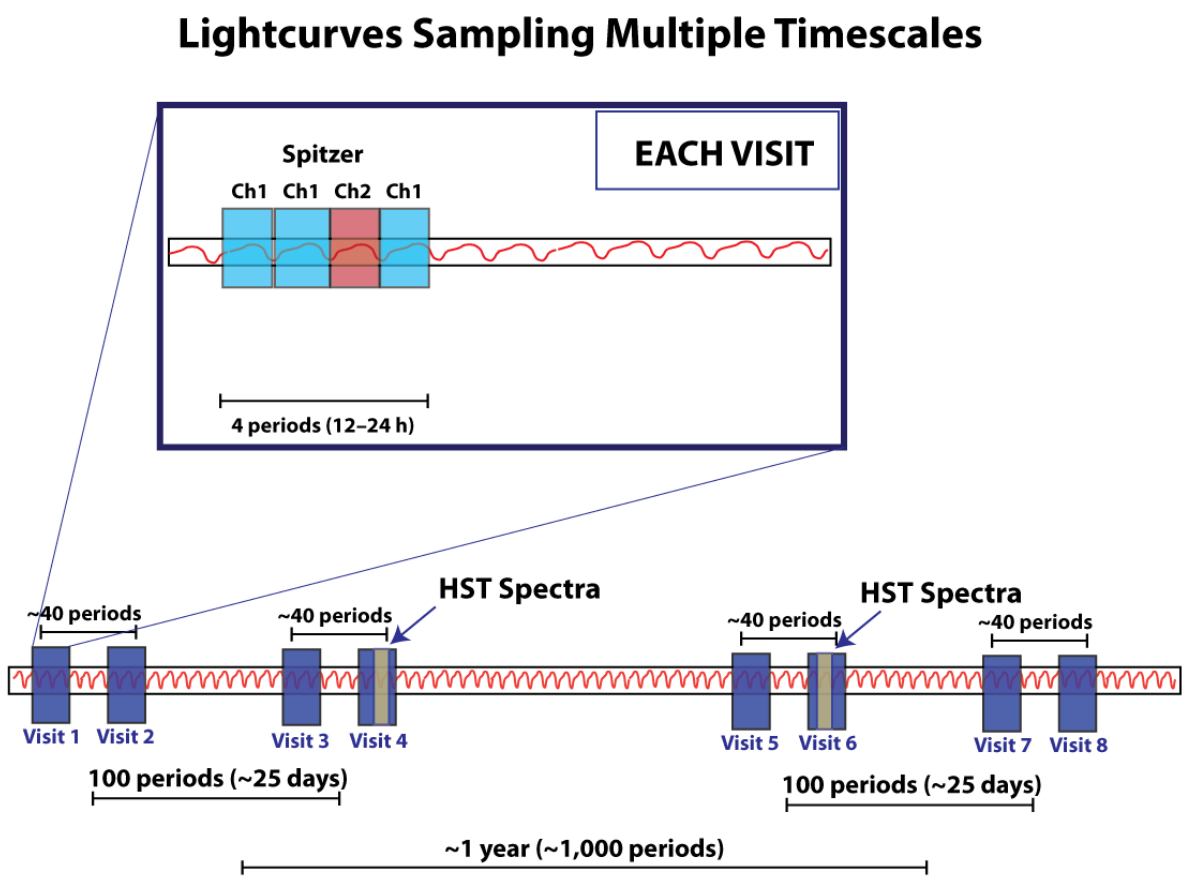

Figure 4: The observing strategy of the *Extrasolar Storms* *Spitzer* Exploration Science program combined long-term monitoring with continuous photometry over four rotational periods. This strategy worked well to probe the lightcurve evolution in brown dwarfs, the timescales of which were unknown. Simultaneous time-resolved spectroscopic observations with *HST* provided additional astrophysical constraints, which helped the interpretation of the short- and long-term lightcurve evolution. From Yang et al. (2015)

The *Extrasolar Storms* results were further enhanced by a few simultaneous, brief ($\sim 6-10$ h long) *HST* time-resolved infrared spectroscopy. The short *HST* spectral scans provided an in-depth picture of the cloud configurations at five pressure levels (typically between 0.1 and 10 bars), which could then be interpreted in the context of the long-term evolution provided by *Spitzer*. These combined observations revealed, among other results, evidence for high-altitude particles in L–type brown dwarfs (Yang et al., 2015), and longitudinal-vertical cloud patterns typical to objects with a given spectral type ([Yang et al., 2016](#)).

Of course, this first study raised several key questions on atmospheric circulation:
(1) Do earlier and later spectral type, and low-amplitude brown dwarfs harbor the same atmospheric circulation as identified in the three high-amplitude L/T brown dwarfs?
(2) What physical process drives the planetary-scale (k=1, k=2) waves observed?
(3) How does surface gravity influence atmospheric circulation in ultracool atmospheres?
(4) Do deeper atmospheric layers follow the waves observed in the upper atmosphere?

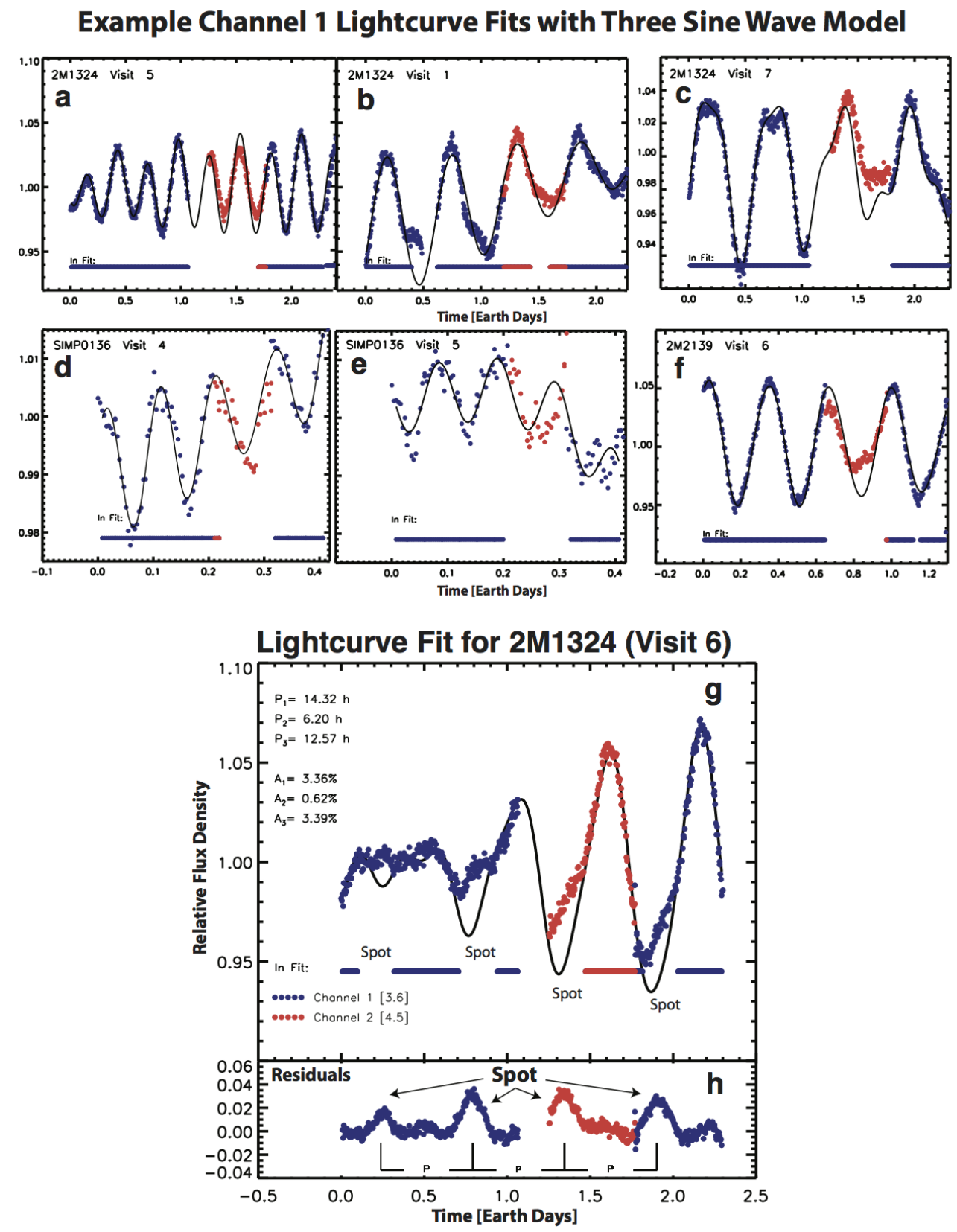

Figure 5: The *Extrasolar Storms* program found that all six brown dwarfs displayed lightcurve evolution over timescales of a few rotational periods, regardless of the rotational period of the object (from 1.4 hours to 13 hours). The observations showed dramatic amplitude evolution. The lightcurve evolution is not irregular but slowly evolving quasi-periodic. From Apai et al. (2017)

Answering the above questions requires continuous or quasi-continuous monitoring of individual brown dwarfs with sub-percent precision over timescales of 5–10 days. The currently available dataset (mainly from *Extrasolar Storms*) demonstrated that lightcurves evolve fast: it is not feasible to recover light curve evolution from fragmented data. No ground-based observatory provides the quasi-continuous, ultra-precise photometric monitoring that is essential for deciphering atmospheric dynamics in ultracool atmospheres; due to operational constraints, *HST* and *JWST* will also be unable to provide the long and very precise datasets that *Spitzer* can.

We envision a Spitzer program in which brown dwarfs of different spectral types are monitored continuously or quasi-continuously for about ten rotations (approximately 40–100 hours per object) in a single channel (Ch1) with the goal of answering questions 1–3 from the above list. These observations would extend the *Extrasolar Storms* sample, which only covered up to four rotations continuously. Furthermore, by restricting the observations to a single channel, the systematics related to switching IRAC channels can be greatly reduced, improving data quality for the fainter and lower-amplitude brown dwarfs. If the sample contains high- and low-gravity brown dwarfs, the impact of surface gravity on atmospheric dynamics can be assessed. Coordinated *HST* or *JWST* spectral maps for one rotation of each object would provide depth-(pressure-)dependent information that would allow exploring the time-evolving waves in the context of longitudinal-vertical cloud structures, similarly to the approach successfully adopted in the *Extrasolar Storms* program. Such a *Spitzer* program could provide unique and novel insights into atmospheric circulation in non-irradiated, ultracool atmospheres. It is likely that the insights gained would greatly enrich the interpretation and modeling of directly imaged exoplanets, as well as transiting, but long-period, gas and ice giants.

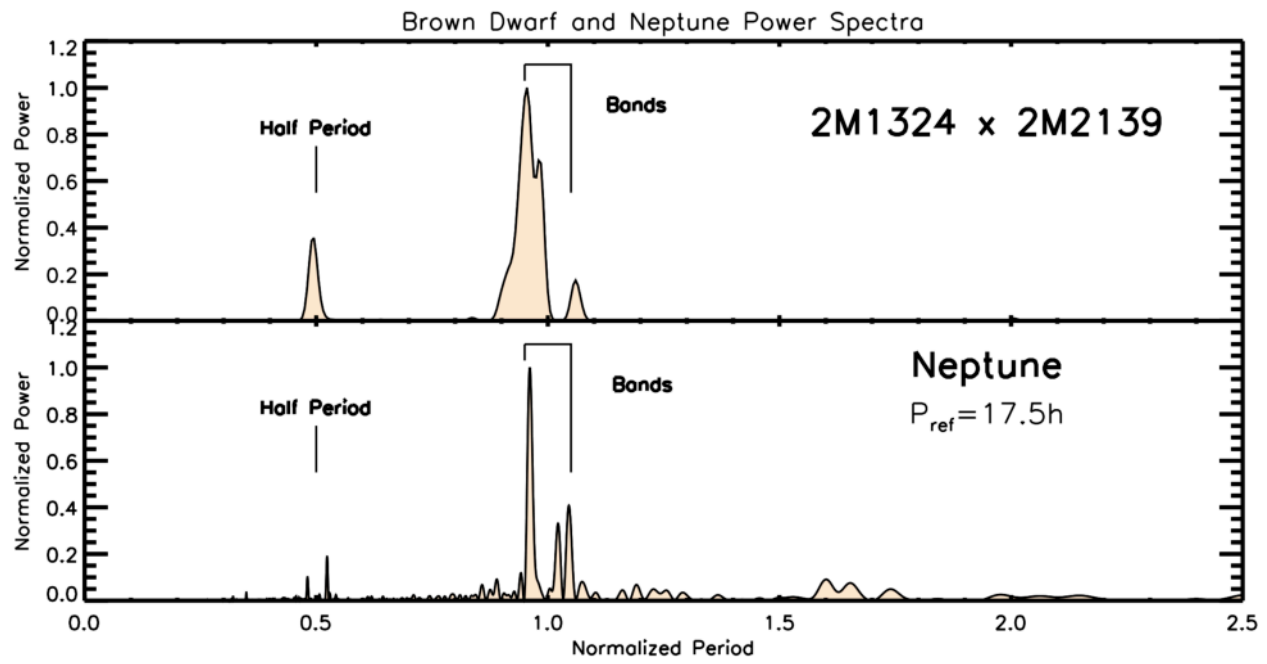

Figure 6: The power spectrum derived for two L/T transition brown dwarfs in the *Extrasolar Storms* program resembles closely that derived for Neptune from *Kepler* scattered light observations. Multiple peaks are seen around the reference rotational period, indicating the presence of atmospheric waves with slightly different observed periods. In Neptune this pattern emerges due to differential rotation of the zones defined by zonal circulation. From Apai et al. (2017).

The Spitzer observations are uniquely valuable for constraining atmospheric dynamics due to the very high photometric precision, sensitivity, *and* continuous monitoring capability. If the data rates degrade due to data downlink limitations, the program will obtain more sparsely sampled lightcurves. This limitation will impact the faintest, lowest-amplitude, and fastest-rotating objects (as their lightcurves would have the lowest completeness) – a factor that could be accounted for in the project planning phase. With thousands of known brown dwarfs, it is very likely that a target sample could be defined – even for seriously degraded data downlink rates – that would allow *Spitzer* observations to answer the science questions identified above.

## 2.4 The Full Functional Form of the Milky Way's Substellar Mass Function

One of the most basic questions in all of astrophysics is how the Universe takes clouds of gas and converts them into stars. The star formation process has been studied in environments both young (moving groups and stellar nurseries) and old (globular clusters and the general field population), and whereas the form of the mass function is well established for higher mass stars, it is far less established for the lowest mass stars and brown dwarfs. Objects of lowest mass may, in fact, have several paths to creation, depending upon their birth environment. Sites replete with high-mass O stars can create brown dwarfs through the ablation, via O star winds, of protostellar embryos that would otherwise have formed bona fide stars. Moreover, protostars in rich clusters with many high mass members may, via dynamical interactions, be stripped from their repository of accreting material, thus artifically stunting their growth. Star forming regions lacking higher mass stars will possess neither of these processes; do they still create brown dwarfs naturally?

Although establishing the diversity of low-mass star formation is critical to understanding the underlying physics in differing environments, we can look at the end results of these various processes using the well-mixed, field population of older low-mass stars and brown dwarfs near the Sun. Burgasser (2004) has shown that the space density of the coldest brown dwarfs can be used both to test various models of the mass function but also to indirectly measure star formation's low-mass cutoff, assuming there is one. With these goals in mind, Kirkpatrick et al. (in prep.) pursued astrometric monitoring programs with *Spitzer*/IRAC in Cycles 7, 8, 9, and 13 to measure the distances to all ≥T6 dwarfs believed to lie within 20 pc of the Sun. The 20-pc limit was chosen because (1) that is within the distance for which *Spitzer* astrometry can deliver parallaxes with ∼10% accuracy over a ∼2-yr baseline and (2) that volume encloses a sufficient number of objects for adequate statistics. An example of one of these *Spitzer* parallax and proper motion measurements is shown in Figure 8.

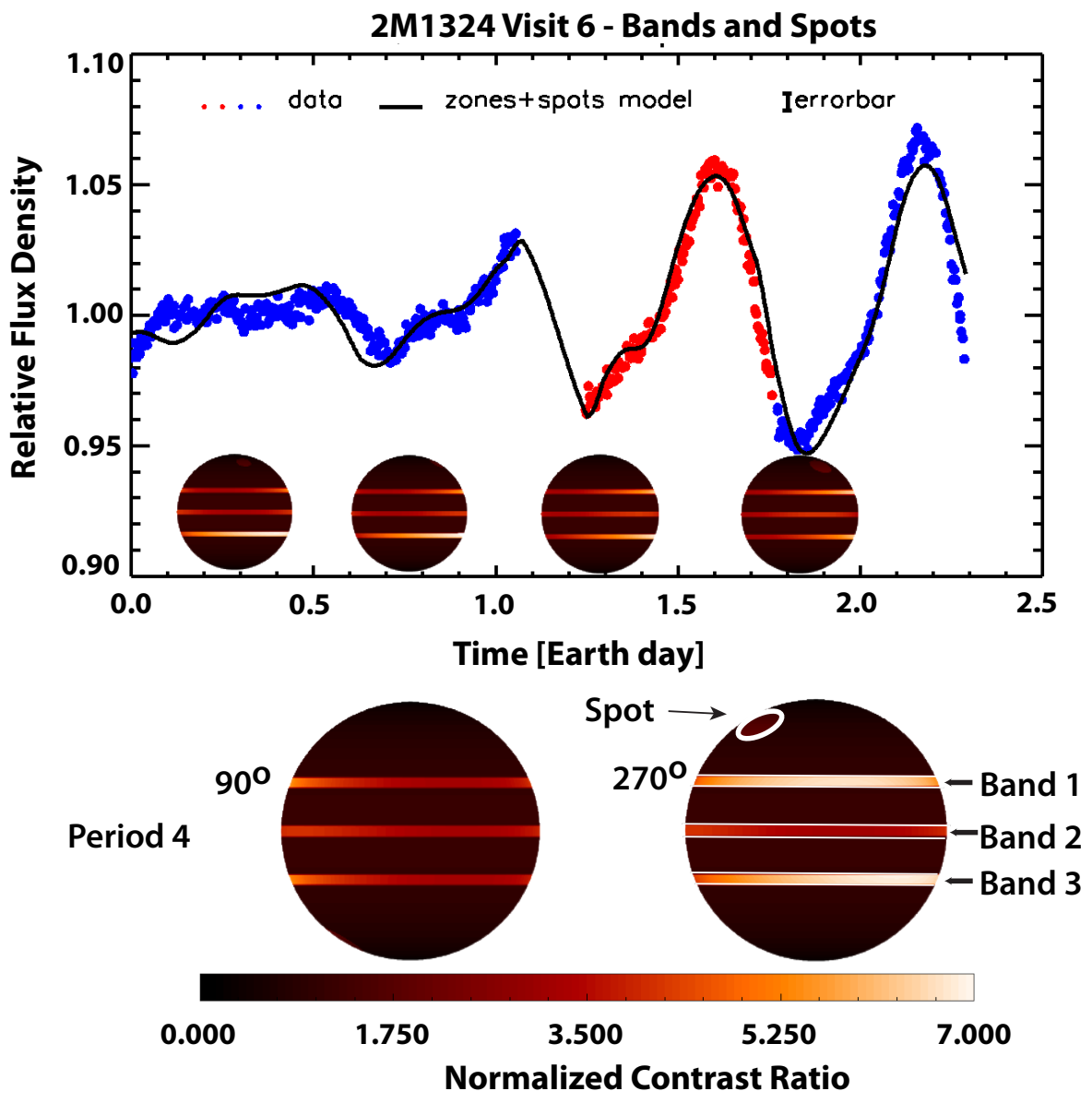

Figure 7: Observed and modeled lightcurve from the *Extrasolar Storms* program. In this particular visit the lightcurve of the L/T transition brown dwarf 2M1324 began as a very low-level variable that rapidly evolved into a high-amplitude (∼10%) lightcurve.

These distance measurements for the 20-pc ≥T6 sample have enabled us to place much better constraints on the functional form of the field mass function, as shown in Figure 9. Despite this triumph, the portion of the field mass function located between the low-mass stars (late-M types) on the far left of this plot and the ≥T6 dwarfs on the far right is poorly constrained. Figure 9 shows these intermediate points in grey. Not only do these points have large error bars or lower limits, but they also have poor resolution in temperature, making comparison to the shape of the predicted mass function problematic.

This $1000 < T_{eff} < 2500$ K region, which roughly covers spectral types from L0 through T6, can be covered by a combination of *Gaia* and the *Spitzer* extension. Figure 10 shows that *Gaia* can provide astrometry for the 20-pc volume out to a spectral type of ∼L5; at colder types, *Gaia* detectability drops sharply. These colder objects are much more easily detected in the infrared, either from ground-based observatories at $J$ or $H$ or, even more easily, by *Spitzer* at 4.5 $\mu$m. This spectral type range contains roughly 200 objects out to 20-pc and of those only ∼40 have ground-based parallaxes good to <10% accuracy despite 15 years of hard-fought observation (see Dahn et al. 2017 and the Database of Ultracool Parallaxes[1]).

Prior IRAC observations have demonstrated that high-quality parallaxes can be obtained for the remaining 160 objects with a modest investment of *Spitzer* time. Observations with 270 s of integration (which currently corresponds to 579 s of clock time, with overheads) will easily enable us to reach SNR = 100 for our targets, as well for as a sufficient number of background stars needed for accurate epoch-to-epoch frame re-registration. An accuracy of 10% is achievable with at least a dozen measurements per source distributed roughly evenly over the 2-yr period of the *Spitzer* extension. Observations will naturally be taken near maximum parallax factor since these coincide with *Spitzer's* visibility windows. The total time request for such a program would be ∼300 hr. Our prior observations, which contained time constrains to place observations within a particular week, were of short duration and often used to efficiently fill gaps in the rest of the observing schedule. A similar methodology could be employed in the *Spitzer* extension.

[1] See http://www.as.utexas.edu/~tdupuy/plx/Database_of_Ultracool_Parallaxes.html

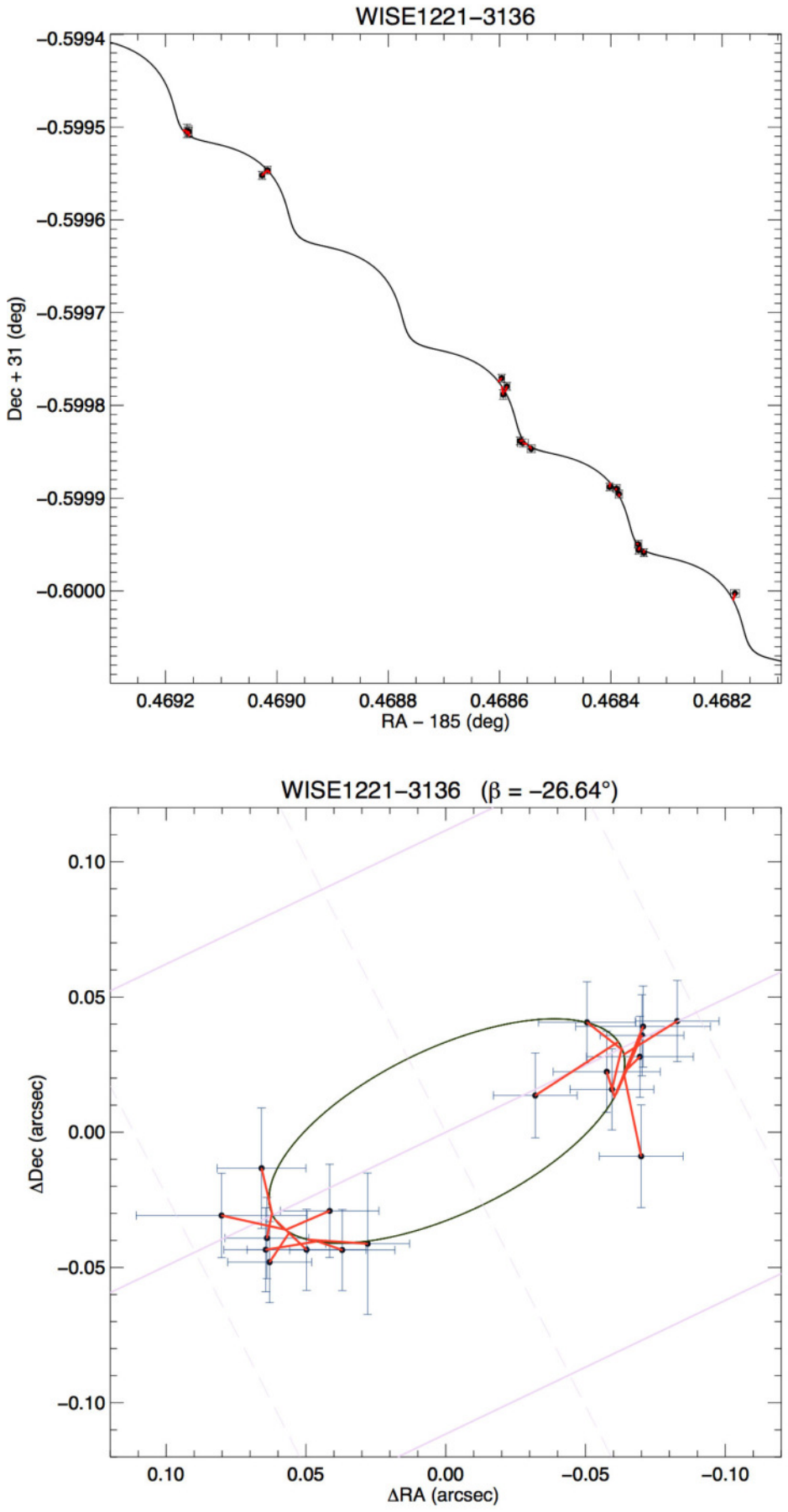

Figure 8: (Top) A plot of sky coordinates showing the best fitting parallax + proper motion solution to the T6.5 dwarf WISE J122152.28−313600.8 using our *Spitzer*/IRAC 4.5 $\mu$m astrometric observations. (Bottom) The resulting fit of the parallactic ellipse once proper motion has been subtracted. The faint grid in the background shows lines of constant ecliptic latitude (solid) and longitude (dashed) on a 100 mas grid spacing. This object is at 14.6 pc.

### 2.5 The Basement of the Initial Mass Function in Young, Nearby Moving Groups

Nearby young moving groups have proven to be a gold mine for uncovering extremely low mass brown dwarfs ($< 13\,M_{Jup}$). Recent years have shown that young associations such as TW Hydrae (5–15 Myr; Weinberger et al. 2013), Beta Pictoris (20–26 Myr; Malo et al. 2014), AB Doradus (110–130 Myr; Barenfeld et al. 2013), and Tucana Horologium (20–40 Myr; Kraus et al. 2014) have a plethora of planetary mass brown dwarfs that are critical to comparative studies with directly imaged exoplanets. These include objects such as PSO 318 which is a 5—9 $M_{Jup}$ member of $\beta$ Pictoris (Liu et al., 2013), WISE 1147 which is a 5–9 $M_{Jup}$ member of TW Hydrae (Schneider et al., 2016), and WISE 1119 which is a 5–9 $M_{Jup}$ resolved near equal mass binary in TW Hydrae (Best et al., 2017).

Within 150 pc there are more than 25 known associations with ages ranging from 5–150 Myr. A handful of these are small enough on the sky for a *Spitzer* IRAC tiling program that would detect the coldest compact sources formed through star formation processes. This includes young free-floating late-type L and T dwarfs that would have masses down to only 2–5 $M_{Jup}$. While such candidates require *JWST* for spectroscopic confirmation, they are also the most easily detectable objects of their mass due to their relatively young ages.

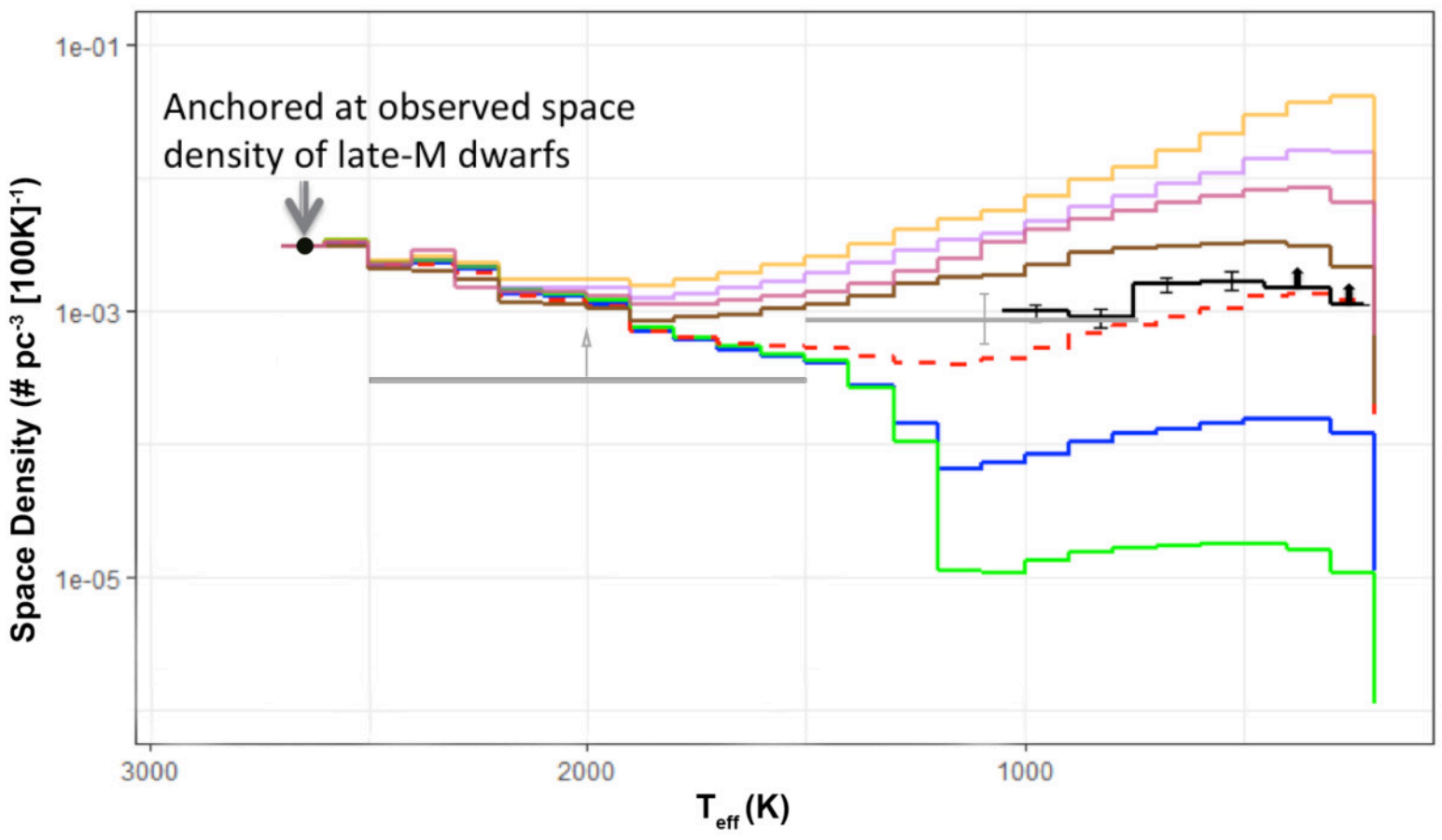

Figure 9: The space density of the 20-pc sample as a function of effective temperature. The black points on the far right represent the space density for ≥T6 dwarfs using known objects with distances measured by *Spitzer* (Kirkpatrick et al., in prep.). These empirical measurements are compared to a series of simulations (colored lines, from A. J. Cayago, priv. comm.) based on various forms of the mass function ranging from simple power laws and the log-normal form favored by Chabrier 2001(colored lines above the dashed curve) to bi-partite power laws favored by Kroupa et al. 2013 (red dashed curve and the ones below it). All simulations are anchored on the left at the well-measured space density of the late-M dwarfs (black dot; see Burgasser 2004). The space density between this point and the late-T dwarfs has never been measured for a volume-limited sample with excellent statistics; our current measurements in this region (from Cruz et al. 2007 and Metchev et al. 2008; shown in grey) are poorly constrained. Measuring the space density in this region can be completed by *Gaia* and the *Spitzer* extension.

Initial mass function studies of nearby star-forming regions have begun probing the basement of the mass function. Spitzer has played a significant role in these studies by identifying candidates in star forming groups such as Chamaeleon I, Taurus, and Perseus (Esplin & Luhman, 2017; Esplin et al., 2017). While these very young associations ($\sim 1$ –3 Myr) and their yield of free floating planet candidates is exciting, the groups suffer from (at times extreme) reddening and are distant ($> 150$ pc), making their lowest mass members difficult to study in detail. Consequently, our current understanding of the initial mass function of young associations into the planetary-mass regime ($\lesssim 13\,M_{\rm Jup}$) is poorly constrained.

Young moving groups in the Solar neighborhood (Zuckerman & Song, 2004; Torres et al., 2008) provide an excellent compromise between ages ($\sim$10–200 Myr) and distances ($\sim$10–150 pc) for the detection and study of planetary-mass objects (e.g., see Gagné et al., 2015b; Faherty et al., 2016). Their relatively older age ensures that the interstellar medium in which young objects are embedded has dissipated. As stated above, studies have shown that nearby young moving groups can provide a treasure trove of extremely low mass substellar mass objects. A dozen planetary-mass objects with model-dependent estimated masses in the range 6–13 $M_{\rm Jup}$ have recently been detected (Liu et al., 2013; Schneider et al., 2014; Gagné et al., 2015a; Schneider et al., 2016; Kellogg et al., 2016; Gagné et al., 2017b), making it possible to obtain the first planetary-mass object spatial densities in nearby young moving groups (Gagné et al., 2015b, 2017a). These first results suggest an over-density of planetary-mass objects compared to the predictions of a fiducial log-normal initial mass function, but they are still based on small-number statistics and are inconsistent with some micro-lensing and open cluster studies (Scholz et al., 2012; Mróz et al., 2017).

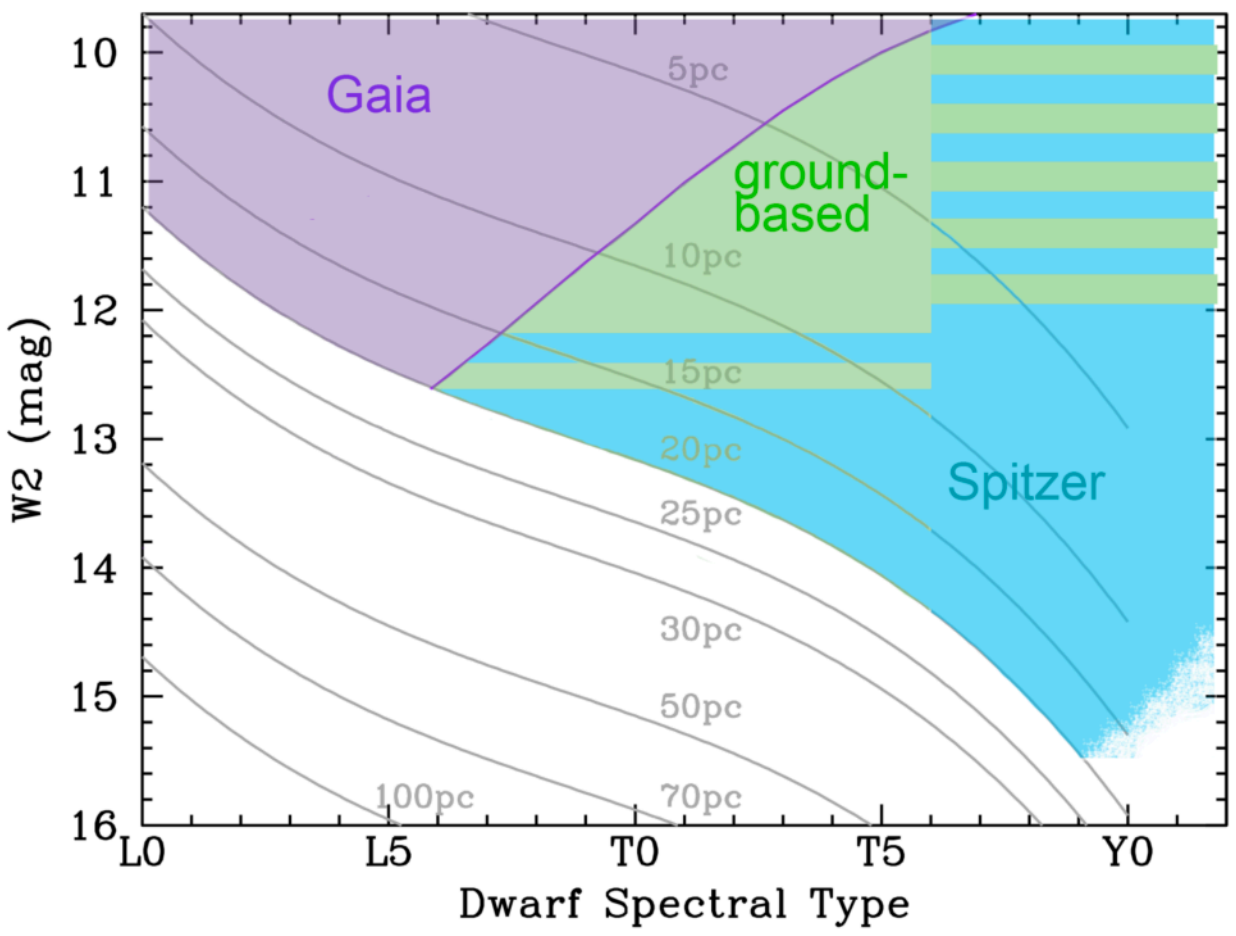

Figure 10: The *WISE* W2 apparent magnitude as a function of spectral type for dwarfs from early-L through early-Y. (*WISE* W2 ≈ IRAC 4.5 $\mu$m.) Lines of constant distance are shown by the grey curves, with distances labeled. The 20-pc sample is shown by the shaded region. Most of the mid-L and earlier dwarfs will have their distances measured by *Gaia*, but later dwarfs quickly become undetectable in *Gaia's* <1 $\mu$m bandpass. Ground-based astrometry at <2.5 $\mu$m is possible for some of the mid-L to mid-T dwarfs, although coverage is spotty and the most distant examples require very long exposures even on 8m-class telescopes. *Spitzer* can easily provide high-quality astrometric monitoring of the rest of the mid-L and later sample because these objects are much brighter in IRAC's 4.5 $\mu$m band.

An extension of the *Spitzer* mission would allow the community to build IRAC infrared imaging tiles around the core of young moving groups such as TW Hya (Kastner et al., 1997), Carina, Alessi 13, IC 2602, the Pleiades, etc. to detect their members with masses down to 2–5 $M_{\rm Jup}$. This would provide a much larger planetary-mass population to build the initial mass functions of young moving groups and determine whether the tentative detection of an over-population is real. This will not only provide information on the formation mechanisms of isolated planetary-mass objects, but it will also provide valuable targets for a detailed atmospheric study with *JWST*.

Given that identifying new members of any known group requires more than just photometry, robust kinematics would be necessary to confirm bona fide members. For the coldest objects detected, we envision two paths to obtaining the required proper motion measurements. An extension of the *Spitzer* mission allows for the possibility of multi-epoch imaging of nearby groups. Alternatively, one could combine a single IRAC epoch with *WISE* catalog data to obtain the required proper motion precision of $\lesssim 60\,\mathrm{mas\,yr^{-1}}$ (Gagné et al., 2015c).

Astrometric precisions of $0.04''$ will be achievable from a single IRAC frame on all objects detected at a good ($\gtrsim 10$) signal-to-noise ratio, compared to the $\approx 0.4''$ precision of *WISE* for the faintest objects (Luhman, 2014). The first method where proper motion is measured by combining a single IRAC epoch with *WISE* benefits from a large temporal baseline of 7 years or more to achieve proper motion precisions of $< 60\,\mathrm{mas\,yr^{-1}}$ in a single IRAC exposure, but is limited to members detected in *WISE*. This method will be preferable for nearby groups that span the largest angular areas that prohibits a full two-epochs coverage, such as TW Hya. The *WISE* $W2$ magnitude detection limit is located between 15.5 and 16.5 depending on ecliptic latitude, which will correspond to a spectral type ≈T2 or a mass of $\approx 3\,M_{\rm Jup}$ at the age of TW Hya.

In the case of more distant associations ($\gtrsim 100$ pc), multi-epoch coverage with IRAC becomes feasible, and will yield proper motion precisions of $\lesssim 40\,\mathrm{mas\,yr^{-1}}$ for temporal baselines above one year. This will make it possible to obtain much deeper imaging to make up for the larger distance of these groups. The recently developed Bayesian classification algorithms (Malo et al., 2013; Gagné et al., 2014) can be utilized to combine photometric information with kinematics and ascertain membership probability.

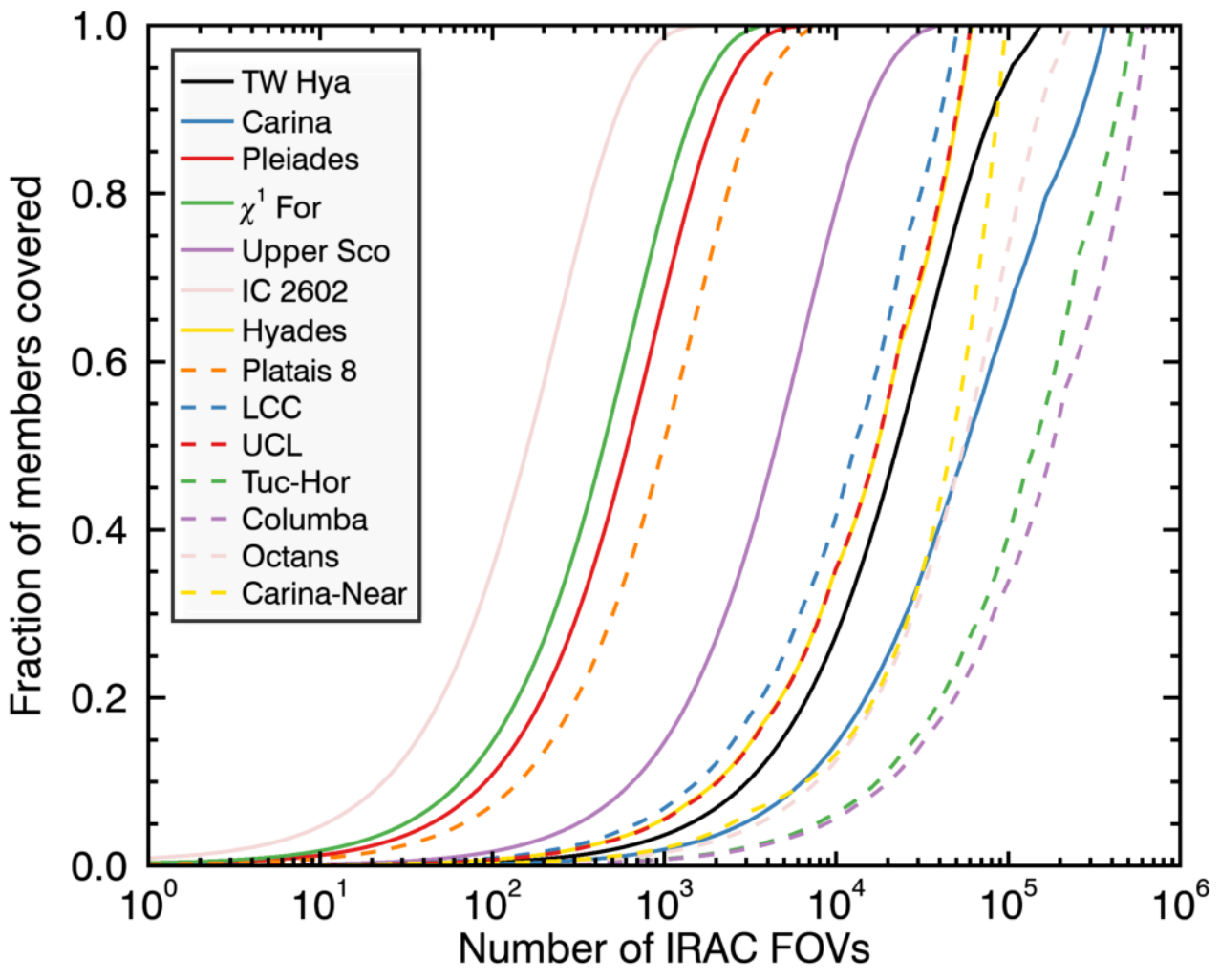

Figure 11: Fractional coverage of association members as a function of the number of IRAC $5.2' \times 5.2'$ tiles. Nearby associations span a larger area of the sky and require more exposures to be fully covered.

Figures 11 and 12 show the number of IRAC exposures necessary to both tile one of several associations and reach a given minimum mass. The fractional coverage of members that can be achieved with IRAC imaging tiles was calculated by drawing $10^6$ synthetic members in $XYZ$ space from the multivariate Gaussian model of each group used in BANYAN $\Sigma$ (Gagné et al., in prep; see Gagné et al. 2017b for a similar model of Carina-Near). The resulting $XYZ$ coordinates were transformed to a 2D ($\alpha \cos \delta$,$\delta$) space where $\alpha$ is the right ascension and $\delta$ is the declination. The full range of the resulting 2D space was divided in a regular grid of $5.2' \times 5.2'$, corresponding to the IRAC field of view, and the number of synthetic members was counted in each grid cell. A search for members was then simulated by counting the synthetic members in each grid cell, starting from the densest regions first, and continuing in decreasing order of density. The resulting fraction of recovered group members as a function of IRAC fields is displayed in Figure 11.

As shown in Figure 12, IC 2602 and $\chi^1$ For are have a combination of distances and ages that makes it possible to reach masses of $\sim 3\,M_{\mathrm{Jup}}$ with a much lower total integration time than other groups ($\sim 200\,\mathrm{hr}$). Some other groups such as TW Hya and Upper Scorpius could in principle be used to reach even lower masses ($\sim 1.5\,M_{\mathrm{Jup}}$) due to their younger age, but obtaining a full coverages at two epochs with IRAC would require a much larger total exposure time ($\sim 5000\,\mathrm{hr}$ twice).

## 2.6 The Coldest Wide-separation Siblings to Solar Neighborhood Stars

The discovery of the 2.3-pc distant, ~250 K, 3-10 $M_{Jup}$ brown dwarf WISEA J085510.74071442.5 (Luhman 2014) underscored the fact that the Solar Neighborhood may contain other cold, or even colder, low-mass objects. If WISE 0855−0714 were somewhat more distant, it would not have been as easily detected by WISE; if it were somewhat colder, it would not have been detected at all. Similar objects – with their cold, planet-like atmospheres not complicated by the presence of a nearby host star – would be prime targets for *JWST* since they can be studied directly as tests of exoplanet atmospheric models.

Both *WISE* and *Spitzer* have the ability to observe at wavelengths near 5 $\mu$m where the coldest brown dwarfs emit their peak flux, but uncovering these objects based on their color alone is difficult.

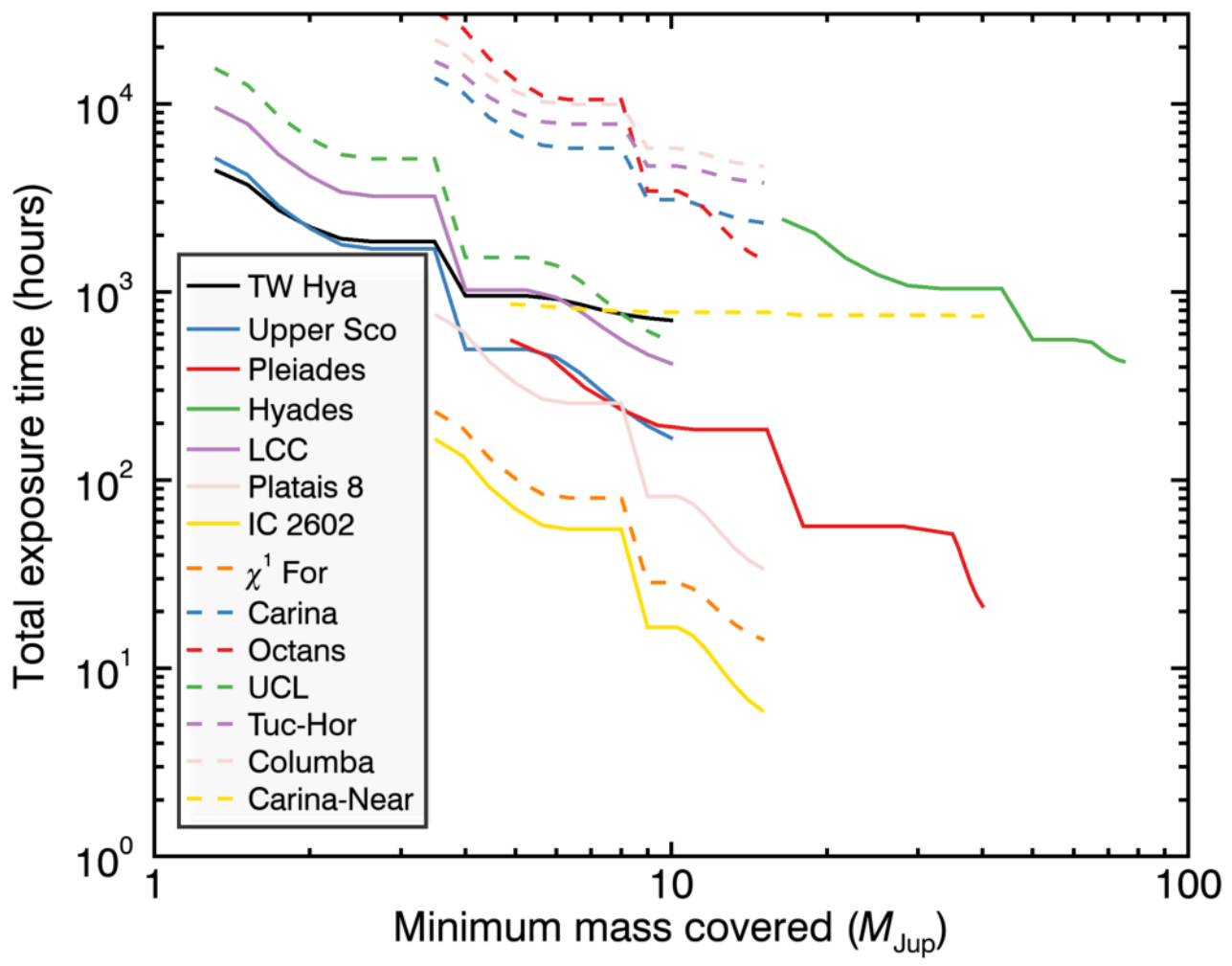

Figure 12: Minimum mass for which $> 90\%$ of members are detected versus the total exposure time needed by *Spitzer* for nearby, young associations. The associations located further away are grouped on a smaller region of the sky and require a smaller amount of IRAC tiles, but longer integration times are needed to detect their faint members.

WISE 0855−0714 is roughly two magnitudes brighter than the limit of the AllWISE Catalog, suggesting that another $15^{+20}_{-11}$ objects have likely already been imaged there (Wright et al. 2014). However, these objects are sufficiently faint in the *WISE* W2 band that their observed W1−W2 colors, which instrinsically are very red due to strong $CH_4$ absorption in the W1 band, are typically lower limits that are simply not red enough to be flagged photometrically. There is an effort underway called CatWISE (Peter Eisenhardt, PI) that is using the *WISE* multi-epoch data to identify these colder brown dwarfs via their proper motion.

*Spitzer* has the ability to probe much deeper than *WISE* and CatWISE in this same wavelength regime but is limited by a small field of view. Tiling an area covering the gravitational tidal radius around relatively massive nearby stars, however, has the potential to find very low mass, widely separated companions, and this requires only that a limited amount of sky be observed. In fact, what is likely the second coldest brown dwarf known, WD 0806−661B, was found via just such a *Spitzer* survey targeting nearby stars with prior imaging data in the *Spitzer* Heritage Archive (Luhman et al. 2011). There was no possibility of completing a volume-, magnitude-, or mass-limited sample in that survey, unfortunately, given that the second epoch was dependent on the heterogeneous nature of the first (see, e.g., Figure 13).

The *Spitzer* extension affords us the opportunity to ameliorate this issue. To maximize the probability of finding far-flung, low-mass companions, only those primaries with the largest gravitational potential wells will be targeted. For example, if we restrict the primaries to white dwarfs and stars of spectral type K or earlier (corresponding to $M > 0.5\ M_\odot$) and to the 8-pc local volume, then there are 52 such objects to survey (Kirkpatrick et al. 2012). A radius of 0.1 pc (∼20,000 AU) has historically been used as the limit at which a Sun-like star can retain a bound, low-mass object for billions of years in the presence of stellar encounters (Weinberg et al. 1987), although several candidate common-proper-motion systems with separations up to several pc have recently been identified (Caballero 2009; Shaya & Olling 2011).

In order to identify common-proper-motion companions, only one photometric channel need be targeted, and the obvious choice is the 4.5 $\mu$m band because that is where cold brown dwarfs are brightest. We can achieve 25 mas astrometric uncertainty per axis for objects with SNR $> 100$ using imaging like that obtained with the *Spitzer* parallax program (see above section). However, such precision is overkill for this program because the motions of stars in the 8-pc sample are very high, generally $> 1$ arcsec yr$^{-1}$. Hence, lower SNR limits on potential companions will suffice.

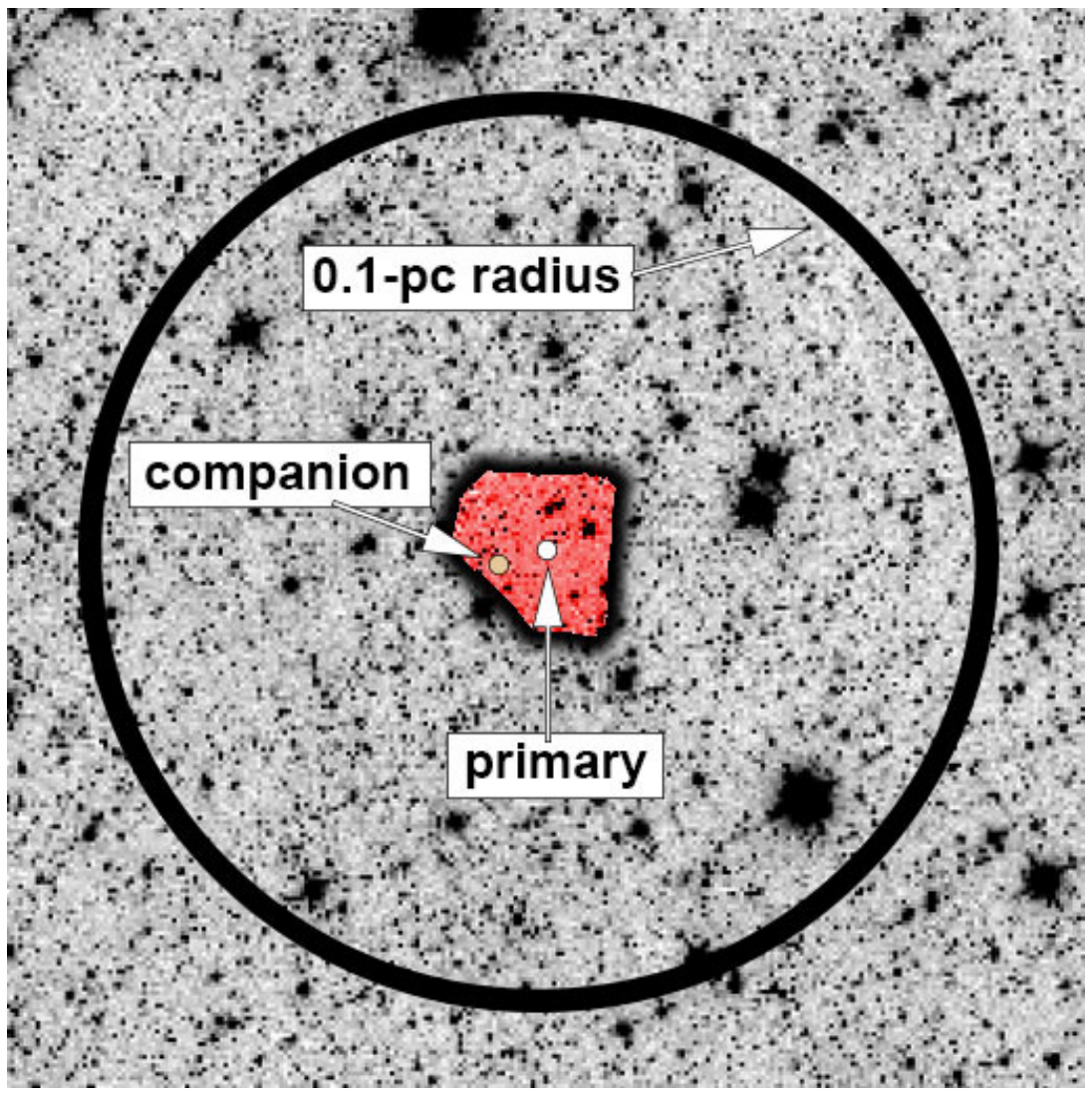

Figure 13: An image of the *Spitzer*/IRAC imaging (red) in the vicinity of WD 0806−661A (white circle) superposed on a background image at the same wavelengths from *WISE* (greyscale). The *Spitzer* footprint has an irregular shape because it is the intersection of two separate mosaics made at widely separated times and with different roll angles. The brown dwarf WD 0806−661B (brown circle) was discovered near the left edge of the *Spitzer* coverage. The 20,000 AU search zone around the white dwarf primary, located 19.2 pc away, is shown by the black circle. Note that the extant *Spitzer* data cover only a small fraction of the area in which possible companions to this object could have been found.

WISE 0855−0714 has an absolute IRAC ch2 magnitude of 17.1 mag (Schneider et al. 2016), so our goal is to reach this magnitude at an SNR $\approx$ 30 (or higher, as needed) in order to measure with $5\sigma$ certainty a motion comparable to that of the primary. (Monet et al. 2010 relate astrometric accuracy achieved as a function of SNR using a formula that we have calibrated using our own *Spitzer* astrometric measurements.) A further restriction is placed on frame times, for which we assume that the minimum frame time during the Spitzer extension is 30s. With these caveats in place, we find that the full 0.1-pc survey around A, F, G, K, and white dwarfs within 8 pc would require around 700 hours of actual integration time (350 hours per epoch).

In the foreseeable future, *Spitzer* imaging is the only resource available to perform such a wide search around the nearest stars for companions below 'room temperature' – a missing piece of phase space between the bulk of WISE brown dwarf discoveries (primarily > 350K; Cushing et al. 2011) and the planet Jupiter (~124K; Hanel et al. 1981). Without this ability, these discoveries will remain hidden and we will miss the opportunity afforded by *JWST* to scrutinize them at wavelengths where they are most easily studied.

# Section 5 – Galactic Science with the Spitzer Extension

Robert A. Benjamin

University of Wisconsin-Whitewater

October 10, 2017

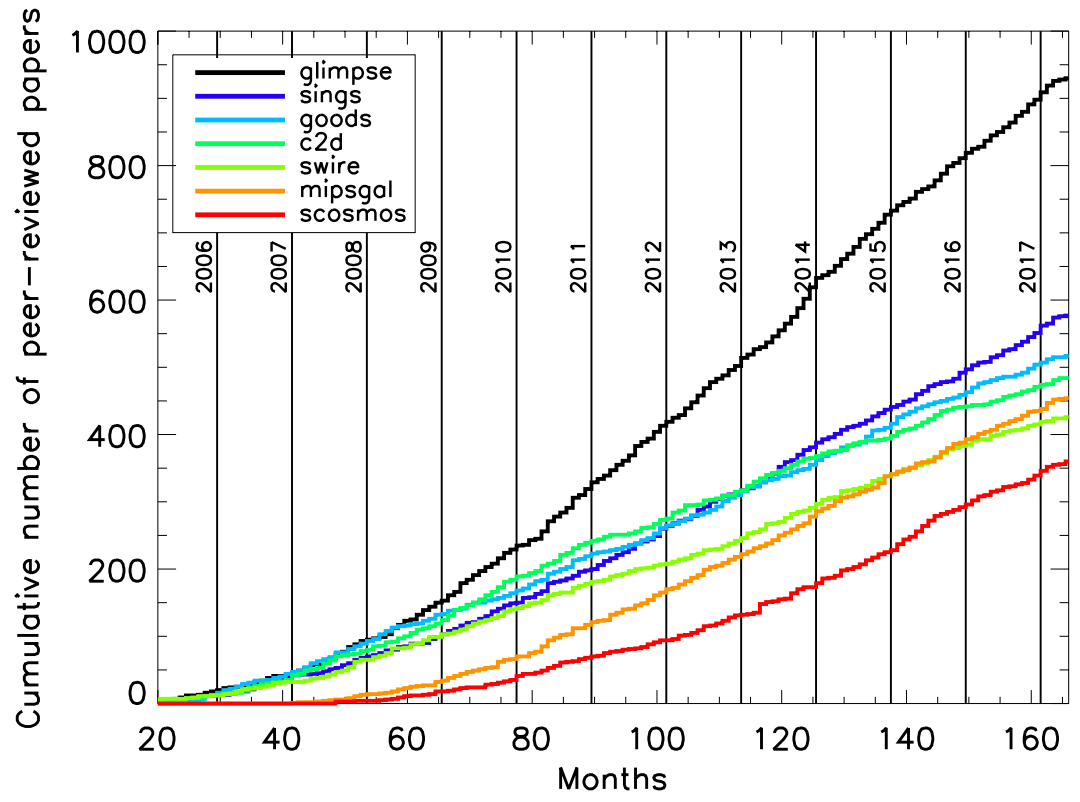

Figure 1: Cumulative number of peer-reviewed publications using data from the top Spitzer Space Telescope Legacy programs.

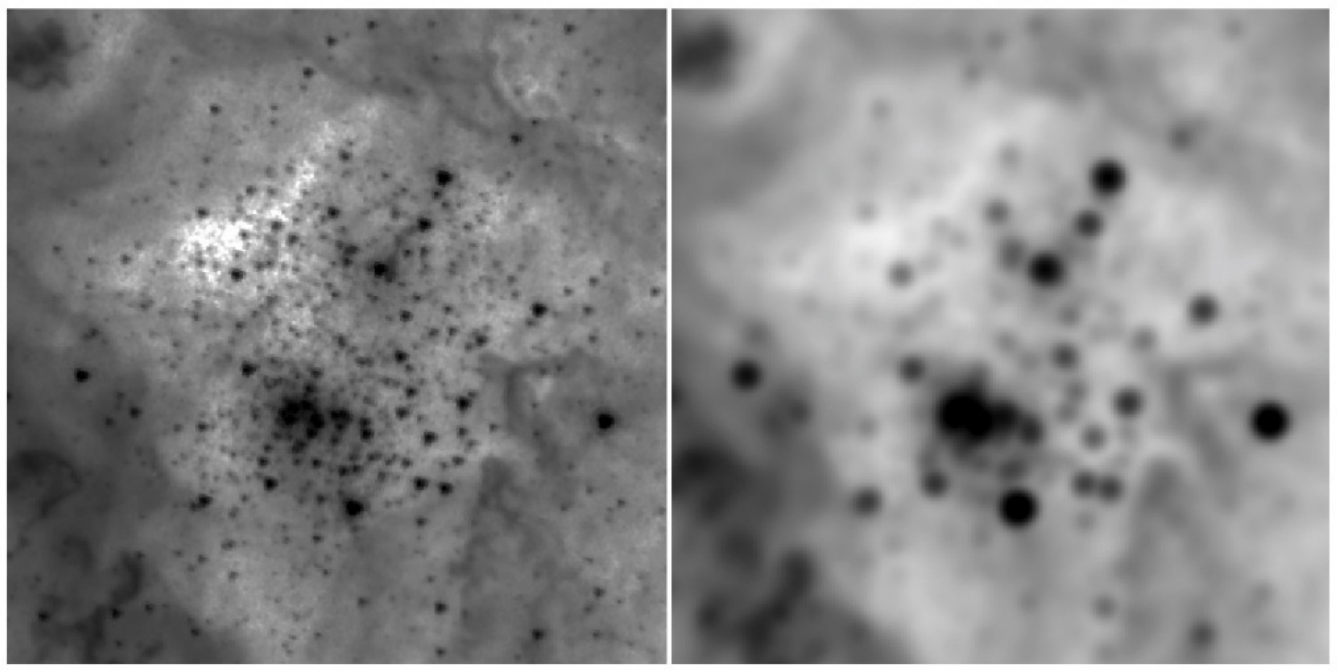

Figure 2: [Left] IRAC [3.6] image (reverse greyscale) at $l = 284.27°, b = -0°.32$ (RCW49/Westerlund 2). [Right] The same field with $6''$ resolution as in the *WISE* survey. *WISE* observations suffer source confusion over a large portion of the cluster.

# 1 Deep observations of the luminous star forming complexes of the Milky Way

The impact of *Spitzer Space Telescope* observations on our our understanding of the structure and star-formation of the Milky Way Galaxy has been significant. The combination of reduced extinction at infrared wavelengths and the bright infrared emission from embedded star formation has led to the identification and characterization of thousands of objects sampling star formation from the local solar neighborhood to the far side of the Galactic disk, a measurement of the current Galactic formation rate (Robitaille & Whitney, 2010) using young stellar objects (YSOs) and new insights on the stellar density structure of the Galactic disk and bar (Benjamin et al., 2005). A demonstration of this impact is the numbers of papers using Spitzer data from two of the original *Spitzer* Galactic Legacy programs (Figure 1). Data from the C2D (Cores to Disk) Legacy program (Evans et al., 2003) have been used in 489 peer-reviewed publications (6.3% of all *Spitzer* publications), data from the GLIMPSE (Galactic Legacy Infrared Midplane Survey Extraordinaire) program (Benjamin et al., 2003; Churchwell et al., 2009), have been used in 948 publications (12.2%).

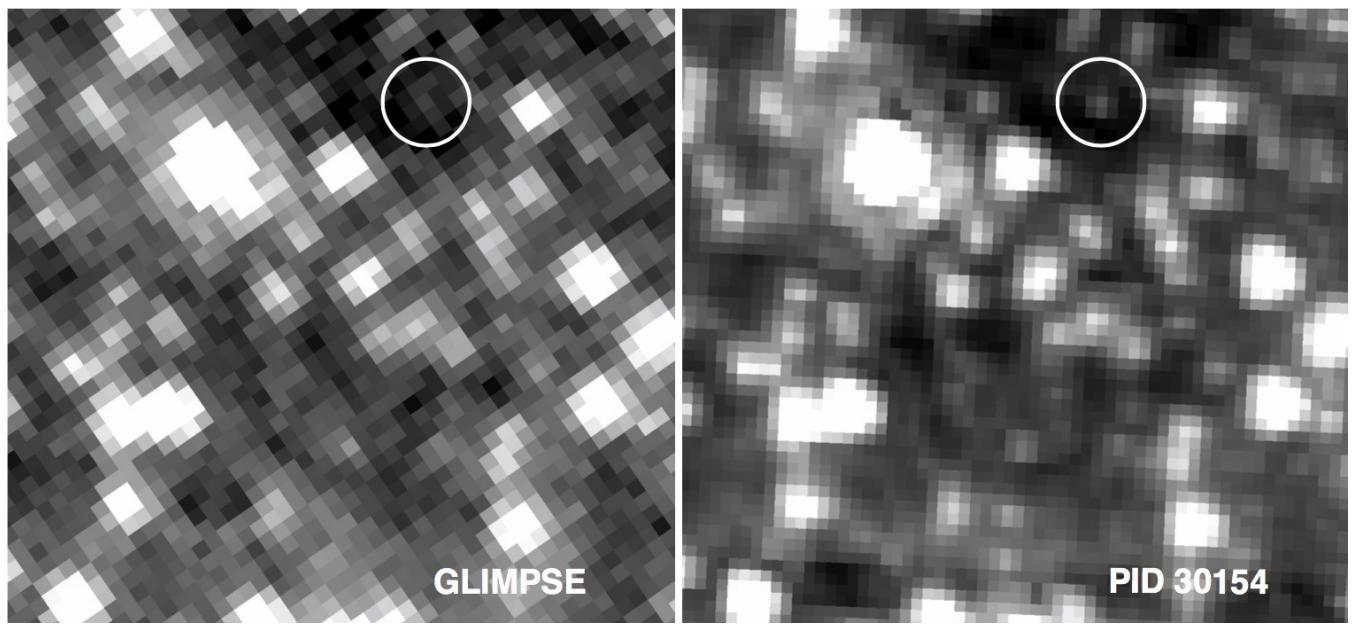

Figure 3: A portion of NGC6334 ($l = 351°.14, b = +00°.46$), as imaged by GLIMPSE (left) and program ID 30154 which used much longer integration times. The circled object is a magnitude 16.5 star in the [4.5] band detected in this program which was undetected in GLIMPSE. This comparison demonstrates the value of deeper observations in resolving and identifying stars of much lower mass in stellar clusters.

The sensitivity and longevity of the *Spitzer Space Telescope* have allowed for large-scale surveys of Galactic star formation. By combining Spitzer high angular resolution 24 micron images with *WISE* 22 micron images, (Anderson et al., 2014) developed a catalog of 8399 HII regions complete to the far side of the Galactic disk; only 12% of these regions had been previously cataloged. Molecular line spectroscopy of 1650 eight-micron selected sources (Urquhart et al., 2014) combined with observations to resolve the kinematic distance ambiguity has led to the identification of the most luminous star forming complexes in the Galaxy, finding that only 10 regions provide one-third of the total luminosity of the Galaxy. A radio-continuum selected sample (Lee et al., 2012) similarly finds 24 complexes produce *half* of the ionizing luminosity of the Galaxy. These complexes are the local analogues of regions that will be studied across cosmic distances by *JWST*.

Ninety percent of these complexes and HII regions lie within regions of the sky previously covered by Spitzer programs; seventy percent lie within the original GLIMPSE survey ($|l| < 60°, |b| < 1°$), but fifty percent of this full sample were only observed with the shallow $2 \times 1.2$ sec observations of the original GLIMPSE program. WISE observations are not sufficient to resolved the stellar populations of these complexes (Figure 2), but deeper observations of these complexes (Figure 3) would allow for the (1) investigation of the lifetimes of disks and envelopes as a function of stellar mass and environment, (2) a comparison of YSO mass functions in different regions, and (3) improved estimates of the total star formation rate of the Galaxy and its variation with Galactocentric radius. By matching coverage depth for the full Galactic plane to the depth used in *Warm Spitzer*, it will be possible to the extend the study of YSOs down to mass limits of 0.1 $M_\odot$ (at 3 kpc) to 2 $M_\odot$ (at 8 kpc). In GLIMPSE, the mass limits were 3 and 8 $M_\odot$, respectively. For example, in W51 ($d = 5.4 \pm 0.3$ kpc), (Kang et al., 2009) found evidence of mass segregation, with the most massive objects concentrated towards the center of the H II regions. In the very young M17 SWex complex ($d = 1.98 \pm 0.14$ kpc) and the more evolved Great Nebula in Carina ($d = 2.3$ kpc), the YSO mass function exhibits a dramatically steeper power law than the Salpeter–Kroupa IMF, providing constrains on disk lifetimes as a function of YSO mass (Povich & Whitney, 2010; Povich et al., 2011). For the 10% of higher latitude star forming complexes that were not previously observed by *Spitzer*, the lower backgrounds and confusion will allow for constraints at even lower stellar masses. Figure 4 shows a color-magnitude diagram (CMD) of the AFGL 490 star forming region compared to a region with no apparent star formation. This shows a large number of YSOs, down to a mass limit of $\sim 0.1 M_\odot$ at a distance of 1 kpc.

# 2 Stellar Proper Motions in the Galactic Disk

The longer the *Spitzer* operates, the more valuable it becomes for studies of stellar proper motions in the Galactic plane. The imminent release of parallax and proper motion data for a billion stars from *Gaia* will revolutionize our understanding of stellar kinematics in the Milky Way. *Spitzer* observations

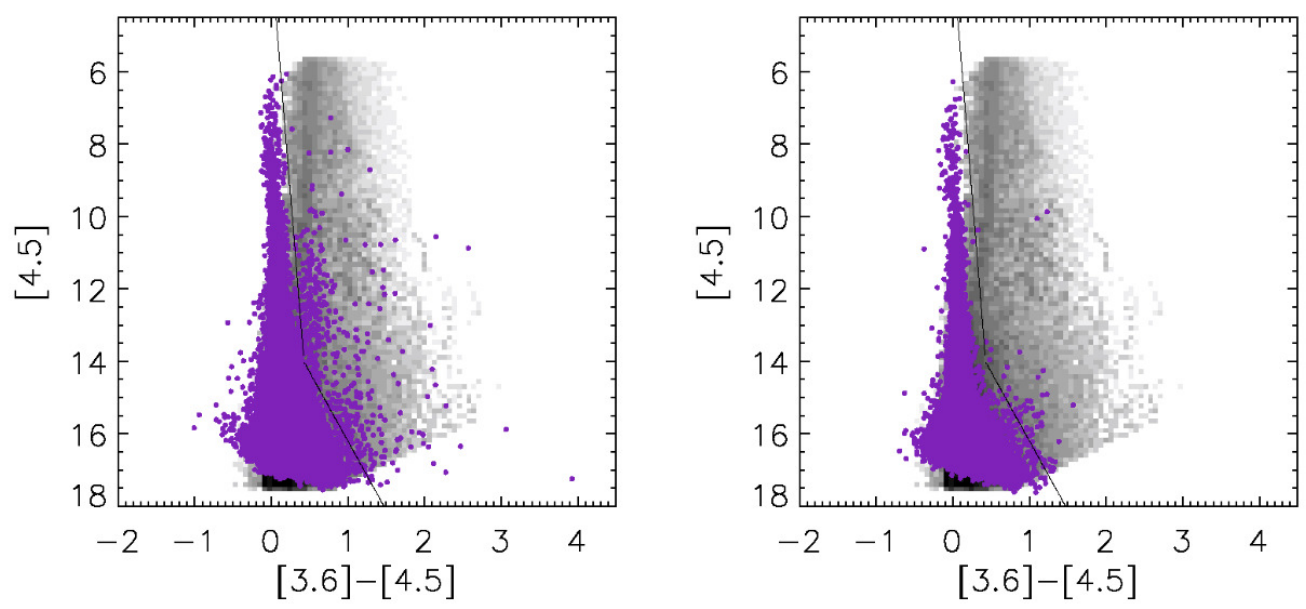

Figure 4: [Left] Color-magnitude plot (4.5 vs 3.6-4.5) of the AFGL 490 region at $l = 142°$, $b = 1.8°$. The purple dots are sources from the GLIMPSE360 catalog. The grey scale is expected colors of YSOs from a grid of models(Robitaille et al., 2006). Sources to the right of the black line are likely YSOs. [Right] Comparison for empty field.

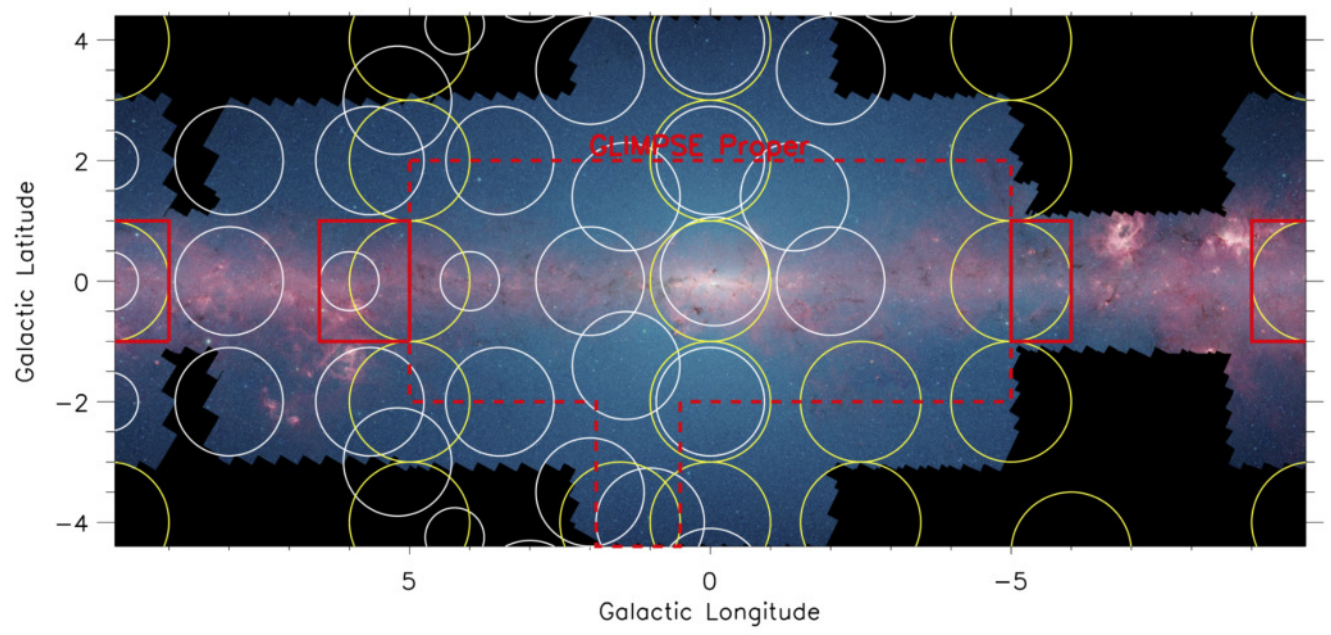

Figure 5: IRAC image of the inner Galaxy showing the location of APOGEE fields (white circles), planned APOGEE-2 fields (yellow circles), and a subset of direction targeted for proper motions studies in Cycle 12 and 13 programs (red boxes). An extended *Spitzer* mission would allow for characterization of stellar proper motions

provide a complementary opportunity to measure the proper motion of distant Galactic disk sources that are off limits to Gaia because of high dust extinction. The foreground *Gaia* sources will be valuable for establishing the astrometric reference frame for these *Spitzer* proper motion investigations as will HST observations of proper motions in the Galactic bulge Clarkson et al. (2008).

Continuing work on characterizing the distortion correction for IRAC currently allows an astrometric accuracy of 50 mas, with a 3rd order polynomial correction, with the prospects of achieving 20 mas (Lowrance et al., 2014). This has been confirmed by Esplin & Luhman (2016) who have developed an independent method to measure proper motions, using GLIMPSE data to validate their results. Over a 10 year baseline, an astrometric precision of 2 milliarcseconds per year for a single source is possible, albeit with significant reprocessing. The large number of sources available will thus allow for measurements of mean proper motion, $\bar{\mu}$, and proper motion dispersion, $\sigma_\mu$, of stellar samples at different depths across the Galaxy. Using much smaller samples than that available with *Spitzer* and optical ground-based telescopes observing from northern sites where the Galactic center is at high air-mass, programs like OGLE-II were still able to measure typical velocity dispersions of $\sigma_\mu = 2.9 \pm 0.1$ mas/year (Rattenbury et al., 2007) and typical streaming motions of $\bar{\mu} = 2.04 \pm 0.7$ mas/year. Cycle 12 and Cycle 13 observations are currently being processed to constrain proper motions towards the inner Galaxy, particularly in the directions probed by APOGEE (Majewski et al., 2017) which provides stellar abundances and radial velocities for a subsample of stars observed by *Spitzer*.

There are three unique measurements that can be done by measuring stellar proper motions in the infrared wavelengths of *Spitzer*. *First*, one can measure the stellar velocity dispersion both parallel and perpendicular to the Galactic plane as a function of (1) direction and (2) depth. By defining samples of red clump giants or red giant branch stars, cf. Nidever et al. (2012), we can probe different depths through the Galactic bulge and disk. Investigations in extinction windows below the Galatic plane indicate that this velocity dispersion changes as a function of direction (Rattenbury et al., 2007), although this has not yet been confirmed, cf. Vieira et al. (2007). *Second*, one can constrain Galactic rotation as a function of depth through the Galaxy and measure non-circular streaming motions in the inner Galaxy due to the Galactic bar, These have been potentially observed (again, well off the plane) by Poleski et al. (2013). Proper motion measurements in the inner Galaxy may also yield useful constraints on the relative contribution of dark and luminous matter in the inner Galaxy. *Third*, brown dwarfs, being nearby, have high proper motions and relatively isotropic distribution on the sky. Because of stellar crowding, the Galactic plane is the hardest place in the sky to look for such objects. As a result, there may still be interesting objects there that can not be detected with *WISE* because of the severe confusion in the inner Galaxy. The discovery of brown dwarfs in the Galactic Plane may make it possible, in special cases, to measure the masses of *single* brown dwarfs (Paczynski, 1995). For brown dwarfs whose astrometric paths can be measured with milliarcsecond accuracy, candidate lensing events of background stars by the brown dwarf can be predicted in advance, cf. Lépine & DiStefano (2012).

These proper motion studies can directly inform the planning for *WFIRST* astrometric studies in the Galactic Bulge fields to search for exoplanet microlensing. *Spitzer* observations will provide a set of test objects against which *WFIRST* pipeline software can test astrometric measurements; the experience gained with *Spitzer* astrometric surveys for best image centroiding practices and proper handling of source crowding will be valuable for *WFIRST* planning.

# Section 6 – The Birth and Evolution of Galaxies

M. A. Malkan[1], G. G. Fazio[2], M. L. N. Ashby[2], P. Capak[3], M. Dickinson[4], H. I. Teplitz[5], and S. P. Willner[2]

[1]Physics and Astronomy Bldg., 3-714 , University of California Los Angeles , Los Angeles, CA 90095-1547, USA
[2]Harvard-Smithsonian Center for Astrophysics , Optical and Infrared Astronomy Division , 60 Garden St., MS-66 , Cambridge, MA 02138, USA
[3]Spitzer Science Center , California Institute of Technology , Mail Code 220-6 , Pasadena, CA, 91125, USA
[4]National Optical Astronomy Observatory , Tucson, AZ, USA
[5]Infrared Processing and Analysis Center , Pasadena, CA 91125, USA

# 1 Introduction

Extending Spitzer's lifetime would enable deep wide extragalactic surveys that go far beyond anything previously contemplated for the study of galaxy evolution. The *Spitzer* Warm Mission 3.6 and 4.5 $\mu$m bands have unique capabilities to identify and characterize very distant galaxies because they sample the galaxy's rest-frame visible light, and can thus yield redshift measurements via photometric techniques, and quantitatively determine the stellar masses and possible AGN contributions. The resulting data would be uniquely powerful for achieving several outstanding scientific goals, including:

(1) understanding the formation and evolution of dusty star-forming galaxies from $1 < z < 4$, and massive galaxies throughout the entire epoch of cosmic reionization;

(2) understanding large-scale cosmic structure, galaxy cluster formation at $z > 1.5$, and environmental effects on star formation;

(3) understanding the origin of the cosmic infrared background radiation (CIBR) through observations of its spatial anisotropies and multi-wavelength correlations; and

(4) identifying rare objects for follow-up by *JWST* and ALMA (and eventually, large ground-based telescopes), including the coldest brown dwarfs and the first generation ($z > 7$) of luminous galaxies and quasars that re-ionized the Universe.

An example baseline survey could encompass a hundred square degrees with $\sim$ 1800 sec per pixel integration time, reaching a sensitivity of $\sim$24 AB mag (5$\sigma$; see Fig. 1) in about 9000 hours of facility time, several magnitudes deeper than *WISE*. This depth is an excellent complement to the deep surveys of near-infrared imaging and spectroscopy to be made with *Euclid* and *WFIRST*, and the combination with IRAC, which measures the red side of the Balmer break, is extremely powerful. The legacy of these joint databases, fully available to the public, would enable a tremendous range of frontier science topics, many of which we can hardly anticipate at the moment, and leave a very important legacy. The high-redshift volume covered would be in the tens of cubic Gpc, with reliable stellar masses measured for millions of galaxies (as well as identifications of hundreds of thousands of new distant AGN). This tremendous advance over current statistics is not merely quantitative. In many cases the new *Spitzer* data will allow the first transformative tests of cosmological models that are not otherwise possible.

The greatest synergies come from observing large connected regions that already have dense multi-wavelength coverage. Covering a hundred square degrees also eliminates the major problem of cosmic variance, which has plagued all deep surveys to date. Cosmic variance means that the galaxy populations measured in small areas are inherently different, depending on the randomness of including regions of high or low density in the selected fields. Surveying one of two large areas will finally provide a representative 'fair sample' of the universe, from rich clusters to voids, and how they are connected at each redshift. In particular, there are some areas, such as the North Ecliptic Pole region, with superb multi-wavelength coverage that lie near the *Spitzer* Continuous Viewing Zone. By virtue of its special accessible location, the NEP region will also be the prime deep field for future space spectroscopy missions including *Euclid*. IRAC integrations of 1800 sec would reach most of the typical $M^*$ galaxies that *Euclid* and *WFIRST* will observe, out to $z = 4$.

A second example would be a shallower but wider survey in the southern sky, where IRAC can make unique contributions to our knowledge of galaxy formation and evolution at the earliest cosmic times by covering the so-called SPT3g survey — contributions not possible with any other observatory now operating or planned. The South Pole Telescope (SPT; Carlstrom et al. 2011) has begun its groundbreaking third-generation survey of 4000 deg$^2$ reaching 4 $\mu$K depth at 90, 150, and 230 GHz, and will discover more than 5000 galaxy clusters via the S-Z effect, 700 of which will lie at $z > 1$ and require near-IR followup. The field is also included in the AT20G survey of the whole southern sky at 20 GHz. By surveying the SPTpol field (500 deg$^2$ within SPT3g for which additional polarization data are available at 150 GHz), IRAC would achieve three main science goals: first, it would probe the evolving relationship between baryonic and dark matter by cross correlating IRAC-selected galaxies at $1 < z < 2$ with the CMB lensing signal detected in the SPTpol data; second, it would improve estimates of the lensing potential by $\sim$50% by allowing up to 3/4 of the lensing B-mode polarization anisotropy to be removed from the SPT maps, greatly enhancing the sensitivity to primordial B-modes from inflation; and third, as a by-product of the survey it would also construct and study a well-defined sample of the first massive galaxy clusters, which form at $z > 1.5$. The potential additions to our knowledge of the B-mode polarization signal in particular are a unique contribution from *Spitzer*. Covering SPT3g to a useful depth (120 sec per pointing at a total cost of order 3000 hours of facility time; Fig. 1) would accomplish all these goals *and* furnish hundreds of valuable high-redshift sources for efficient, targeted followup by *JWST*.

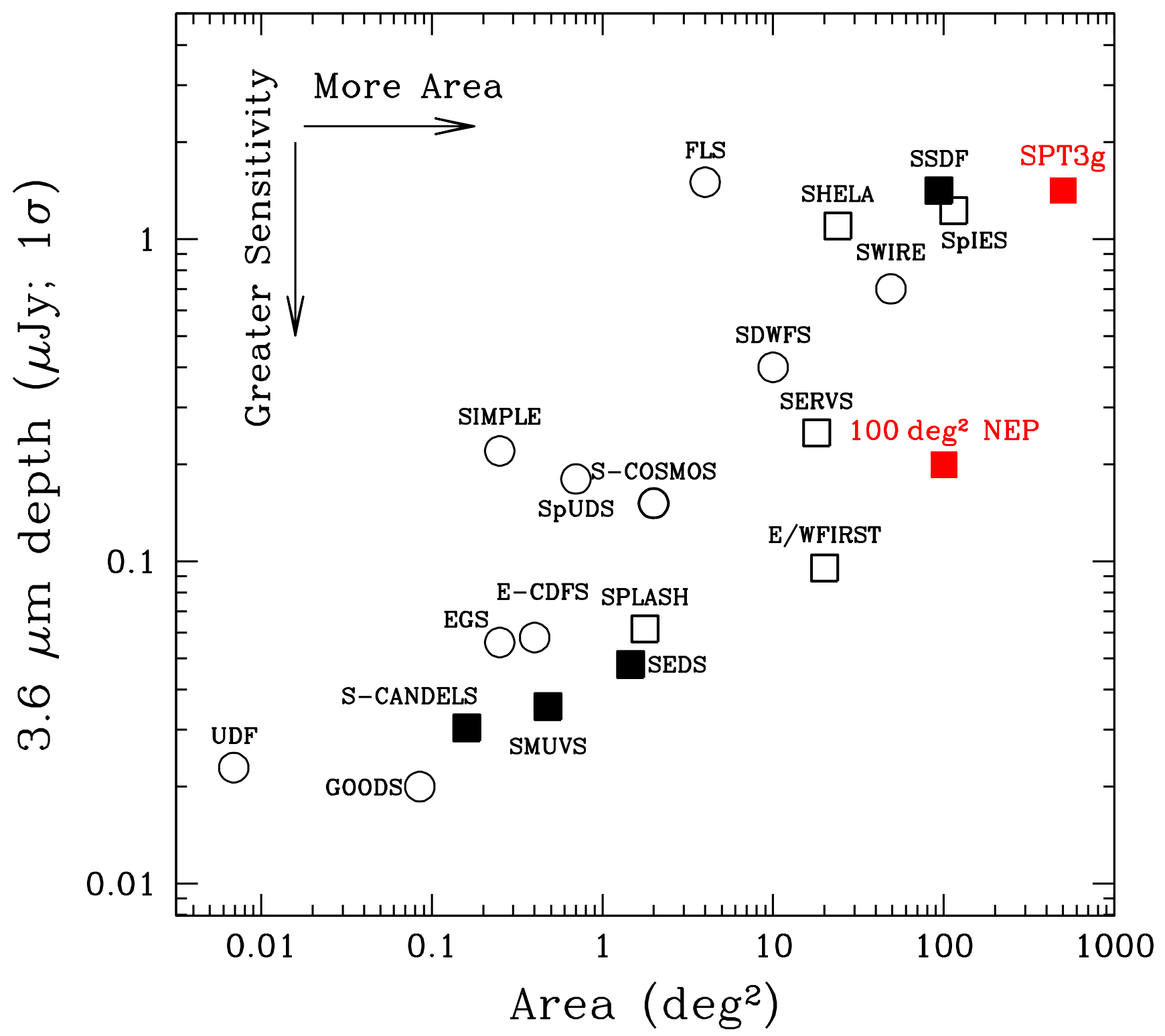

Figure 1: The area-sensitivity parameter space covered by all completed *Spitzer*/IRAC extragalactic surveys to date (black symbols), from M. Ashby et al. (2017), in preparation. The two groundbreaking surveys proposed in the text are shown in red. Circles and squares indicate the 3.6 $\mu$m $1\sigma$ point-source sensitivities for surveys executed during the cryogenic and warm mission phases, respectively. Solid black squares indicate sensitivities calculated with simulations; the other sensitivities shown are either taken from the literature or from the online calculator SENS-PET under low-background conditions. The IRAC surveys carried out include S-COSMOS (Scoville et al. 2007), SPLASH (Steinhardt et al. 2014), SEDS (Ashby et al. 2013a), S-CANDELS (Ashby et al. 2015), SMUVS (Ashby et al. 2017, in preparation), the First Look Survey (FLS; Fang et al. 2004), *Spitzer*-SPT Deep Field (SSDF; Ashby et al. 2013b), *Spitzer*-IRAC Equatorial Survey (SpIES; Timlin et al. 2016), *Spitzer*-HETDEX Exploratory Large-area Survey (SHELA; Papovich et al. 2016), *Spitzer* Wide-area Infrared Extragalactic Survey (SWIRE; Lonsdale et al. 2003, 2004), *Spitzer* Deep, Wide-Field Survey (SDWFS; Ashby et al. 2009), *Spitzer* Extragalactic Representative Volume Survey (SERVS; Mauduit et al. 2012), *Spitzer* IRAC/MUSYC Public Legacy in E-CDFS (SIMPLE; Damen et al. 2011), *Spitzer* Public Legacy Survey of UKIDSS Ultra-deep Survey (SpUDS; Caputi et al. 2011), *Euclid*/WFIRST *Spitzer* Legacy Survey (E/WFIRST; PI Capak), the Extended *Chandra* Deep Field South (E-CDFS; Rix et al. 2004), the Extended Groth Strip (EGS; Barmby et al. 2008), the Ultra-deep Field (UDF; Labbé et al. 2013), and the Great Observatories Origins Deep Survey (GOODS; Lin et al. 2012). SERVS, SWIRE, SEDS and S-CANDELS are multi-field surveys for which the total areas are indicated.

# Section 7 – *Spitzer* Extended Mission: Technical Capabilities of *Spitzer* Operations Beyond March 2019

Sean J. Carey[1], James G. Ingalls[1], Jessica E. Krick[1], Patrick J. Lowrance[1], Carl J. Grillmair[1], and William A. Mahoney[1]

[1]Caltech/IPAC-Spitzer, Pasadena, CA 91125

# 1 Executive Summary

*Spitzer* can support unique and compelling science observations of transiting exoplanets, the high redshift universe, near-Earth objects as well as characterize brown dwarfs and measure the parallaxes of microlensing events. Both *Spitzer* and the IRAC instrument have been extremely stable throughout post-cryogenic operations with no degradation in performance. The hardware is fully redundant and the consumable should last another 14 years. Changes in operations are entirely due to the increased distance and change in geometry with respect to Earth. As a result, the amount of time that can be spent downlinking data is decreased with each calendar year and the data rate drops in 2020. Despite the decreased data volume that can be transmitted, the downlinks are more than adequate to collect the observations necessary to execute the science programs discussed in the white papers that this technical discussion accompanies.

# 2 Introduction

The Spitzer Beyond mission is scheduled to cease data collection in March 2019, ending operations of NASA's Infrared Great Observatory as the James Webb Space Telescope will commence science operations shortly afterward. The science community is exploring the benefits of operating *Spitzer* past March 2019 to support a more focused program of scientific inquiry. To support this effort, a request was made for the Spitzer Science Center to provide technical information regarding expected performance of the *Spitzer Space Telescope* after March 2019.

This document details the *Spitzer* project's best assessment of the technical capabilities of the observatory through December 2020. On-orbit characteristics change with time, so we can only realistically predict the performance of *Spitzer* over a window of approximately 2 years. The information provided can be updated in the future using knowledge as we acquire it in 2018, 2019, and beyond. For a discussion of the *science* potential of operating *Spitzer* beyond March 2019, please refer to the relevant white papers in previous sections.

# 3 Relevant Aspects of the *Spitzer Space Telescope* for Operations Beyond 2019

The post-cryogenic mission phase of *Spitzer* began science operations on 27 July 2009. Operations of the observatory remain the same as in the cryogenic mission with the exception that only the two shortest wavelength science cameras are operational. A more complete description of the telescope and the complete Spitzer mission can be found in Van Dyk et al. (2013) (see also Werner et al. 2004).

The *Spitzer Space Telescope* is a cryogenic 80 cm Ritchey-Chrétien telescope which is passively cooled to a temperature of 27 K. *Spitzer* is in an Earth-trailing solar orbit with a set of solar panels that are pointed towards the Sun to provide power to the spacecraft and instruments and provide thermal shielding of the telescope and instrument chamber. The solar arrays are augmented by a battery that stores 16 Ampere-hours of charge. The battery used for *Spitzer* has a recommended maximum depth of discharge of 50%. The telescope and instrument chamber are thermally isolated from the solar array and spacecraft bus and supported by $\gamma-$Aluminum struts. The performance of the thermal shielding is excellent and the temperatures of the cryogenic components (telescope and instrument) have not varied to within our ability to measure them during the entire post-cryogenic mission to date.

The spacecraft is constrained in pitch and roll to avoid sunlight on the cryogenic system and to simultaneously provide power to the solar arrays (see Figure 1). Because of these constraints the visibility of a source is a function of its Ecliptic latitude and position relative to the *Spitzer*-Sun vector. The minimum visibility for any source is two 38 day windows spaced 6 months apart, which holds for targets in the Ecliptic plane. Within 6.3 degrees of the Ecliptic poles, sources are continuously visible. The constraint on the roll angle around the *Spitzer* boresight is $\pm 2$ deg, while the pitch angle constraint confines the angle between the boresight and the Sun to be from 82.5 deg to 120 deg. In the remainder of this document, we define pitch angle as the angle between the normal vector of the solar array and the Sun direction. This is the pitch of the boresight minus 90 degrees. Negative angles have the boresight pointing towards the Sun, positive angles have the boresight in the anti-Sun direction. The power delivered by the solar arrays is proportional to $\cos^2(Pitch)$.

The operational instruments during the post-cryogenic mission are the Infrared Array Camera (IRAC, Fazio et al. 2004) and the Pointing Control Reference Sensor (PCRS, Mainzer & Young 2004). The PCRS is a peakup camera in the optical that provides routine calibrations of the offset between the star tracker and the telescope boresight, and allows accurate placement of sources on the IRAC arrays to facilitate high-precision photometry (Ingalls et al., 2016). IRAC has two sensitive In:Sb 256x256 arrays covering fields of view 5.1 x 5.1 arcminutes in size. The two arrays cover passbands centered on 3.6 and 4.5 $\mu m$. The images are moderately undersampled and have resolutions of $\sim 1.4$ arcseconds. In one hour of integration, the IRAC arrays reach a $5\sigma$ depth of 720 and 1040 nJy for low background regions, respectively. For high-precision photometric monitoring, IRAC routinely achieves near-photon limited performance (after decorrelating systematics) with precisions down to 30 ppm.

# 4 Variations in Observatory Properties and Performance as a Function of Time

The primary driver for changes in the performance and operations of the *Spitzer Space Telescope* with time is the increasing distance of *Spitzer* from the Earth. Like the Earth, *Spitzer* orbits around the sun, but its semi-major axis is slightly greater than the Earth's. As a result, the *Spitzer*-Earth distance increases as a function of time as shown in Figure 2.

## 4.1 Communication using Low-Gain Antenna

The increased distance primarily affects communication with *Spitzer*. For initial communication during anomalies, the two low-gain antennas (LGA) are used for initial acquisition and commanding. For the Spitzer Beyond mission segment, the LGA data rate during anomalies was reduced from 40 bits-per-second (bps) to 10 bps in 2014 and that data rate is supported for anomaly recovery through 2020. Anomaly recovery beginning in 2019 will require contact passes with the ground station at higher elevations (30 degrees instead of 20 degrees) and will require using a 70 meter antenna in combination with a 34 meter antenna. The consequence of higher elevations for contact passes and additional antennas needed to communicate is that recovery from anomalies is likely to be slower as the mission continues.

## 4.2 Communication using the High-Gain Antenna

*Spitzer* primarily communicates with the Earth using a body-fixed high-gain antenna (HGA) which is located on the bottom of the spacecraft bus. The HGA is used for most commanding and all data downlinks. With the increasing distance from Earth, Spitzer has to pitch to large angles away from the Sun to point the HGA back to Earth. Figure 3 shows the pitch angle for HGA downlink as a function of time.

Pitch angles greater than 30 deg are outside of the nominal science pointing zone for *Spitzer* and no science observations can be conducted at these angles. For pitch angles greater than 30 deg, the incidence of light on the solar panels decreases such that the solar array power is not sufficient to power the observatory. During downlinks in the Spitzer Beyond phase and continuing past March 2019, downlink durations are limited by the amount of power drawn from the battery during the downlink. As the necessary pitch angle increases due to the increased distance of the spacecraft from Earth, the available downlink duration shortens. For operational purposes, the available downlink durations are not specified as a smooth function but have transitions at defined dates such that appropriate power and state of charge on the battery are maintained with adequate margins.

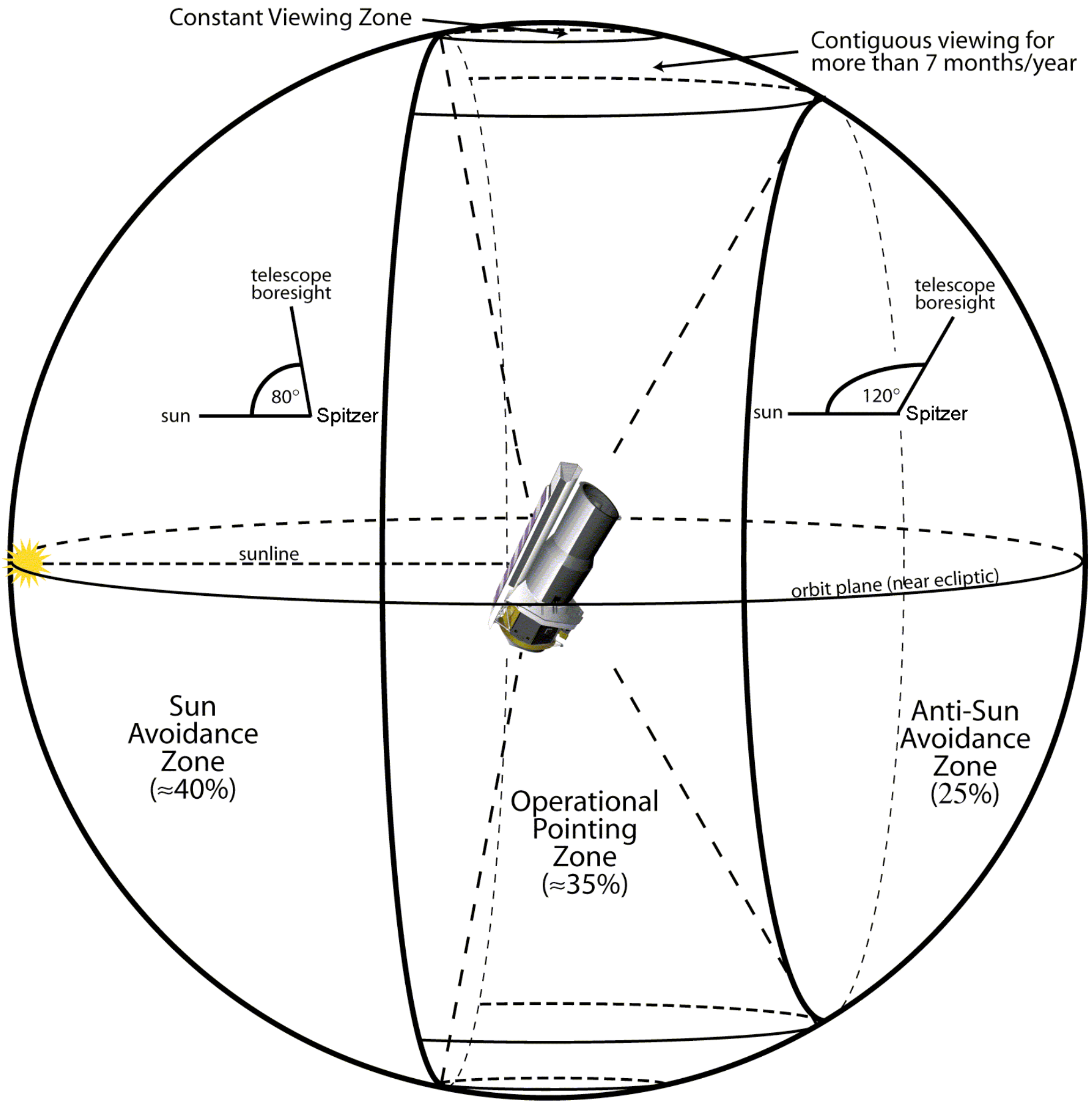

Figure 1 Operational pointing zone of *Spitzer* relative to Sun in Ecliptic coordinates. Approximately 35% of the sky is visible to *Spitzer* at any given time and the same regions are visible six months later.

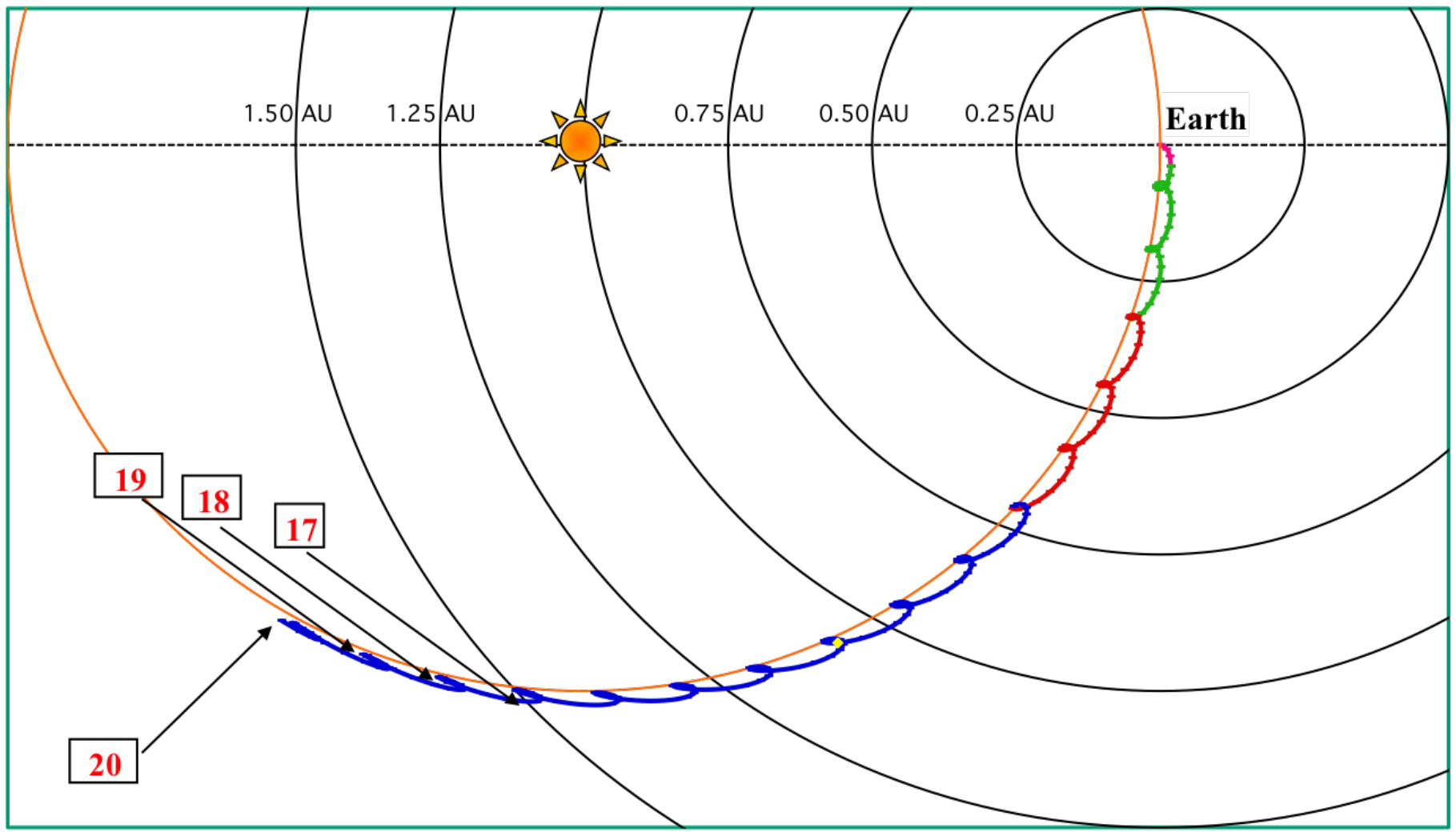

Figure 2 Orientation of *Spitzer* with respect to Earth and Sun as a function of date. *Spitzer* is receding from the Earth at $\sim 0.1 au\ year^{-1}$. The arrows and corresponding date boxes represent the position of *Spitzer* relative to the Earth on launch anniversaries, 25 August 2017-2020, respectively. The red and green lines represent the cryogenic mission while blue is the post-cryogenic phase. The concentric circles give the range of *Spitzer* from the Earth. The Sun obstructing Spitzer's communications with Earth is not a concern until well after 2025.

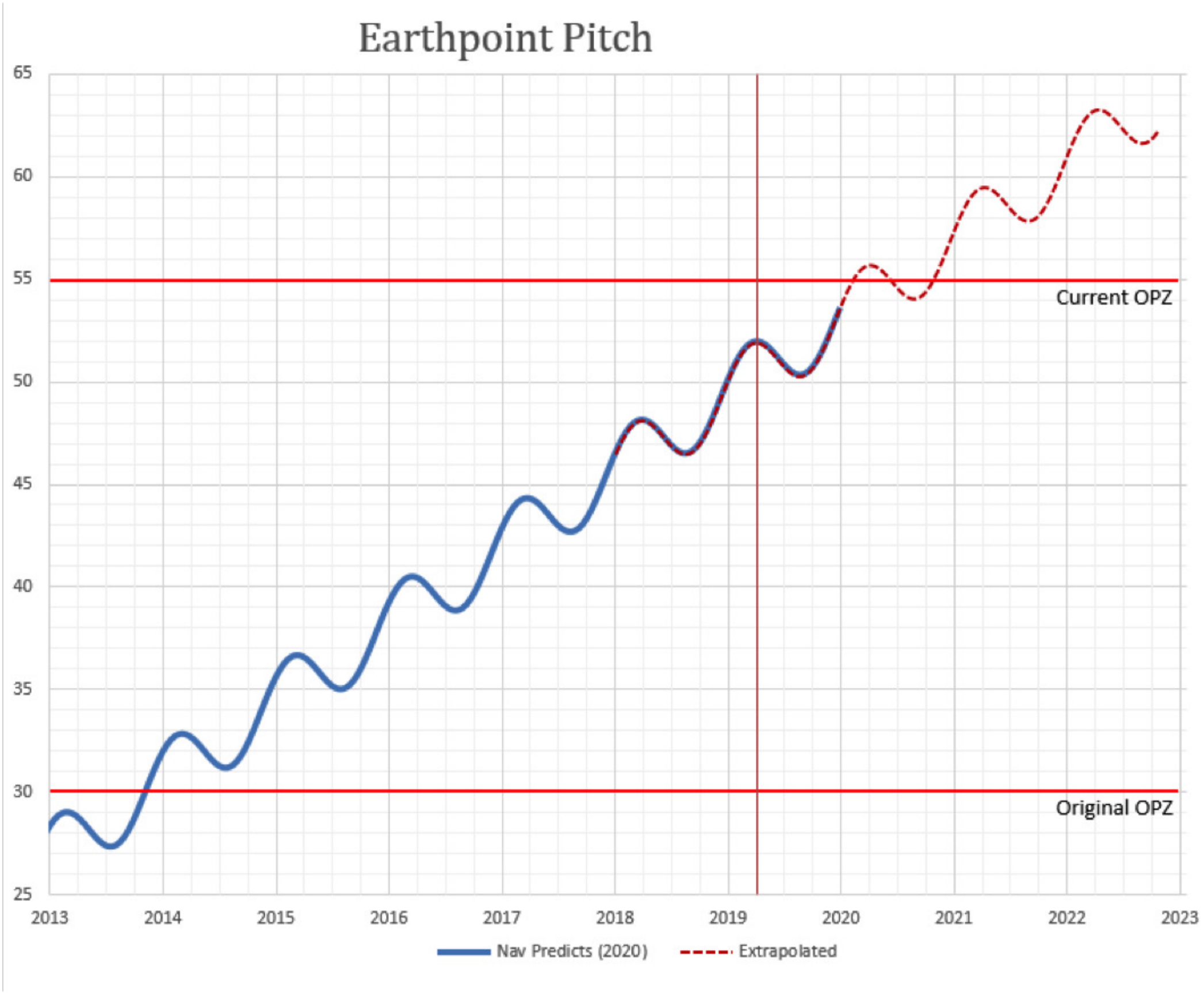

Figure 3 Pitch angle of observatory in degrees for downlink attitude on the HGA as a function of date in calendar years. The pitch angle is the angle between the normal vector of the solar array and the sun direction vector as seen by Spitzer. Zero degrees is when the solar panels are directly facing the Sun. Observations are conducted at pitch angles between -7.5 and 30 degrees. Positive angles indicate that the boresight is pointed in the anti-Sun direction. The lines at constant angle represent the operational pointing zone for *Spitzer* outside of which a fault protection response is triggered. The line at 30 degrees is the original OPZ prior to the Beyond mission phase. The 55 degree line is the current OPZ which provides enough margin for operations through March 2019. For continued operations, the OPZ will be extended to $\sim 60$ degrees.

# 5 Impact of Downlink Duration on Science

The amount of data downlinked, hence, the science that can be supported is a function of both maximum downlink duration and downlink rate. For most of the mission and through 2017, the maximum downlink duration has been set to 4 hours which is more than enough to support all science conducted with *Spitzer*. Starting in 2018 in response to power considerations during downlinks, the maximum downlink duration has been reduced to 3.5 hours. In 2019 and 2020, the maximum duration will be 3 and 2.5 hours, respectively. Though a full discussion of operations beyond 2020 is not the domain of this document, downlinks in 2021 would continue to allow durations of 2.5 hours (with slightly less margin than the 2020 downlinks) and in 2022 the downlink duration would be reduced to 2 hours. These maximum downlink durations are designed to include appropriate power margin to ensure that the battery is not discharged to an unsafe level.

The current data rate of 550 kilobits per second (kbs) can be supported through 2019 with a combination of higher elevation DSN passes and use of a 70 meter and 34 meter antenna arrayed together. For calendar year 2020, the data rate will need to drop to 275 kbs as the increased distance precludes the 550 kbs data rate for most realistic combinations of track elevation and antenna configuration. 275 kbs should be able to be supported through 2022.

Not all of the available downlink time can be used for transmitting science data. The first thirty minutes at downlink attitude must be used for lockup and commanding. Spacecraft and instrument housekeeping telemetry take 25 minutes to downlink at 550 kbs and 50 minutes at 275 kbs, assuming downlinks are spaced every 24 hours so a full day's worth of telemetry is accumulated. Thus, the time available for science data downlink in 2019 will be 125 minutes, and 70 minutes will be available in 2020. Assuming a standard image compression ratio of 0.45, $\sim$ 6900 images can be downlinked every 24 hours in 2019 and $\sim$ 1900 in 2020.

## 5.1 Example Science Cases Supported By Downlink Capability

Potential science cases as described in previous sections for extending *Spitzer* include deep imaging of extragalactic fields in particular to support *WFIRST* and studies of transiting exoplanet systems such as those identified by TESS or ground based searches such as SPECULOOS. The deep extragalactic imaging uses full array dithered maps with 100 second frametimes. We use Program #13058, The Euclid/WFIRST Spitzer Legacy Survey (PI Peter Capak) as a proxy for future extragalactic investigations. For transiting exoplanet studies with host stars [4.5] $>$ 7.5, we use the 2 second subarray staring mode observations of Program # 13067 (PI Michael Gillon) which studied the TRAPPIST-1 system as proxy for future exoplanet programs. In addition, it is likely that observations of Near-Earth Asteroids and temporal monitoring of brown dwarfs will also be conducted. These science investigations will also primarily use the 100 second frametimes and 2 second subarray staring mode observations, respectively. In addition, additional campaigns of the *Spitzer* microlensing parallax program and proper motion/parallax measurements of nearby brown dwarfs are likely to be conducted. These science cases typically take short, well dithered (5-6 dithers) observations of many targets typically using the 30 second full array frametime.

Table 1 compares the expected science data downlink capability through 2020 with the number of data frames of the representative extragalactic and exoplanet programs if each was assumed to be occupy all of the time between downlinks. The table also provides the total number of images observed as would be done for a microlensing program per downlink. Both the deep extragalactic science case and the exoplanet/brown dwarf monitoring programs are easily supported by the available downlink bandwidth. Microlensing parallax can be supported as well though 2020 with a small decrease in observing efficiency. As both NEOs and brown dwarf targets for parallax are spaced farther apart on the sky than the microlensing sources in the Galactic bulge, these programs can also be accommodated with the available downlink capability.

Observations of exoplanet host stars brighter than [4.5] $<$ 7.5 will need to use shorter frametimes than the 2 second subarray. Data rates may have to be limited depending on the length of observation. The data rates can be arbitrarily throttled back using instrument engineering requests (IERs) that would have to be developed by the Spitzer Science Center. IERs are currently used for a handful of

Table 1. Expected Downlink Volumes and Science Yield

| Year | Pitch Angle Degrees | Hours | Rate kbs | Number of Images Downlinked | XGal | ExoP | $\mu$lens |
|---|---|---|---|---|---|---|---|
| 2019 | 52.5 | 3 | 550 | 6944 | 1512 | 588 | 1930 |
| 2020 | 56.5 | 2.5 | 275 | 1944 | 1548 | 602 | 2068 |

Note. — Downlinked is the maximum number of images that can be downlinked. XGal is the maximum number of deep extragalactic images that are collected if downlinks are spaced every 24 hours. ExoP is the maximum number of staring mode data frames using 2 second subarray that are collected between downlinks. $\mu$lens is the number of images collected if time between downlinks is spent just performing microlensing parallax observations. Microlensing observations take an average of 235 seconds per observation (including initial slew) which corresponds to 344 observations per downlink period with 6 images per each observation.

high data volume, long duration observations of sources like 55 Cnc.

# 6 Thermal Properties

One potential concern during extended operations is changes in the thermal properties of the observatory due to light shining on surfaces that have not been illuminated at lower pitch angles. Thermal variations could manifest in three ways: 1) Heat input to the cold assembly, 2) Excess heating of the spacecraft bus and other warm components of the observatory, and 3) fluctuations in the pointing stability of the observatory due to thermal changes, Based on our current knowledge of the thermal stability and the surfaces likely to be illuminated at the higher pitch angles during downlinks in 2019 and 2020, it is deemed highly unlikely that there will be heat input to the cold assembly and it is expected that the temperatures of the telescope and instrument focal plane arrays will remain as they have since the start of post-cryogenic operations. Secondly, the temperatures of the warm components are likewise expected to remain within limits. Variations in pointing due to thermal fluctuations are discussed in Section 7.

# 7 Pointing Variations and Considerations for High-Precision Photometry

*Spitzer* has several types of pointing variations during high-precision staring mode observations. These pointing variations coupled with intra-pixel sensitivity variations are the dominant systematic in the high-precision observations and need to be decorrelated. Grillmair et al. (2014) provide some details of the types of pointing variations seen in staring mode observations. There are two basic types of pointing variation, drift which is roughly linear in each axis and wobble which had originally manifested as a sawtooth pattern with a period of $\sim 60$ minutes. The wobble was traced to a cycling of a heater used to keep the battery at its operating temperature when it was not being used. With increased pitch angle during downlinks, the battery heater is being used less as the battery when discharging is self-heating. As a result, the wobble has changed behavior and is much less signficant in amplitude. The self-heating of the battery does not produce the same thermo-mechanical variation that the heater cycling did. Figure 4 displays the distribution of wobble frequency and amplitude throughout the post-cryogenic mission. For 2016 and 2017, the wobble is almost non-existent. The large spread in the measured periods is due to the low amplitude of the wobble. Our expectation is that pointing wobble will not be a significant systematic in 2018-2020.

The dominant term in the pointing drift is due to an idiosyncrasy of the pointing control system; the correction for stellar aberration is not updated during a staring mode observation in response to the motion of *Spitzer* in its orbit. The resulting drift is $\sim 0.36\ arcsec\ day^{-1}$ in the -y direction on an image. Thermal relaxation of components between the telescope boresight and the star tracker providing the pointing reference also contributes drift. The largest variation in heating of the spacecraft bus occurs when there is a large change in spacecraft pitch angle such as slewing from a downlink to a science target. Figure 5 displays the distribution of measured drifts as a function of time. Most of the time the drift is the expected -0.36 arcsec day in the y-direction. There is some drift in the -x direction as well for 2017.

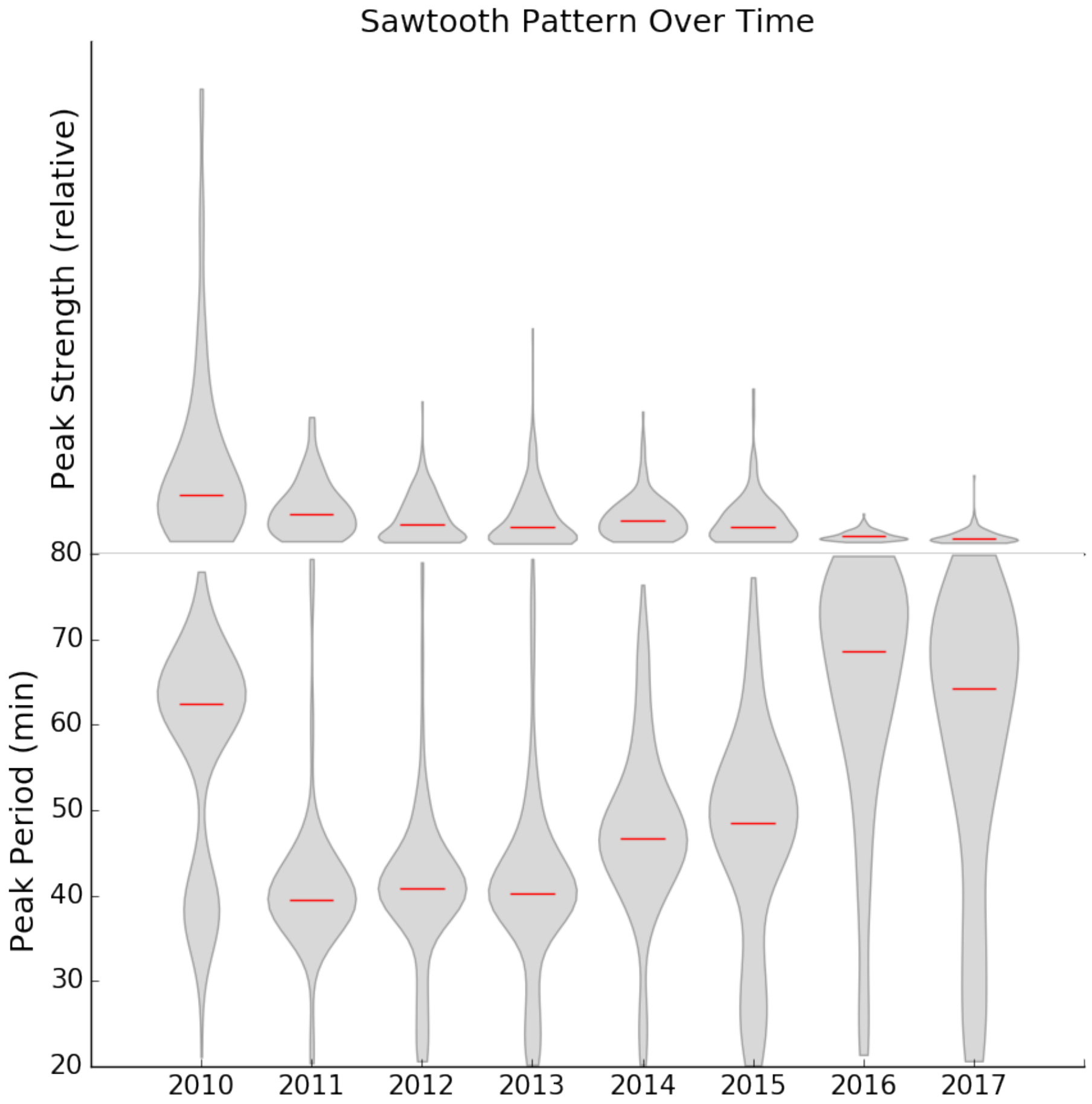

Figure 4 Distribution as a function of calendar year of the amplitude (top) and period (bottom) of the pointing wobble in all the post-cryogenic staring mode observations. The periods and amplitudes were measured by applying a Lomb-Scargle Periodgram to the measured centroids after fitting the long-term drift. The red lines are the median amplitude and period and the width of the grey shaded area gives the likelihood of a given value.

To test the pointing stability for operations in 2018, we performed a single staring mode observation with duration 10 hours immediately after a 3 hour dwell at a pitch angle of 48.5 degrees. The idea was to mimic a downlink at the most extreme pitch angle that would be used in 2018. Similar to the 2017 results, an increased drift in the -x direction was seen. The resulting light curve was decorrelated to near-photon limited precision using pixel-level decorrelation (Deming et al., 2015). If there is increased drift in 2019/2020, we expect that the systematic can be removed through decorrelation and an increase in the frequency of the re-acquisition peakup used. Currently for observations longer than 12-16 hours, the source is reacquired using the PCRS peakup mode to recenter the source on the IRAC sweet-spot and effectively remove the pointing drift (Ingalls et al., 2012; Krick et al., 2017).

## 8 Instrument Performance

IRAC has been remarkably stable throughout both the cryogenic and warm missions. The bias, gain and photometric zero-point have shown no significant variations. There is a slight (0.1% per year decrease in the IRAC 3.6 $\mu$m throughput which is probably due to increased Rayleigh scattering in the camera lenses due to radiation damage. Figures 6 and 7 display the photometric calibration stability. Figure 8 shows the excellent bias stability of the arrays. The gain maps have not varied to our ability to measure them ($< 0.1\%$).

## 9 Consumables and Lifetime Analysis

All components of the spacecraft and instrument that were designed to be redundant are still redundant. Only the IRAC focal plane arrays and associated cold electronics are single string. All primary systems are entirely nominal and have been extremely stable. The only consumable is the $N_2$ gas that *Spitzer* uses as propellant for the reaction control system which is used to dump excess momentum from the reaction wheels. Currently $>$ 7 kg of the initial 15 kg supply are available.

## 10 Summary

Continued operation of *Spitzer* through 2020 is viable. The hardware and ground system can support continued operations without any degradation in data quality. The most serious constraint are the decreased time that can be spent downlinking data due to the spacecraft being power negative at downlink orientation in 2017-2020 with increased drain on the battery with increasing date. In 2020, the downlink rate will also be cut in half due to the increased distance from Earth. Even with a smaller volume of data that will be able to be downlinked, *Spitzer* operations will support the most compelling science cases: studies of transiting and microlensing exoplanets, synoptic monitoring of brown dwarfs and improved parallax measurements, characterization of Near-Earth objects, properties and structure of the Milky Way galaxy, and deep surveys of the extragalactic sky.

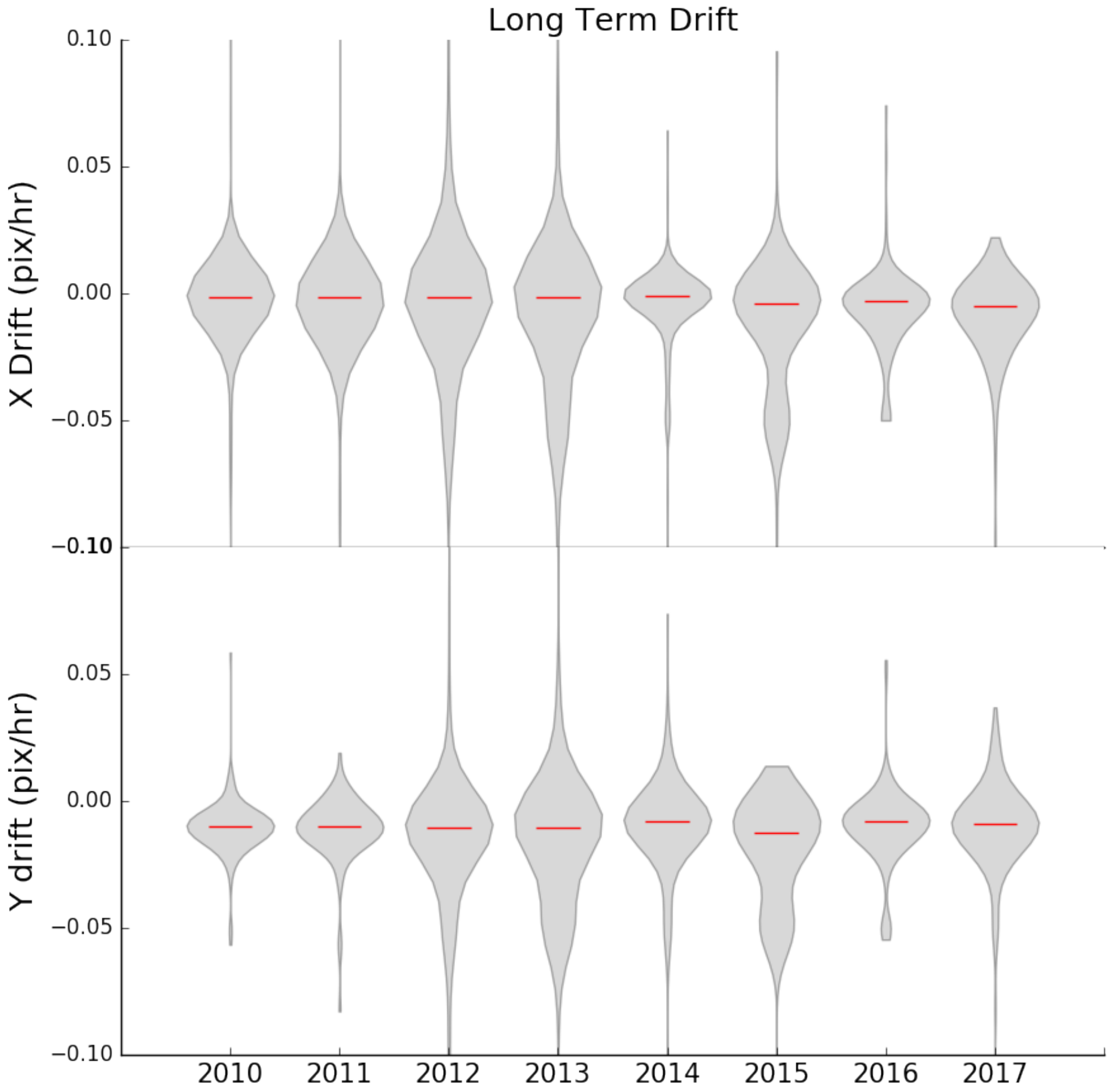

Figure 5 Distribution as a function of calendar year of the long-term drift for all staring mode observations taken in the post-cryogenic mission. The long-term drift is determined by fitting slopes to the centroid positions of all data after the first 30 minutes of an observation. The red lines are the median drift values. The width of the grey shaded area shows the likelihood of a given drift value. The drift in the x-direction is slightly more negative for the most recent year and is likely due to increased thermal variation. The drift in the y-direction is still dominated by the correction for the aberration of starlight.

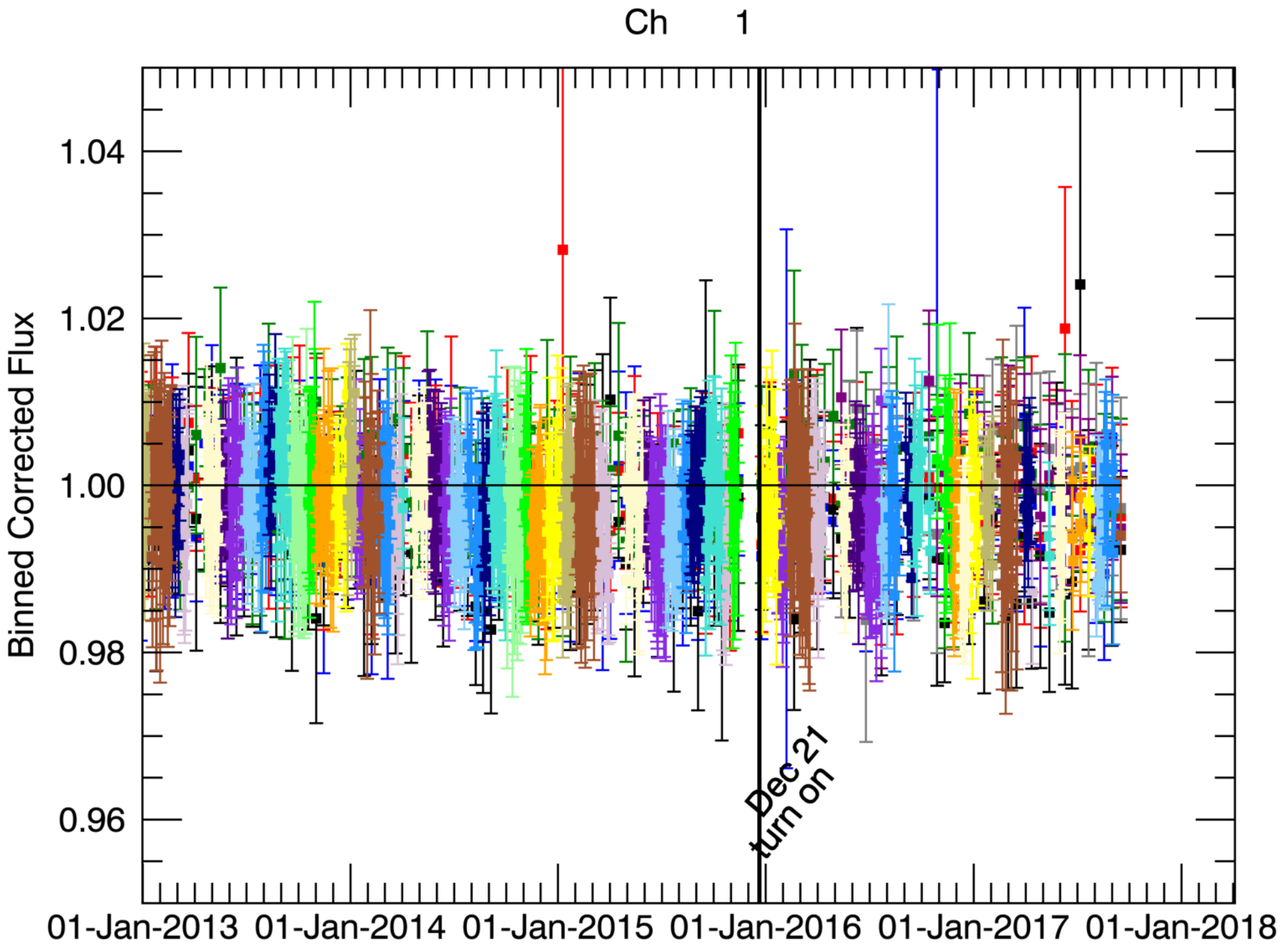

Figure 6 Normalized 3.6 $\mu$m measurements of all primary and secondary calibrators for the entirety of the post-cryogenic mission. Each individual star is normalized to the median value for that star.

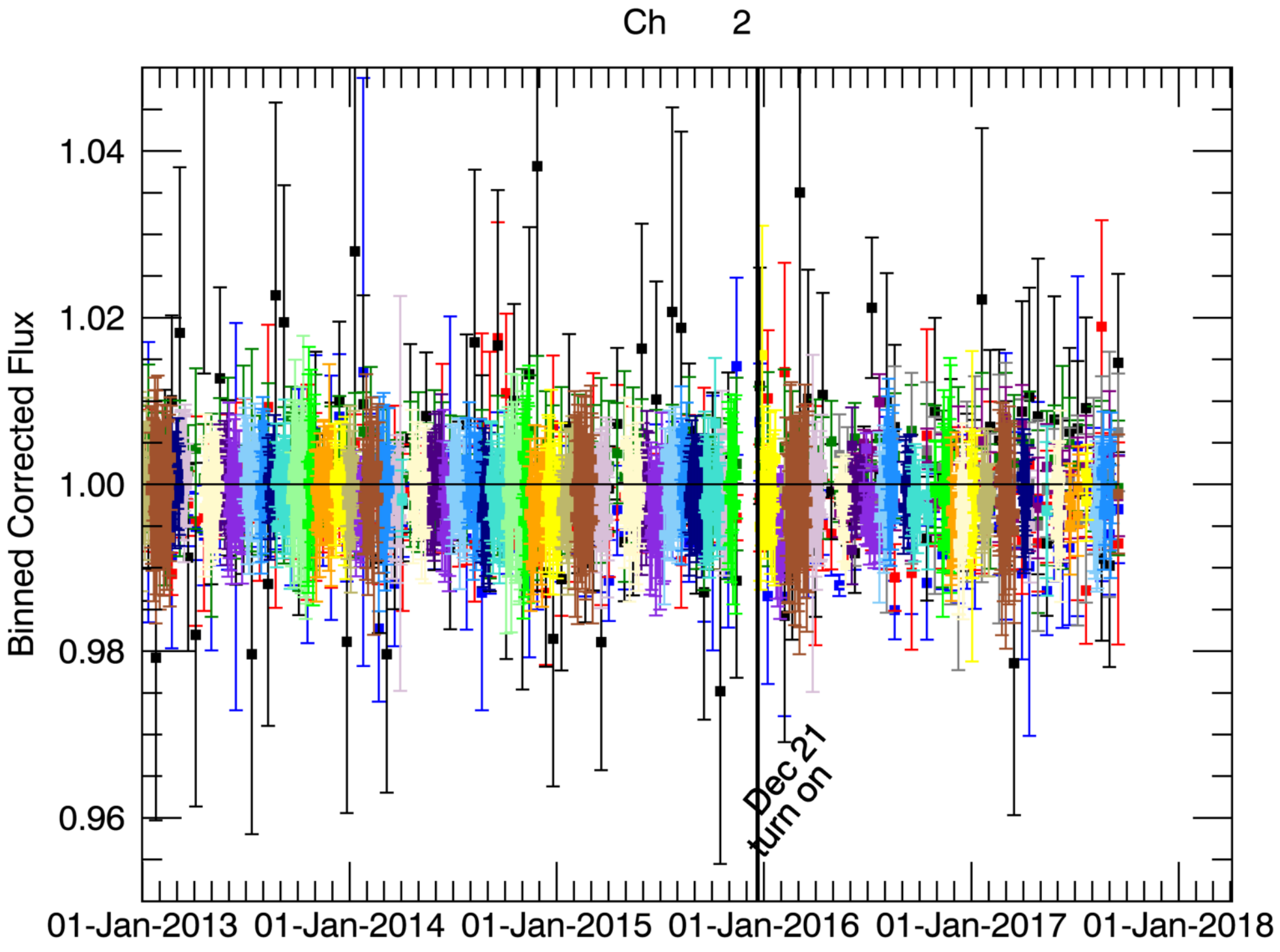

Figure 7 Normalized 4.5 $\mu$m measurements of all primary and secondary calibrators for the entirety of the post-cryogenic mission. Each individual star is normalized to the median value for that star.

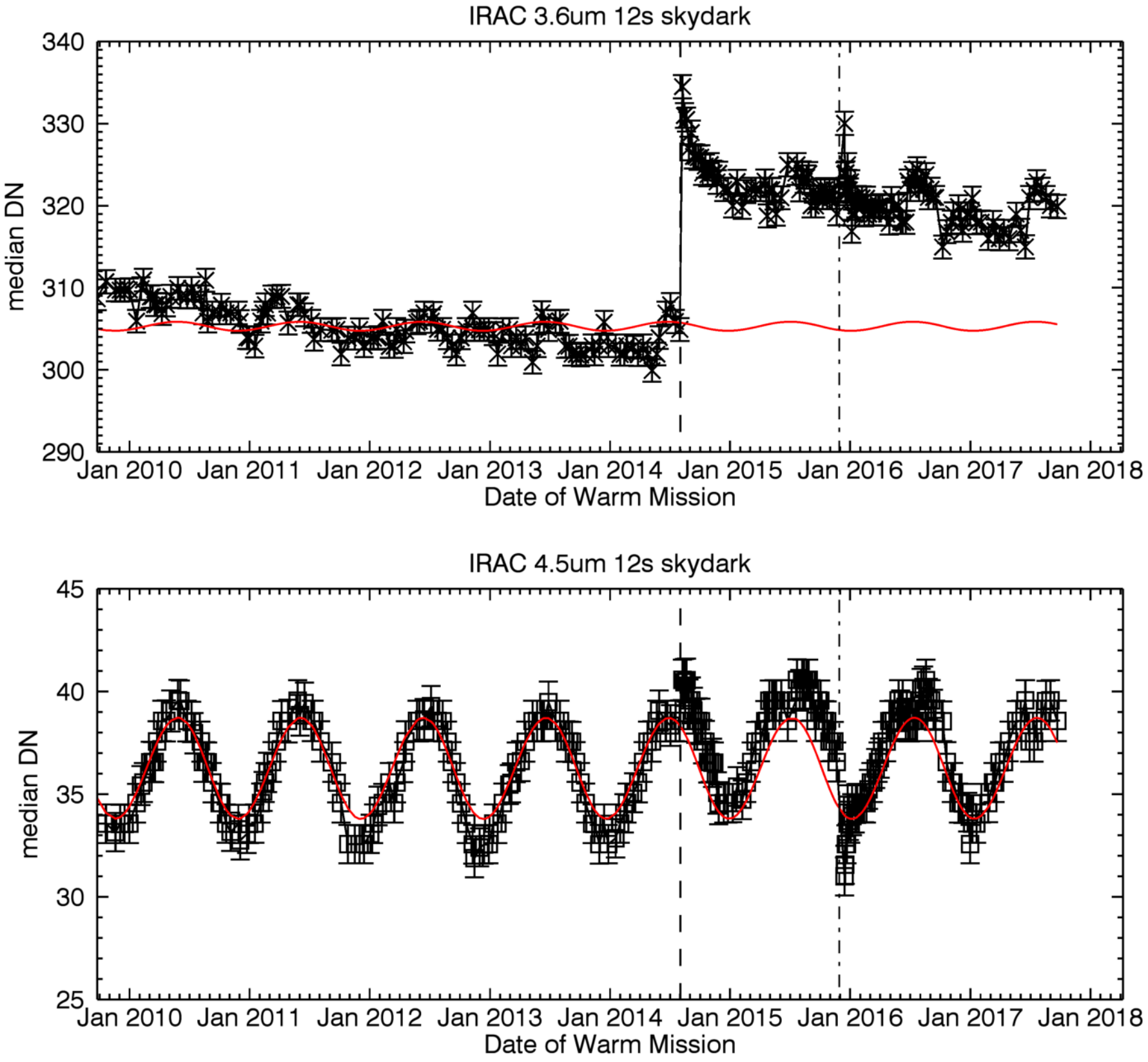

Figure 8 The median skydark level across the 3.6 $\mu$m array (top) and the 4.5 $\mu$m array (bottom) throughout the post-cryogenic mission. The red line in each figure is the contribution to the skydark due to Zodiacal light at the North Ecliptic Pole. The Zodiacal light contribution varies with time due to Spitzer's orbit through the Zodiacal dust cloud. The small variation in the 3.6 $\mu$m bias is due to the voltage settings for the array being re-initialized at the beginning of August 2014 with a smaller jump due to IRAC being power cycled in December 2015.